\documentclass{jfm}

\usepackage{graphicx}
\usepackage{natbib}
\usepackage{math_pack}
\usepackage{color}
\usepackage{subfigure}
\usepackage{amsmath}
\usepackage{amssymb}
\usepackage{math_pack}

\ifCUPmtlplainloaded \else
  \checkfont{eurm10}
  \iffontfound
    \IfFileExists{upmath.sty}
      {\typeout{^^JFound AMS Euler Roman fonts on the system,
                   using the 'upmath' package.^^J}%
       \usepackage{upmath}}
      {\typeout{^^JFound AMS Euler Roman fonts on the system, but you
                   dont seem to have the}%
       \typeout{'upmath' package installed. JFM.cls can take advantage
                 of these fonts,^^Jif you use 'upmath' package.^^J}%
       \providecommand\upi{\pi}%
      }
  \else
    \providecommand\upi{\pi}%
  \fi
\fi

\ifCUPmtlplainloaded \else
  \checkfont{msam10}
  \iffontfound
    \IfFileExists{amssymb.sty}
      {\typeout{^^JFound AMS Symbol fonts on the system, using the
                'amssymb' package.^^J}%
       \usepackage{amssymb}%
         \let\leq=\leqslant
         
      }{}
  \fi
\fi

\ifCUPmtlplainloaded \else
  \IfFileExists{amsbsy.sty}
    {\typeout{^^JFound the 'amsbsy' package on the system, using it.^^J}%
     \usepackage{amsbsy}}
    {\providecommand\boldsymbol[1]{\mbox{\boldmath $##1$}}}
\fi

\newcommand\omegab{\boldsymbol{\omega}}
\newcommand\taub{\boldsymbol{\tau}}

\newsavebox{\astrutbox}
\sbox{\astrutbox}{\rule[-5pt]{0pt}{20pt}}

\definecolor{cinnamon}{rgb}{0.82, 0.41, 0.12}
\title[Sedimentation of finite-size spheres]{Sedimentation of finite-size spheres in quiescent and turbulent environments}

\author[W. Fornari, F. Picano and L. Brandt]%
{Walter Fornari$^1$%
  \thanks{Email address for correspondence: fornari@mech.kth.se},\ns
Francesco Picano$^2$\break
and Luca Brandt$^1$}

\affiliation{$^1$Linn\'e Flow Centre and Swedish e-Science Research Centre (SeRC), KTH Mechanics,
SE-10044 Stockholm, Sweden\\[\affilskip]
$^2$Department of Industrial Engineering, University of Padova, Via Venezia 1, 35131 Padua, Italy}

\pubyear{2010}
\volume{650}
\pagerange{119--126}
\date{?; revised ?; accepted ?. - To be entered by editorial office}
\begin{document}

\maketitle

\begin{abstract}
Sedimentation of a dispersed solid phase is widely encountered in applications and environmental flows, yet
little is known about the behavior of finite-size particles in
homogeneous isotropic turbulence. 
To fill this gap, we perform Direct Numerical Simulations of sedimentation in quiescent and turbulent environments using an Immersed Boundary Method to account
for the dispersed rigid spherical particles. The solid volume fractions considered are $\phi=0.5-1\%$,
while the solid to fluid density ratio $\rho_p/\rho_f=1.02$.
The particle radius is chosen to be approximately 6 Komlogorov lengthscales. 
The results show that
the mean settling velocity is lower in an already turbulent flow than in a quiescent fluid. 
The reduction with respect to a single particle in quiescent fluid is about 12\% and 14\%  for the two volume fractions investigated. 
The probability density function of the particle velocity is almost Gaussian in a turbulent flow, whereas it displays large positive tails in quiescent fluid. These tails are associated to the intermittent fast sedimentation of particle pairs in drafting-kissing-tumbling motions.
The particle lateral dispersion is higher in a turbulent flow, whereas the vertical one is, surprisingly, of comparable magnitude as a consequence of the highly intermittent behavior observed in the quiescent fluid.
Using the concept of mean relative velocity we estimate the mean drag coefficient from empirical formulas and show that non stationary  effects, related to vortex shedding, explain the increased reduction in mean settling velocity in a turbulent environment.

\end{abstract}


\section{Introduction}
The gravity-driven motion of solid particles in a viscous fluid is a relevant process 
in a wide number of scientific and engineering applications \citep{guazzelli2011}. Among these we recall fluvial
geomorphology, chemical engineering systems, as well as pollutant transport in underground water and settling
of micro-organisms such as plankton.\\
The general problem of sedimentation is very complex  due to the high number of factors from
which it depends. Sedimentation involves large numbers of particles
settling in different environments. The fluid in which the particles are suspended may be
quiescent or turbulent. Particles
may differ in size, shape, density and stiffness. The range of spatial and temporal scales involved is wide and the global
properties of these suspensions can be substantially different from one case to another. Because of these
complexities, our general understanding of the problem is still incomplete.\\
\subsection{Settling in a quiescent fluid}
One of the earliest investigations on the subject at hand is Stokes' analysis of the sedimentation of a
single rigid sphere through an unbounded quiescent viscous fluid at zero Reynolds number. This led to the
well-known formula that links the settling velocity to the sphere radius, the solid to fluid density ratio
and the viscosity of the fluid that bears his name. 
Later, the problem was studied both theoretically and experimentally.
\citet{hasimoto1959} obtained expressions for the drag force exerted by the fluid on three different cubic
arrays of rigid spheres. These relate the drag force only to the solid volume fraction, but were
derived under the assumption of very dilute suspensions and Stokes flow. The formulae have later been revisited
by \citet{sangani1982}. 
A different approach was instead pursued by \citet{batchelor1972} who found a relation between
the mean settling velocity and the solid volume fraction by using conditional probability arguments.
When the Reynolds number of the settling particles ($Re_t$) becomes finite,
the assumption of Stokes flow is less acceptable (especially for $Re_t > 1$). The fore-aft symmetry of
the fluid flow around the particles is broken and wakes form behind them. Solutions should be derived using the
Navier-Stokes equations, but the nonlinearity of the inertial term makes the analytical treatment of such problems
extremely difficult. For this reason theoretical investigations have progressively given way to experimental and
numerical approaches.\\
The first remarkable experimental results obtained for creeping flow were those by 
\citet{richardson1954}. These authors proposed an empirical formula relating the mean settling velocity of a suspension
to its volume fraction and to the settling velocity of an isolated particle. This formula is believed to be accurate
also for concentrated suspensions (up to a volume fraction $\phi$ of about $25\%$) and for low Reynolds numbers. Subsequent
investigations improved the formula so that it could also be applied in the intermediate Reynolds numbers regime
\citep{garside1977,di1999}.\\
Efficient algorithms and sufficient
computational power have become available only relatively recently and since then many different numerical methods have been used to improve our understanding of the problem \citep{AndreaFoF}.
Among others we recall the dynamical simulations performed by \citet{ladd1993}, the finite-elements simulations of
\citet{johnson1996}, the force-coupling method simulations by \citet{climent2003}, the
Lattice-Boltzmann simulations of \citet{yin2007}, the Oseenlet simulations by \citet{pignatel2011}, and the
Immersed Boundary simulations of \citet{kempe2012} and \citet{uhlmann2014}. Thanks to the most recent techniques 
it has become feasible to gain more insight on the interactions among the different phases and the resulting microstructure of the
sedimenting suspension \citep{yin2007,uhlmann2014}. \citet{uhlmann2014}, most recently, simulated the settling of dilute suspensions with particle Reynolds numbers in the range $140-260$ and
studied the effect of the Archimedes number (namely the ratio between gravitational and viscous forces) on the microscopic
and macroscopic properties of the suspension. These authors observe an increase of the settling velocity at higher Archimedes, owing to particle clustering in a regime 
where the flow undergoes a steady bifurcation to an asymmetric wake.
Settling in stratified environments has also been investigated experimentally, i.e. by \citet{bush2003}, and numerically, i.e. by \citet{doost2015}.
\subsection{Sedimentation in an already turbulent flow}
The investigations previously reported consider the settling of particles in quiescent or uniform flows. There are
many situations though, where the ambient fluid is in fact nonuniform or turbulent. As in the previous case, the first
approach to this problem was analytical. In the late 40's and 50's \citet{tchen1947} and later \citet{corrsin1956} 
proposed an equation for the motion of a small rigid sphere settling in a nonuniform flow. In the derivation,
they assumed the particle Reynolds number to be very low so that the viscous Stokes drag for a sphere could be applied.
The added mass and the augmented viscous drag due to a Basset history term were also included. \citet{maxey1983} 
corrected these equations including also the Faxen forces due to the unsteady Stokes flow.\\
In a turbulent flow many different spatial and temporal scales are active.
Therefore the behaviour and motion of one single particle does not depend only on its dimensions and characteristic response time,
but also on the ratios among these and the characteristic turbulent length and time scales. The turbulent quantities
usually considered are the Kolmogorov length and time scales which are related to the smallest eddies. Alternatively,
the integral lengthscale and the eddy turnover time can also be used. It is clear that a
 particle smaller than the
Kolmogorov legthscale will behave differently than a particle of size comparable to the energetic flow structures. A sufficiently large particle with a characteristic time scale 
larger than the timescale of the velocity fluctuations will definitely be affected by and affect  the turbulence. 
A smaller particle with a shorter relaxation time will more closely follow
the turbulent fluctuations. When particle suspensions are considered, the situation becomes even more complicated. 
If the particles are solid, smaller than the Kolmogorov lengthscale and dilute, the turbulent flow
field is unaltered (i.e., one-way coupling). Interestingly, the turbulent dynamics is instead altered by microbubbles.  The presence of these microbubbles leads to relevant drag reduction  in boundary layers and shears flows (e.g.\ 
Taylor-Couette flow)\citep{sugi2008,ceccio2010}.
If the mass of the dispersed phase is similar to that of the carrier
phase, the influence of the solid phase on the fluid phase cannot be ignored (i.e., two-way coupling).  Interactions among particles (such as collisions) must also be considered in
concentrated suspensions. This
last regime is described as four-way coupling \citep{elgo1991,balach2010}.

Because of the difficulty of treating the problem analytically, the investigations of the last three
decades have mostly been either experimental or numerical. In most of the numerical studies heavy and small particles were
considered. The reader is referred to \citet{toschi2009} for a more detailed review than the short summary reported here. \citet{wang1993} studied the settling of dilute heavy particles in homogeneous
isotropic turbulence. The particle Reynolds number based on the relative velocity was assumed to be much less than unity
so that Stokes drag force could be used to determine the particle motion. These authors show that heavy particles smaller than the Kolmogorov
lengthscale tend to move outward from the center of eddies and are often swept into regions of downdrafts (the
so called preferential sweeping later renamed fast-tracking). In doing so, the particle mean settling velocity is 
increased with respect to that in a  quiescent fluid. A series of studies 
confirmed and extended these results examining particle clustering \citep{bec2014,gustavsson2014}, preferential concentration 
\citep{aliseda2002}, the effects of the particle shape, orientation and collision rates \citep{siewert2014}, as well the effects of one- or two-way coupling algorithms \citep{bosse2006},   
to mention few aspects. Numerous experimental studies were also performed in order to confirm these results and to study the turbulence
modulation due to the presence of particles \citep{hwang2006}.

The results on the mean settling velocities of particles of the order or larger than the Kolmogorov scale are 
not conclusive. \citet{good2014} studied particles smaller than the Kolmogorov scale and with density ratio $\mathcal O(1000)$, whereas 
Variano (experiments; private communication) and \citet{byron2015} studied finite-size particles at density ratios comparable to ours. \citet{good2014} found that 
the mean settling velocity is reduced only when nonlinear drag corrections are considered in a one-way coupling approach when particles 
have a long relaxation time (a linear drag force would always lead to a settling velocity enhancement). For finite-size almost neutrally-buoyant 
particles, Variano and \citet{byron2015} observe instead that the mean settling velocity is smaller than in a quiescent fluid. 
In relative terms, the settling velocity decreases more and more as the ratio between the turbulence fluctuations and the terminal velocity of a single particle in a quiescent fluid increases. 
It is generally believed that the reduction of settling speed is due to the non-linear relation between the particle drag and the Reynolds number.
Nonetheless, unsteady and history effects may also play a key role \citep{olivieri2014,bergougnoux2014motion}. 
\citet{tunstall1968} demonstrated already in
1968 that the average settling velocity is reduced in a flow oscillating about a zero mean, due to the interactions of the particle
inertia with a non-linear drag force. \citet{stout1995} tried to motivate these findings in terms of the relative motion 
between the fluid and the particles. When the period of the fluid velocity fluctuations is smaller than the particle response 
time, a significant relative motion is generated between the two phases. Due to the drag non-linearity, appreciable 
upward forces can be produced on the particles thereby reducing the mean settling velocity.\\
Unsteady effects may become important when considering suspensions with 
moderate particle-fluid density ratios, as suggested by \citet{mordant2000} and \citet*{sobral2007}. The 
former studied experimentally the motion of a solid sphere settling in a quiescent fluid
and explain the transitory oscillations of the settling velocity found at $Re \approx O(100)$ by the 
presence of a transient vortex shedding in the particle wake. The latter, instead, analyzed an equation similar 
to that proposed by \citet{maxey1983}, and suggested that unsteady hydrodynamic drags might become 
important when the density ratio approaches unity.

\subsection{Fully resolved simulations}

As already mentioned, most of the numerical studies of settling in turbulent flows used either one or two-way coupling algorithms. In order to properly
understand the microscopical phenomena at play, it would be ideal to use fully resolved simulations. An algorithm often used to accomplish this is the 
Immersed Boundary Method with direct forcing for the simulation of particulate flows 
originally developed by \citet{uhlmann2005}. 
The code was later used to study the clustering of non-colloidal particles settling in a quiescent environment \citep{uhlmann2014}. With a similar method \citet{lucci2010} studied the 
modulation of isotropic turbulence by particles of Taylor length-scale size.
Recently, \citet{homann2013} used an Immersed Boundary
Fourier-spectral method to study finite-size effects on the dynamics of single particles in turbulent flows. These authors found that the drag force on a particle suspended in 
a turbulent flow increases as a function of the turbulent intensity and the particle Reynolds number.  We recently used a similar algorithm to examine  turbulent channel flows of particle suspensions \citep*{picano2015}.

The  aim of the present study is to  
simulate the sedimentation of a
suspension of particles larger than the Kolmogorov lengthscale in homogeneous isotropic turbulence with a
finite difference Immersed Boundary Method. We focus
on particles slightly denser than the suspending fluid ($\rho_p/\rho_f=1.02$) and investigate particle and fluid
velocity statistics, non-linear and unsteady contributions to the overall drag and turbulence modulation. The suspensions
considered in this work are dilute ($\phi=0.5-1\%$) and monodispersed. 
The same simulations are also performed in absence of turbulence to appreciate differences
of the particle velocity statistics in the two different environments. Due to the size of the particles considered it has
been necessary to consider very long computational domains in the settling direction, especially for the quiescent environment.
In the turbulent cases, smaller domains provide converged statistics since the particle wakes
are disrupted faster.
The parameters of the simulations have been inspired by the  
experiments by Variano, \citet{byron2015} and co-workers at UC Berkeley. These authors investigate
 Taylor-scale particles in turbulent aquatic environments using Refractive-Index-Matched Hydrogel particles to measure particle linear and angular velocities.

Our results  show that the mean settling velocity is lower in an already turbulent flow than in a quiescent fluid. The reduction is
about $12\%$ and $14\%$  for the two volume fractions investigated. 
By looking at probability density functions $pdf$ of the settling velocities, we observe that
the $pdf$ is well approximated by a Gaussian function centered around the mean in the turbulent cases. In the
laminar case instead, the $pdf$ shows a smaller variance and a larger skewness, indicating that it is more probable to find
particles settling more rapidly than the mean value rather than more slowly. These events are associated to particle-particle interactions, in particular to the drifting-kissing-tumbling motion of particle pairs. We also calculate mean relative velocity fields
and notice that vortex shedding occurs around each particle in a turbulent environment. Using the concept of mean relative velocity we
calculate a local Reynolds number and the mean drag coefficient from empirical formulas to quantify the
importance of unsteady and history effects on the overall drag, thereby explaining the reduction in mean settling velocity. In
fact, these terms become important only in a turbulent environment.

\section{Methodology}\label{sec:method}
\subsection{Numerical Algorithm}
Different methods have been proposed in the last years to perform Direct Numerical Simulations of multiphase flows.
The Lagrangian-Eulerian algorithms are believed to be the most appropriate for solid-fluid suspensions {\citep{ladd2001,zhang2010,lucci2010,uhlmann2014}. In the present study, simulations
have been performed using a tri-periodic version of the numerical code originally developed by \citet{breugem2012} that models the coupling between the solid and fluid phases. The Eulerian fluid phase is evolved according to the incompressible
Navier-Stokes equations,
\begin{equation}
\label{div_f}
\div \vec u_f = 0
\end{equation}
\begin{equation}
\label{NS_f}
\pd{\vec u_f}{t} + \vec u_f \cdot \grad \vec u_f = -\frac{1}{\rho_f}\grad p + \nu \grad^2 \vec u_f + \vec f
\end{equation}
where $\vec u_f$, $\rho_f$ and $\nu=\mu/\rho_f$ are the fluid velocity, density and kinematic viscosity respectively ($\mu$ is
the dynamic viscosity), while $p$ and $\vec f$ are the pressure and the force field used to mantain turbulence and model the presence of particles. The particles centroid
linear and angular velocities, $\vec u_p$ and $\vec \omegab_p$ are instead governed by the Newton-Euler Lagrangian equations,
\begin{align}
\label{lin-vel}
\rho_p V_p \td{\vec u_p}{t} &= \rho_f \oint_{\partial \mathcal{V}_p}^{} \vec \taub \cdot \vec n\, dS + \left(\rho_p - \rho_f\right) V_p \vec g\\
\label{ang-vel}
I_p \td{\vec \omegab_p}{t} &= \rho_f \oint_{\partial \mathcal{V}_p}^{} \vec r \times \vec \taub \cdot \vec n\, dS
\end{align}
where $V_p = 4\upi a^3/3$ and $I_p=2 \rho_p V_p a^2/5$ are the particle volume and moment of inertia, with $a$ the particle radius;
$\vec g$ is the gravitational acceleration;
$\vec \taub = -p \vec I + 2\mu \vec E$ is the fluid stress, with $\vec I$ the identity matrix and $\vec E = \left(\grad \vec u_f + \grad \vec u_f^T \right)/2$ the
deformation tensor; $\vec r$ is the distance vector from the center of the sphere while $\bf{n}$ is the unit vector normal to the
particle surface $\partial \mathcal{V}_p$. Dirichlet boundary conditions for the fluid phase are enforced on the particle
surfaces as $\vec u_f|_{\partial \mathcal{V}_p} = \vec u_p + \vec \omegab_p \times \vec r$.\\
In the numerical code the coupling between the solid and fluid phases is obtained by using an Immersed Boundary Method. The boundary
condition at the moving particle surface (i.e. $\vec u_f|_{\partial \mathcal{V}_p} = \vec u_p + \vec \omegab_p \times \vec r$) is
modeled by adding a force field on the right-hand side of the Navier-Stokes equations. The problem of re-meshing is therefore avoided
and the fluid phase is evolved in the whole computational domain using a second order finite difference scheme on a staggered mesh. The
time integration is performed by a third order Runge-Kutta scheme combined with a pressure-correction method at each sub-step. The
same integration scheme is also used for the Lagrangian evolution of eqs.~(\ref{lin-vel}) and (\ref{ang-vel}). The forces exchanged by
the fluid and the particles are imposed on $N_L$ Lagrangian points uniformly distributed on the particle surface. The force $\vec F_l$
acting on the $l-th$ Lagrangian point is related to the Eulerian force field $\vec f$ by the expression $\vec f(\vec x) = \sum_{l=1}^{N_L} 
\vec F_l \delta_d(\vec x - \vec X_l) \Delta V_l$. In the latter $\Delta V_l$ represents the volume of the cell containing the $l-th$
Lagrangian point while $\delta_d$ is the Dirac delta. This force field is obtained through an iterative algorithm that mantains 
second order global accuracy in space. Using this IBM force field eqs.~(\ref{lin-vel}) and ~(\ref{ang-vel}) are rearranged as follows 
to maintain accuracy,
\begin{align}
\label{lin-vel-ibm}
\rho_p V_p \td{\vec u_p}{t} &= -\rho_f \sum_{l=1}^{N_l} \vec F_l \Delta V_l + \rho_f \td{}{t} \int_{\mathcal{V}_p}^{} \vec u_f\, dV + \left(\rho_p - \rho_f\right) V_p \vec g \\
\label{ang-vel-ibm}
I_p \td{\vec \omegab_p}{t} &= -\rho_f \sum_{l=1}^{N_l} \vec r_l \times \vec F_l \Delta V_l + \rho_f \td{}{t} \int_{\mathcal{V}_p}^{} \vec r \times \vec u_f\, dV 
\end{align}
where the second terms on the right-hand sides are corrections to account for the inertia of the fictitious fluid contained
within the particle volume. In eqs.~(\ref{lin-vel-ibm}),(\ref{ang-vel-ibm}) $\vec r_l$ is simply the distance from the center of a particle.
Particle-particle interactions are also considered. When the gap distance between two particles is smaller
than twice the mesh size, lubrication models based on Brenner's asymptotic solution \citep{brenner1961} are used to correctly
reproduce the interaction between the particles. A soft-collision model is used to account for collisions among particles with
an almost elastic rebound (the restitution coefficient is $0.97$). These lubrication and collision forces are
added to the right-hand side of eq.~(\ref{lin-vel-ibm}). More details and validations of the numerical code can be found in \citet{breugem2012}, \citet{lambert2013},
\citet{lashgari2014} and \citet{picano2015}.
 
\subsection{Parameter setting}
Sedimentation of dilute particle suspensions is considered in an unbounded computational domain with periodic
boundary conditions in the $x$, $y$ and $z$ directions. Gravity is chosen to act in the positive $z$ direction. 
A zero mass flux is imposed in the simulations. A cubic
mesh is used to discretise the computational domain with eight points per particle radius, $a$. Non-Brownian 
rigid spheres are considered with solid to fluid density ratio $\rho_p/\rho_f=1.02$. Hence, we consider
particles slightly heavier than the suspending fluid. 
Two different solid volume fractions $\phi=0.5\%$ and $1\%$ are considered. In addition 
to the solid to fluid density ratio $\rho_p/\rho_f$ and the solid volume fraction $\phi$, it is necessary to introduce another 
nondimensional number. This is the Archimedes number (or alternatively the Galileo number $Ga=\sqrt{Ar}$),
\begin{equation}
\label{arch}
Ar = \frac{\left(\frac{\rho_p}{\rho_f}-1\right) g (2a)^3}{\nu^2}
\end{equation}
a nondimensional number that quantifies the importance of the gravitational forces
acting on the particle with respect to viscous forces. In the present case the Archimedes number  $Ar=21000$. 
Using  the particle terminal velocity $v_t$ we define the Reynolds
number $Re_t=2a v_t/\nu$. This can be related  by empirical relations to the drag coefficient of an
isolated sphere when varying the Archimedes number, $Ar$. Different versions of these empirical relations giving the drag coefficient as a function of $Re_t$ and $Ar$ have been proposed.
As \citet{yin2007} we will use the following relations,
\begin{equation}
\label{Cd_re}
  C_D = \left\{
    \begin{array}{ll}
      \frac{24}{Re_t}\left[1+0.1315 Re_t^{(0.82-0.05 log_{10} Re_t)}\right], & 0.01 < Re_t \leq 20\\[2pt]
      \frac{24}{Re_t}\left[1+0.1935 Re_t^{0.6305}\right], & 20 < Re_t <260
  \end{array} \right.
\end{equation}
since $C_D = {4}{Ar}/(3{Re_t^2})$ \citep{yin2007} we finally write,
\begin{equation}
\label{arc_re}
  Ar = \left\{
    \begin{array}{ll}
      18 Re_t\left[1+0.1315 Re_t^{(0.82-0.05 log_{10} Re_t)}\right], & 0.01 < Re_t \leq 20\\[2pt]
      18 Re_t\left[1+0.1935 Re_t^{0.6305}\right], & 20 < Re_t <260
  \end{array} \right.
\end{equation}
The Reynolds number calculated from eq.~(\ref{arc_re}) is approximately $188$ for $Ar=21000$.

In order to generate and sustain an isotropic and homogeneous turbulent flow field, a random forcing is applied to the first shell of wave 
vectors. We consider a $\delta$-correlated in time forcing of fixed amplitude $\hat f_0$ \citep{vincent1991,zhan2014}. The turbulent field, alone,  is characterized by a Reynolds number based on the 
Taylor microscale, $Re_{\lambda}=\lambda u'/\nu$, where $u'$ is the fluctuating velocity and $\lambda=\sqrt{15\nu u'^2/\epsilon}$ is the transverse Taylor length scale. This is about $90$ in our simulations. 
The ratio between the grid spacing 
and the Kolmogorov lengthscale $\eta=(\nu^3/\epsilon)^{1/4}$ (where $\epsilon$ is the energy dissipation) is approximately $1.3$ while 
the particle diameter is circa $12\eta$. Finally, the ratio among the expected mean settling velocity and the turbulent velocity fluctuations 
is $v_t/u'=3.4$. The parameters of the turbulent flow field are summarized in table~\ref{tab:turb}. For the definition of these parameters, 
the reader is referred to \citet{pope2000}.

\subsection{Validation}\label{sec_val}

\begin{figure}
  \centerline{\includegraphics[scale=0.4]{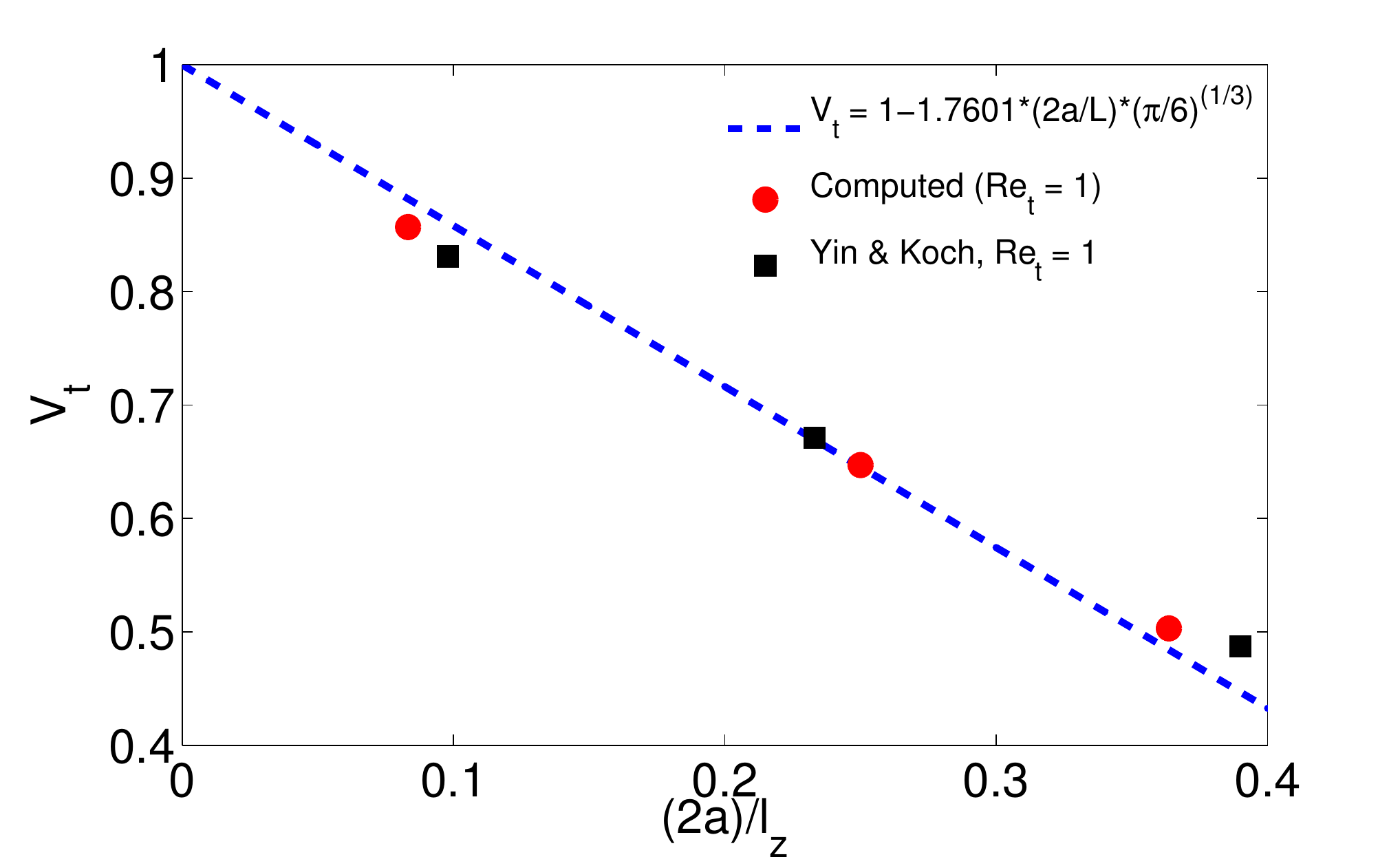}}%
  %\centerline{\includegraphics[scale=0.4]{FIG/acr_paper.eps}}%
  \caption{Terminal velocity of a periodic array of spheres.
  Present results $Re_t=1$ denoted by red circles are compared with the ones obtained by Yin and Koch by black 
squares at the same $Re_t$ and with the analytical solution for $Re_t=0$ of eq.~\eqref{acr_sol}.
\label{fig:acr_fig}}
\end{figure}

To check the validity of our approach we performed simulations of a single sphere settling in a cubic lattice in boxes of different sizes. Since
this is equivalent to changing the solid volume fraction, we compared our results to the analytical formula
derived by \citet{hasimoto1959} and \citet{sangani1982},
\begin{equation}
\label{acr_sol}
|\vec V_t| = \frac{|\vec v_t|}{|\vec V_s|} = 1-1.7601 \phi^{1/3}
\end{equation}
where
\begin{equation}
\label{stk-v}
|\vec V_s| = \frac{2}{9}\frac{\left(\frac{\rho_p}{\rho_f} -1\right) g a^2}{\nu}
\end{equation}
is the Stokes settling velocity. In terms of the size of the computational domain, $l_z$, eq.~(\ref{acr_sol}) can also be rewritten as
$V_t=1-1.7601(2a/l_z)(\upi/6)^{1/3}$. In figure~\ref{fig:acr_fig} we show the results obtained with our code for $Re_t=1$ together
with the data by \citet{yin2007} and the analytical solution. Although the analytical solution was
derived with the assumption of vanishing Reynolds numbers, we find good agreement among the various results.

The actual problem arises when considering particles settling at relatively high Reynolds numbers. If the computational box is not
sufficiently long in the gravity direction, a particle would fall inside its own wake (due to periodic boundary conditions), thereby accelerating unrealistically.
Various simulations of a single particle falling in boxes of different size were preliminarily carried out, in particular
$48a\times48a\times48a$, $32a\times32a\times96a$ and $32a\times32a\times320a$. The first two boxes turn out to be unsuitable for our
purposes. We find a terminal Reynolds number  $Re_t=200$ in the longest domain considered, which corresponds to a difference 
of about 6\%
with respect to the value obtained from the empirical relations (\ref{arc_re}).
As reference velocity we use the value obtained from simulations
performed in the largest box at
a solid volume fraction two orders of magnitude smaller than the cases under investigation, $\phi=5 \cdot 10^{-5}$ \cite[as in][]{uhlmann2014}, corresponding to a terminal velocity such that $Re_t=195$, 4\% larger than the value from the empirical relations (\ref{arc_re}).
Further increasing the length in the $z$ direction would
make the simulations
 prohibitive. 
Note also that simulations in a turbulent environment turn out to be less demanding as turbulence disrupts and decorrelates the flow structures induced by the particles.
The final choice was therefore a computational box
of size $32a\times32a\times320a$ with $256\times256\times2560$ grid points, $391$ particles for $\phi=0.5\%$ and $782$ particles for $\phi=1\%$. In all cases, the particles 
are initially distributed randomly in the computational volume with zero initial velocity and rotation.

\begin{table}
  \begin{center}
\def~{\hphantom{0}}
  \begin{tabular}{cccccccc}
      $\eta/(2a)$  & $u'$   &   $k$ & $\lambda/(2a)$ & $Re_{\lambda}$ & $\epsilon$ & $T_e$ & $Re_{L_0}$ \\[3pt]
       0.084   & 0.30 & 0.13 & 1.56 & 90 & 0.0028 & 46.86 & 1205 \\
  \end{tabular}
  \caption{Turbulent flow parameters in particle units, where $k$ is the turbulent kinetic energy, $\lambda$ is the Taylor microscale, $T_e=k/\epsilon$ is 
the eddy turnover time and $Re_{L_0}$ is the Reynolds number based on the integral lengthscale $L_0=k^{3/2}/\epsilon$.}
  \label{tab:turb}
  \end{center}
\end{table}

 A snapshot of the suspension flow for $\phi=0.5\%$ 
is shown in figure~\ref{fig:turb_in}. The instantaneous velocity component perpendicular to gravity is shown on different orthogonal 
planes.

The simulations were run on a Cray XE6 system at the PDC Center for High Performance Computing at the KTH, Royal Institue of Technology. The fluid phase is evolved for approximately $6$ eddy turnover times before adding the solid phase. The simulations 
for each solid volume fraction were performed for both quiescent fluid and turbulent flow cases in order to compare the results. Statistics are collected 
after an initial transient phase of about $4$ eddy turnover times for the turbulent case and $15$ relaxation times ($T_p=2\rho_pa^2/(9\rho_f\nu)$) for the quiescent case. 
Defining as reference time the time it takes for an isolated particle to fall over a distance equal to its diameter, $2a/v_t$, the initial transient corresponds to approximately $170$ units. Statistics are collected over a time interval of 500 and 300 in units of $2a/v_t$ for the quiescent and turbulent cases, respectively. Differences between the statistics presented here and those computed from half the samples is below 1\% for the first and second moments.

\begin{figure}
  \centerline{\includegraphics[scale=0.45,angle=270]{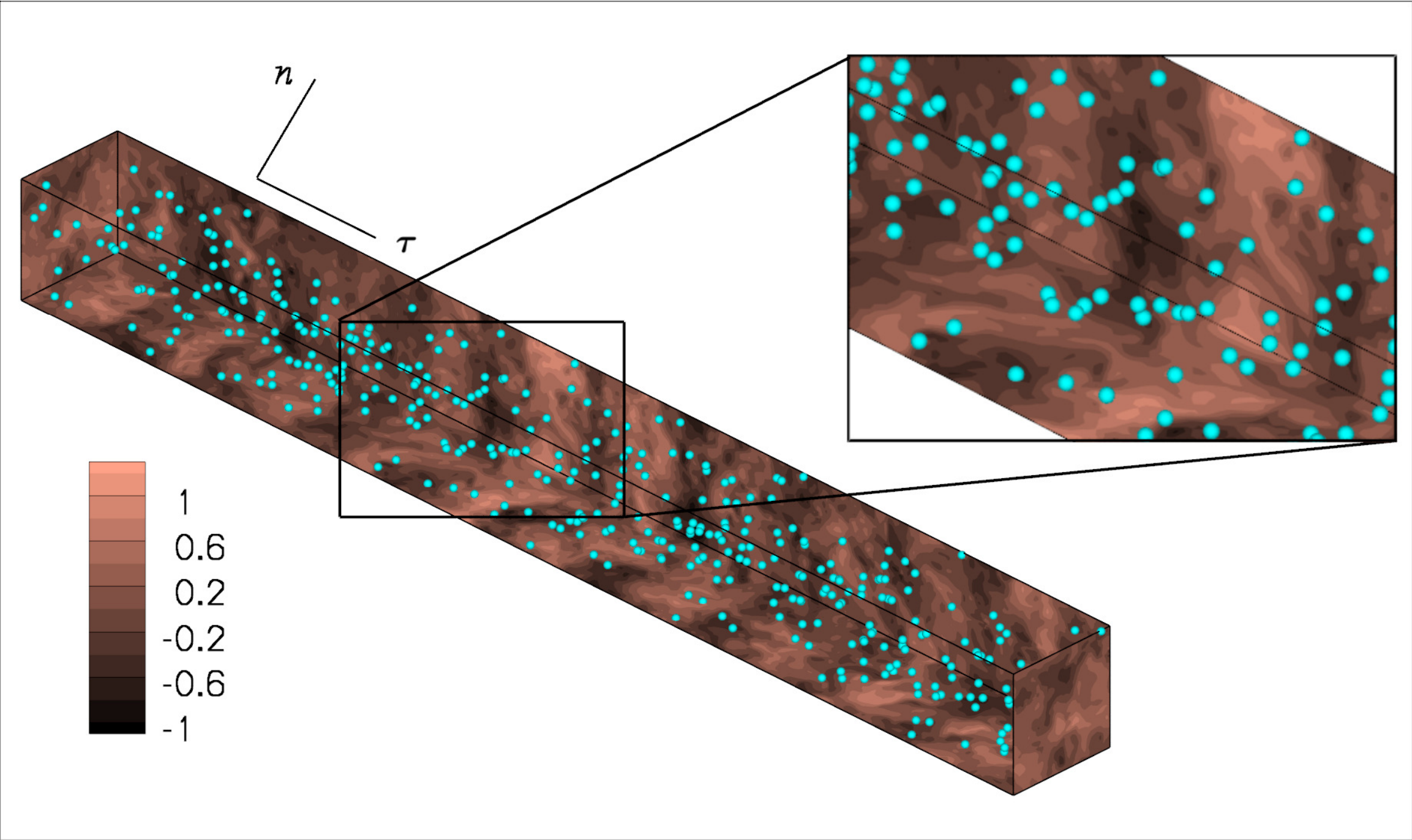}}% 
  \caption{Instantaneous snapshot of the velocity component perpendicular to gravity on different orthogonal planes, together with the corresponding 
particle position for $\phi=0.005$. A zoomed view of a particular section is also shown.} 
\label{fig:turb_in}
\end{figure}
\begin{figure}
  \centering
  \subfigure{%
    \includegraphics[scale=0.35]{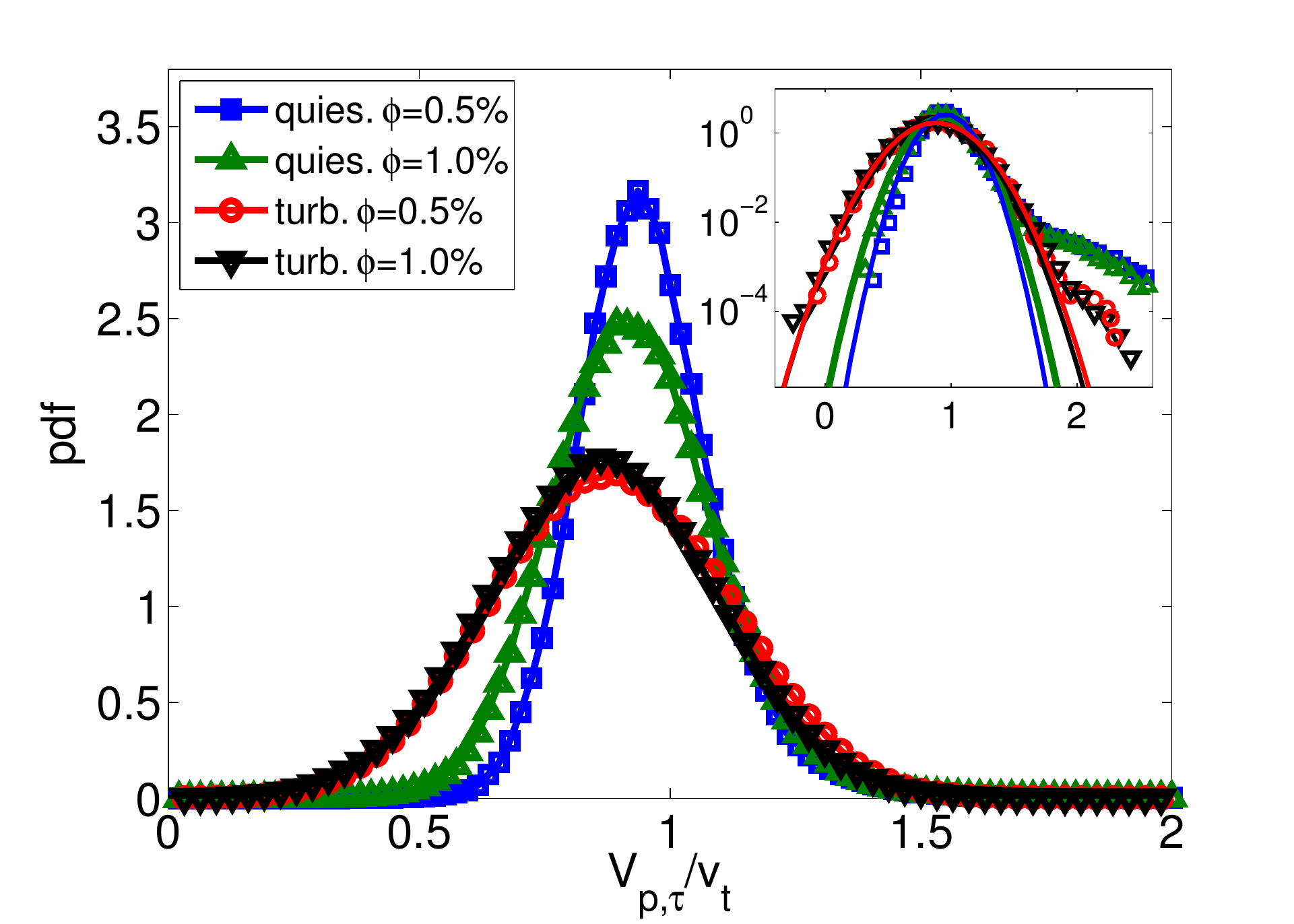}
    \put(-195,110){{\large a)}}
}%
  \subfigure{%
    \includegraphics[scale=0.35]{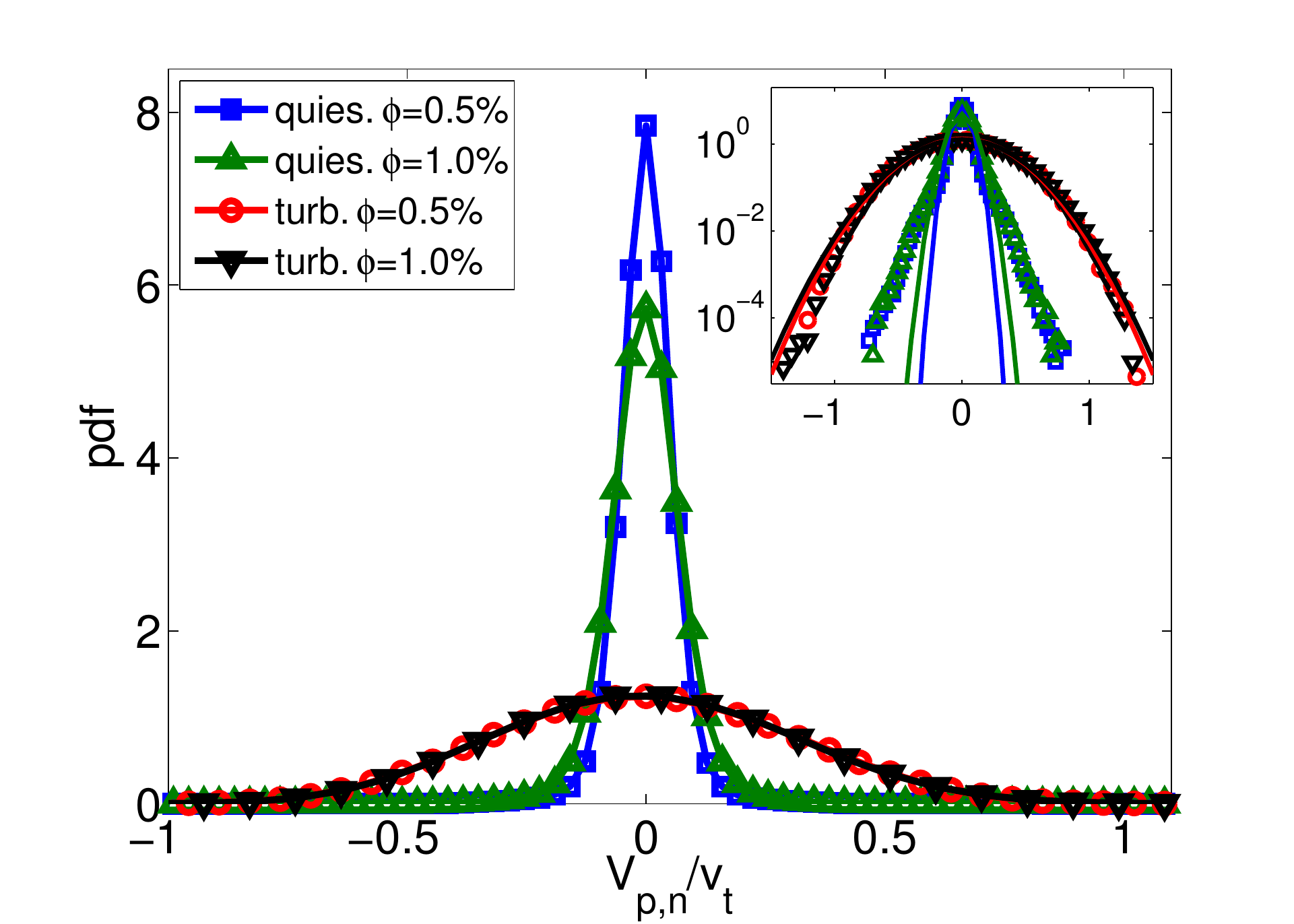}    \put(-190,110){{\large b)}}
}
\caption{Probability density functions of the particles velocities along the directions  a) parallel $V_{p,\tau}$ and b) perpendicular 
$V_{p,n}$ to gravity for $\phi=0.5\%$ and $1\%$. 
The velocities are normalized by $v_t$. 
Quiescent fluid cases denoted by blue curve with squares for $\phi=0.5\%$ and green curve with triangles for $\phi=1\%$, 
turbulent flow data by red curve with circles for $\phi=0.5\%$ and black curve with downward-pointing triangles for $\phi=1\%$. In the insets we show the $pdf$s in lin-log scale.}
\label{fig:pdf_vp}
\end{figure}

\section{Results}\label{sec:results}

We investigate and compare the behavior of a suspension of buoyant particles in a quiescent and turbulent environment.
The behaviour of a suspension in a turbulent flow depends on both the particles and turbulence characteristic time- and length-scales: 
 homogeneous and isotropic turbulence is defined by the dissipative, Taylor and integral scales, whereas 
the particles are characterized by their settling velocity $v_t$ and 
by their Stokesian relaxation time $T_p=2\rho_p a^2/(9\rho_f\nu)=11.1$ (time is scaled by $(2a/v_t)$ throughout the paper). 
A comparison between characteristic  time scales is given by 
the Stokes number,
i.e.\ the ratio between the particle relaxation time and a typical flow time scale $St_f=T_p/T_f$. 
In the present cases, the Stokes number based on the dissipative scales (time and velocity) 
is $St_\eta=T_p/T_K=8.1$ so the particles are inertial on this scale. 
In addition, because the particles are about 12 times larger than the Kolmogorov length and fall about 16 times faster than the Kolmogorov velocity scale, we can 
expect that motions at the smallest scales weakly affect the particle dynamics. 

Considering therefore the large-scale motions, we introduce an integral-scale Stokes number $St_{T_e}=T_p/T_e=0.24$. 
This value of $St_{T_e}$ reflects the fact that the particles are about 20 times smaller than the integral length scale $L_0$.
The strong coupling between particle dynamics and turbulent flow field occurs at scales of the order of the Taylor scale for
the present cases. Indeed, the Taylor Stokes number is $St_\lambda=T_p/T_\lambda=2.1$ with $T_\lambda=\lambda/u'$. 
It should be noted that the Taylor length is slightly larger than the particle size, $3.1a$, while 
 particles fall around 3.4 times faster than typical fluid velocity fluctuations, $v_t/u'=3.4$. Hence particles are strongly influenced 
by the fluid fluctuations occurring at scales of the order of $\lambda$.

\subsection{Particle statistics}
We start by comparing the single-point flow and particle velocity statistics for the two cases studied, i.e.\ quiescent and turbulent flow. 
The results are collected when a statistically steady state is reached. Due to the axial symmetry with respect to the direction of 
gravity, we consider only two velocity components for both phases, the components parallel and perpendicular to gravity, $V_{\alpha,\tau}$ and $V_{\alpha,n}$ respectively, where $\alpha=f,\, p$ indicates the solid and fluid phases.

In figure~\ref{fig:pdf_vp} we report the probability density function of the particle velocities for both volume fraction investigated here, $\phi=0.5\%$ and $1\%$; the moments extracted from these distribution are summarized in  table~\ref{tab:pdf_vp_tab}. The data in~\ref{fig:pdf_vp}a) show the 
$pdf$ for the component of the velocity aligned with gravity $V_{p,\tau}$ normalized by the settling
 velocity for $\phi \to 0$ in a quiescent environment, $v_t$. 
 This is extracted from the simulation of a very dilute suspension discussed in section~\ref{sec_val}.

In the quiescent cases, the mean settling velocity slightly reduces when increasing the volume fraction $\phi$, in agreement with the findings of \citet{richardson1954} and \citet{di1999} among others. 
The sedimentation velocity decreases to 0.96 at $\phi=0.5\%$ and 0.93 at $\phi=1\%$.
\citet{di1999} investigated experimentally the settling velocity of dilute suspensions of spheres ($\phi=0.5\%$) with density ratio 1.2 in quiescent fluid, for a large range of terminal Reynolds numbers (from $0.01$ to $1000$). 
Following the empirical fit proposed in \citet{di1999},
we obtain $\langle V_{f,\tau} \rangle \simeq 0.98$ at $\phi=0.5\%$, approximately $1\%$ larger than our result. 
On the other hand, the formula suggested for the intermediate regime in \cite{di1999} and \cite{yin2007}, $Re_t<150$, 
would give an estimated value of 0.88,  $6\%$ lower than our result.

The mean settling velocity for the quiescent case at $\phi=0.5\%$ is instead close to the estimated value from eq.~(\ref{arc_re}).
This is in agreement with what reported in \citet{uhlmann2014}. 
These authors found that the particle mean settling velocity of a dilute suspension increases above the reference value only when the Archimedes number $Ar$ is larger than approximately $24000$ and clustering occurs. 
In our case $Ar \simeq 21000$ and no clustering is noticed accordingly with their findings.

Interestingly, we observe an additional  non-negligible reduction when a turbulent
background flow is considered, in our opinion the main result of the paper. 
The reduction of the mean settling velocity $\langle V_{p,\tau} \rangle$ 
is $12\%$ at $\phi=0.5\%$ and $14\%$ at $\phi=1\%$,
see table~\ref{tab:pdf_vp_tab}. 
This result unequivocally shows that the turbulence reduces the settling velocity of a 
suspension of finite-size buoyant particles, in agreement 
with the experimental findings by Variano and 
co-workers (Variano, private communication) and \citet{byron2015}. 
We also note that the reduction of the settling velocity with the volume fraction is less important for the turbulent cases.

Looking more carefully at the $pdf$s we note that fluctuations are, as expected, larger in a turbulent environment. In addition, 
 the vertical particle velocity fluctuations are the largest component in a quiescent fluid, whereas in a turbulent flow the fluctuations are largest in the horizontal directions, as summarized in table~\ref{tab:pdf_vp_tab}.
In the quiescent case
the rms of the 
tangential velocity $\sigma_{V_{p,\tau}}$  is 0.15 and 0.17 for $\phi=0.5\%$ and $\phi=1\%$ respectively, while it increases up to $0.23$ in the corresponding turbulent cases.
The difference in the width of the $pdf$ is particularly large in the directions normal to gravity where the rms of the variance $\sigma_{V_{p,n}}$ is 0.3 for both turbulent cases, while it
is 5 and 4 times smaller for the quiescent flows at $\phi=0.5\%$ and $\phi=1\%$.  We believe that the  
interactions among the particle wakes, mainly occurring in the settling direction, promote the higher vertical velocity fluctuations 
found in the quiescent cases.  
The shape of the $pdf$ is essentially Gaussian for the turbulent cases, showing vanishing skewness and normal flatness. Interestingly
 an intermittent and skewed behaviour is exhibited in the quiescent cases. The flatness $K$ is around $9$ for both components at
 $\phi=0.5\%$ and slightly reduces to $5.5$ at $\phi=1\%$. The settling velocity of the quiescent cases is skewed towards intense
fluctuations in the direction of the gravity. The skewness $S$ is higher for the more dilute case, being $1.26$ at $\phi=0.5\%$ and
$0.7$ at $\phi=1\%$. 

We interpret the intermittent behavior suggested by values of $K > 3$ by the collective dynamics of the particle suspension. 
The significant tails of the $pdf$s shown in figure~\ref{fig:pdf_vp}a) are indeed associated to a specific behavior:
as particles fall, they tend to be 
accelerated by the wakes of other particles, before showing drafting-kissing-tumbling behavior \citep{fortes1987}. Snapshots of the drafting-kissing-tumbling behavior between two 
spherical particles are shown in figure~\ref{fig:kissing}. When this close interaction occurs, particles are found to fall with velocities that can be more than twice the mean settling velocity 
$\langle V_{p,\tau} \rangle$. In the quiescent case the fluid is still, the wakes are the only perturbation present in the field and 
 are long and intense so their effect can be felt far away from the reference particle. The more dilute the suspension the more intermittent the particle velocities are.
On the contrary, when the flow is turbulent the wakes are disrupted quickly 
and therefore fewer particles feel the presence of a wake. The velocity disturbances are now mainly due to turbulent eddies 
of different size that interact with the particles to increase the variance of the velocity homogeneously along all directions leading to the almost perfect normal 
distributions shown above, with variances similar to those of the turbulent fluctuations.
The features of the particle wakes will be further discussed  in this manuscript 
to support the present explanation.

We also note that the sedimenting speed in the quiescent fluid
is determined by two opposite contributions. The 
excluded volume effect that contributes to a reduction of the mean settling velocity with
respect to an isolated sphere  and the pairwise interactions (the drafting-kissing-tumbling), increasing the mean velocity of the two particles involved in
the encounter.
To try to identify the importance of the drafting-kissing-tumbling effect, we
fit the left part (where no intermittent behaviour is found) of the $pdf$ pertaining the quiescent case at 
$\phi=0.5\%$  with a gaussian function. 
The mean of $V_{p,\tau}$ is reduced to about $0.93$ (value due only to the hindrance effect) instead of 0.96 in the full simulation; thereby the increment in mean settling velocity due to  drafting-kissing-tumbling can be estimated to be of about $3\%$.

\begin{figure}
  \centering
  \subfigure{%
    \includegraphics[scale=0.20,angle=270]{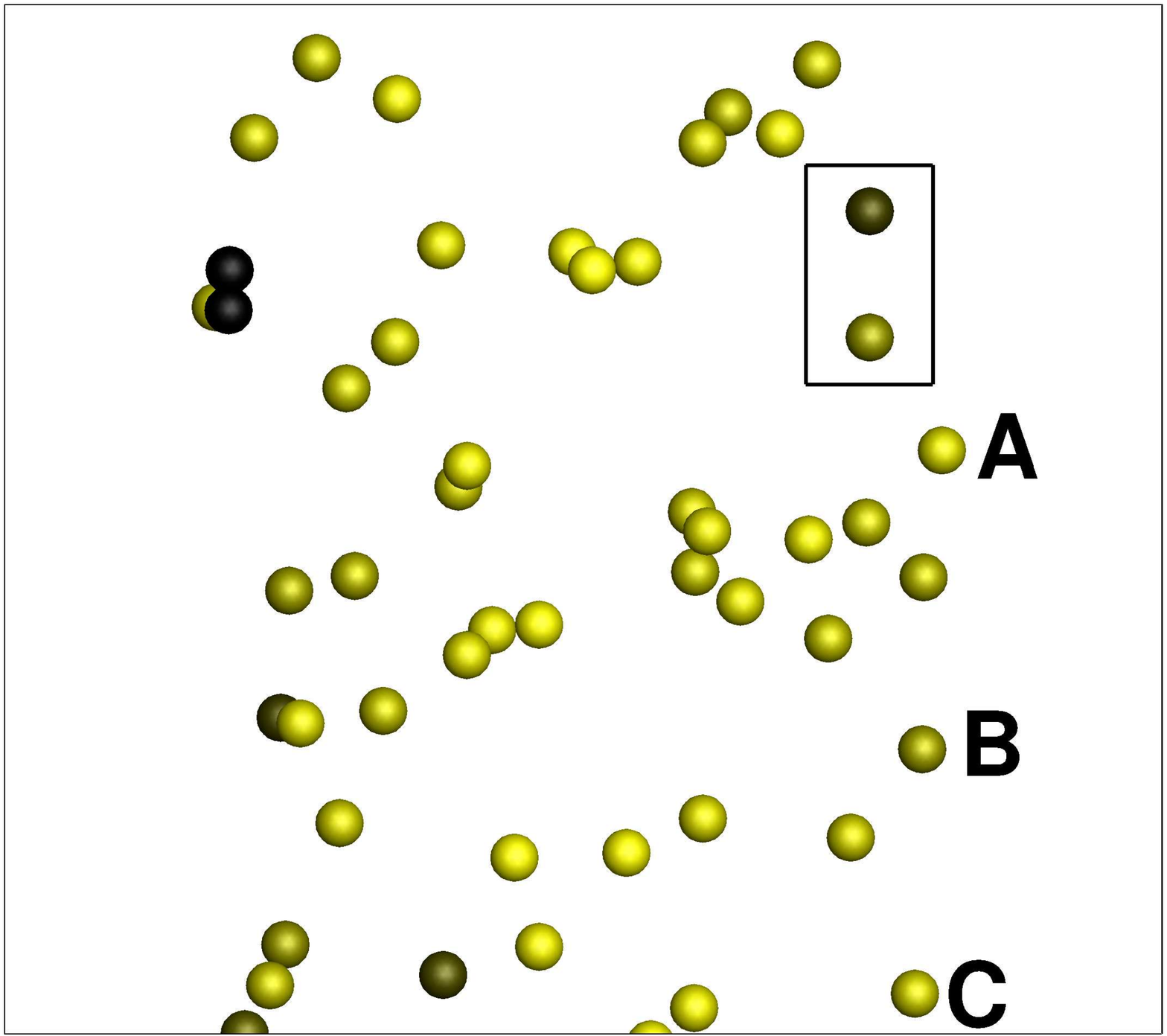}   
     \put(-205,-20){{\large a) t=1789}}
}%
\\
  \subfigure{%
    \includegraphics[scale=0.20,angle=270]{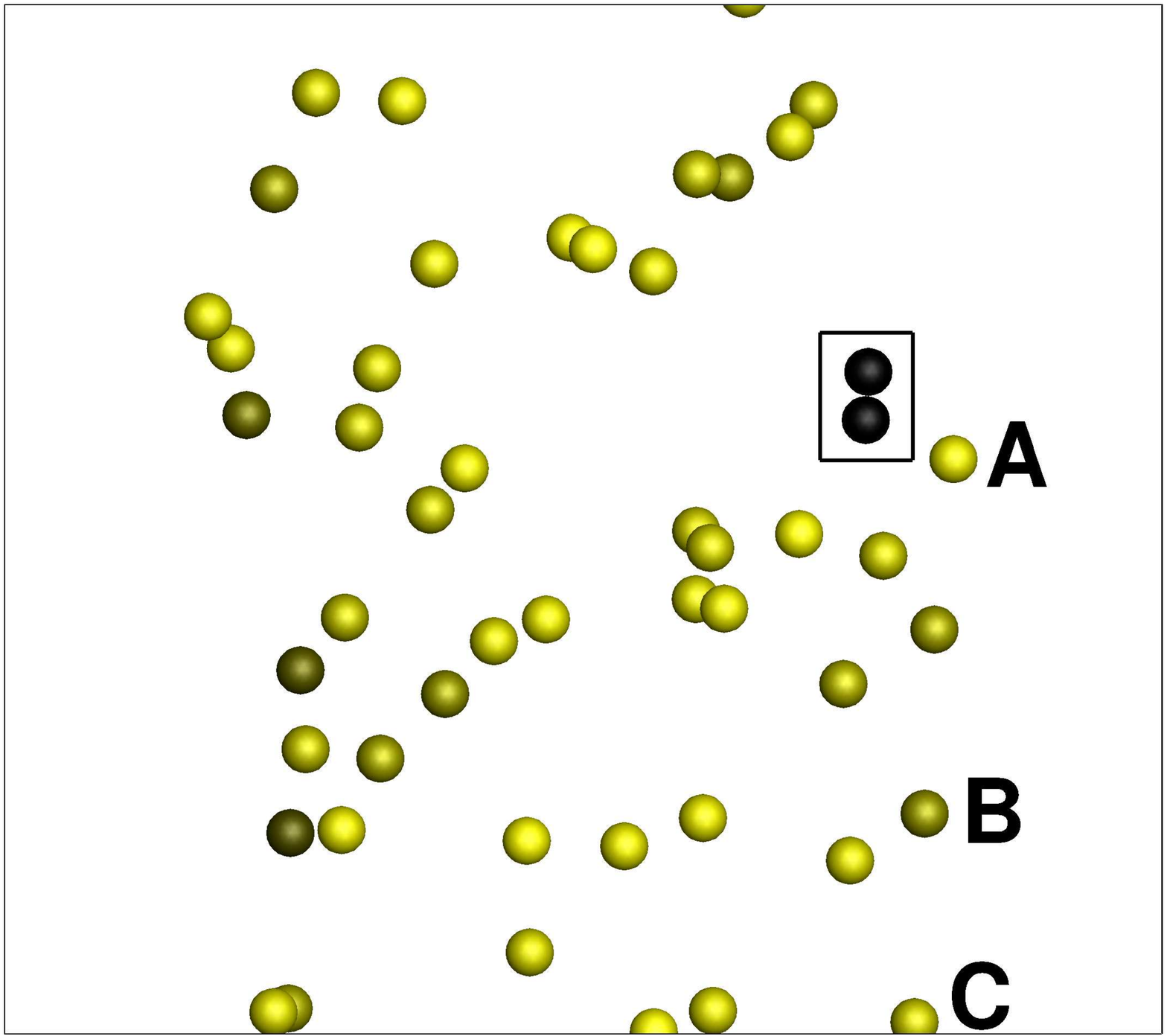}
     \put(-205,-20){{\large b) t=1794}}
     }%
\\%
  \subfigure{%
    \includegraphics[scale=0.20,angle=270]{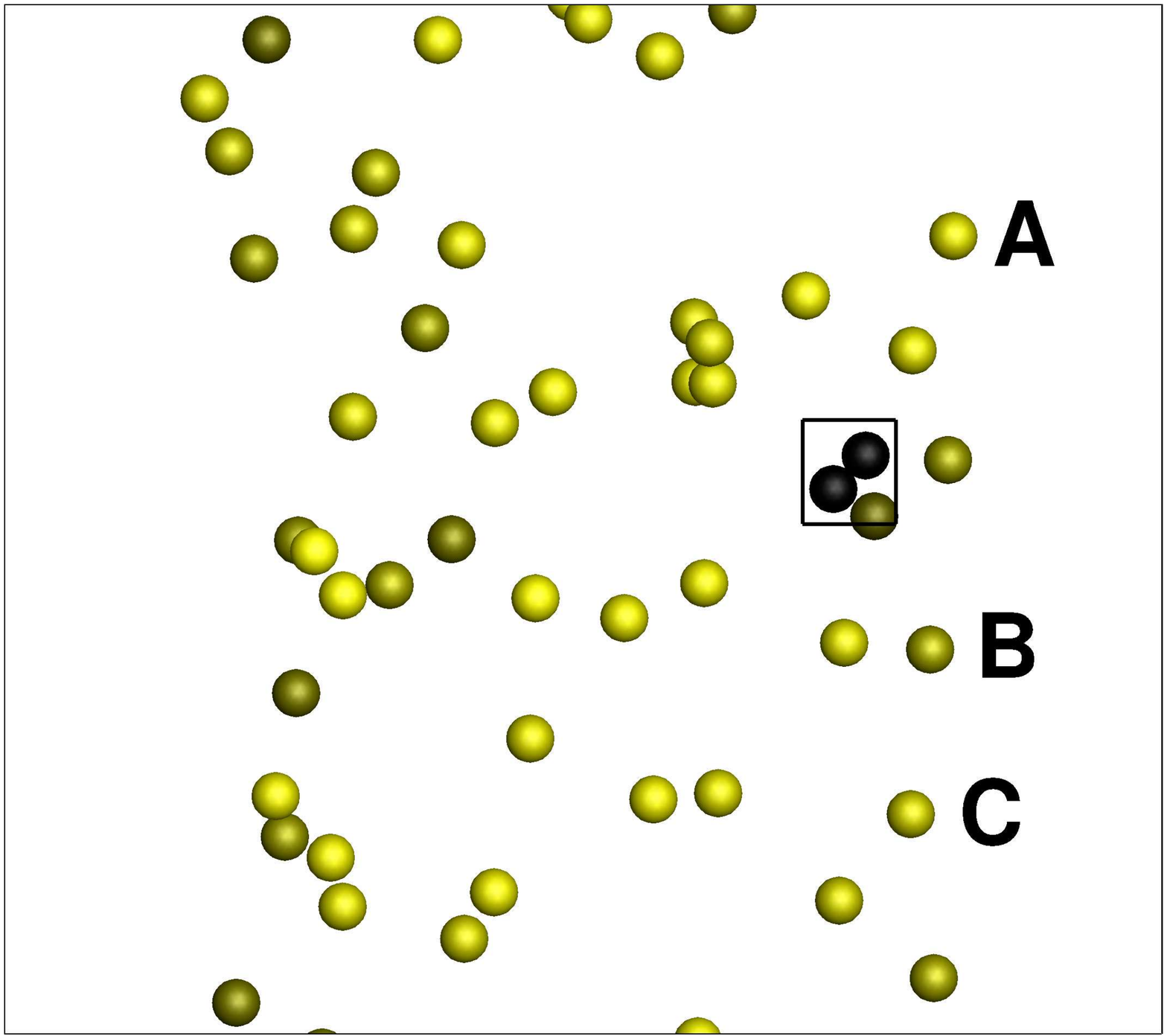}
         \put(-205,-20){{\large c) t=1799}}}%
\\
  \subfigure{%
    \includegraphics[scale=0.20,angle=270]{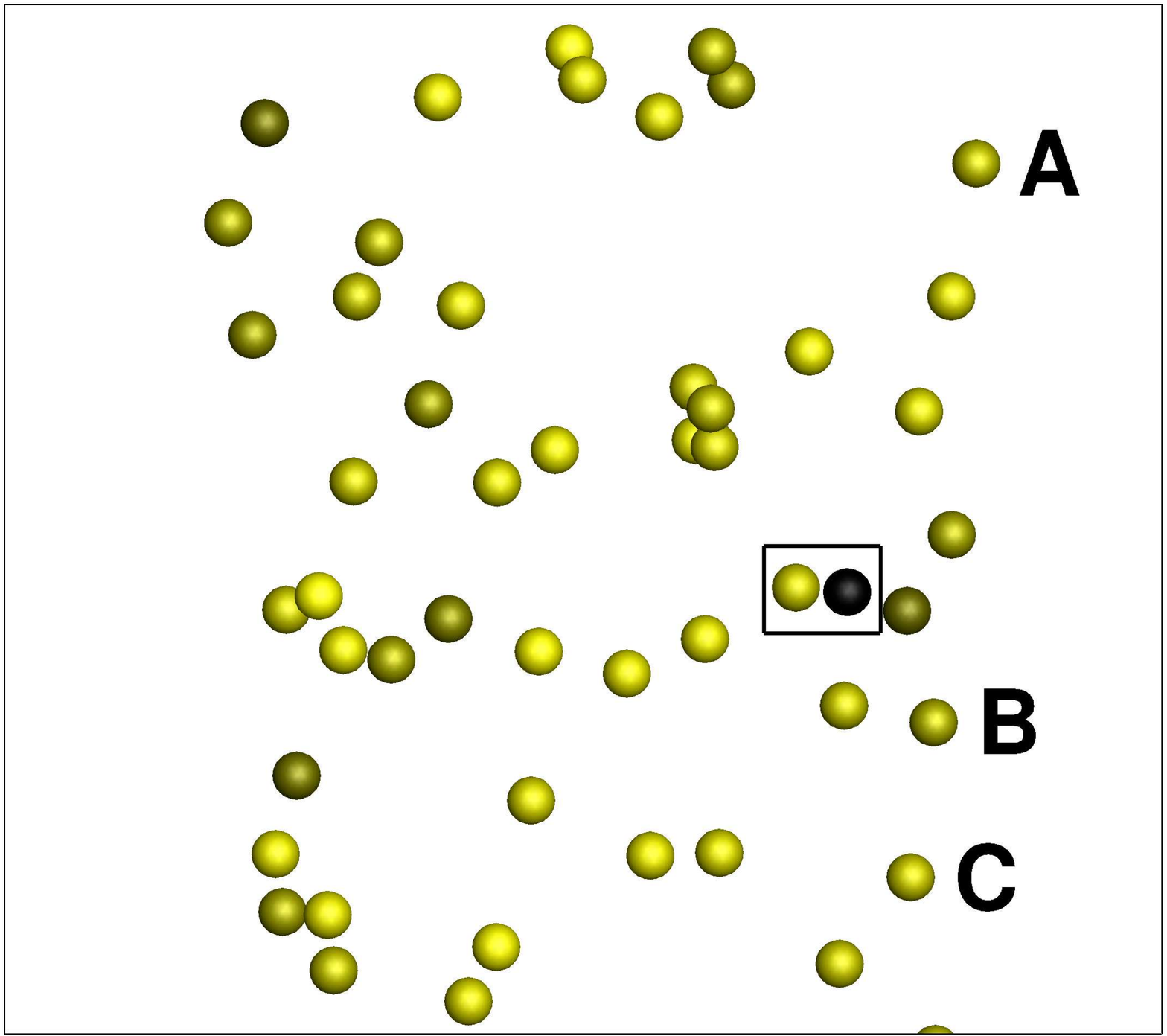}
         \put(-205,-20){{\large d) t=1801}}}%
\caption{Drafting-kissing-tumbling behavior among two spherical particles in the quiescent case with $\phi=0.5\%$. The particles are coloured with the absolute value of their velocity component in the direction of 
gravity. Three particles are labeled with "A,B,C" in order to show how accelerated the two interacting particles are compared to the others.}
\label{fig:kissing}
\end{figure}

\begin{table}
  \begin{center}
\def~{\hphantom{0}}
  \begin{tabular}{ccccc}
                   & Quiescent $\phi=0.5\%$ & Turbulent $\phi=0.5\%$ & Quiescent $\phi=1\%$ & Turbulent $\phi=1\%$ \\
  $\langle V_{p,\tau} \rangle$    & $+0.96$                & $+0.88$                & $+0.93$                & $+0.86$ \\
  $\sigma_{V_{p,\tau}}$           & $+0.15$                & $+0.23$                & $+0.17$                & $+0.23$ \\
  $S_{V_{p,\tau}}$                & $+1.26$                & $+0.01$                & $+0.70$                & $+0.01$ \\
  $K_{V_{p,\tau}}$                & $+9.65$                & $+2.92$                & $+6.01$                & $+3.15$ \\
   \hline
  $\langle V_{p,n} \rangle$    & $+0.33\cdot10^{-4}$    & $-1.93\cdot10^{-3}$       & $-8.78\cdot10^{-4}$    & $-0.97\cdot10^{-3} $ \\
  $\sigma_{V_{p,n}}$           & $+0.06$                & $+0.31$                & $+0.08$                & $+0.31$ \\
  $S_{V_{p,n}}$                & $-1.22\cdot10^{-3}$    & $+0.04$                & $-3.17\cdot10^{-3}$    & $+0.07$ \\
  $K_{V_{p,n}}$                & $+8.95$                & $+2.78$                & $+5.55$                & $+2.80$ \\
  \end{tabular}
  \caption{First four central moments of the probability density functions of $V_{p,\tau}$ and $V_{p,n}$ normalized by $v_t$. $S_{V_{p,\tau}}$ ($S_{V_{p,n}}$) and $K_{V_{p,\tau}}$ ($K_{V_{p,n}}$) are 
respectively the skewness and the flatness of the $pdf$s.}
  \label{tab:pdf_vp_tab}
  \end{center}
\end{table}

\begin{figure}
  \centering
  \subfigure{%
    \includegraphics[scale=0.35]{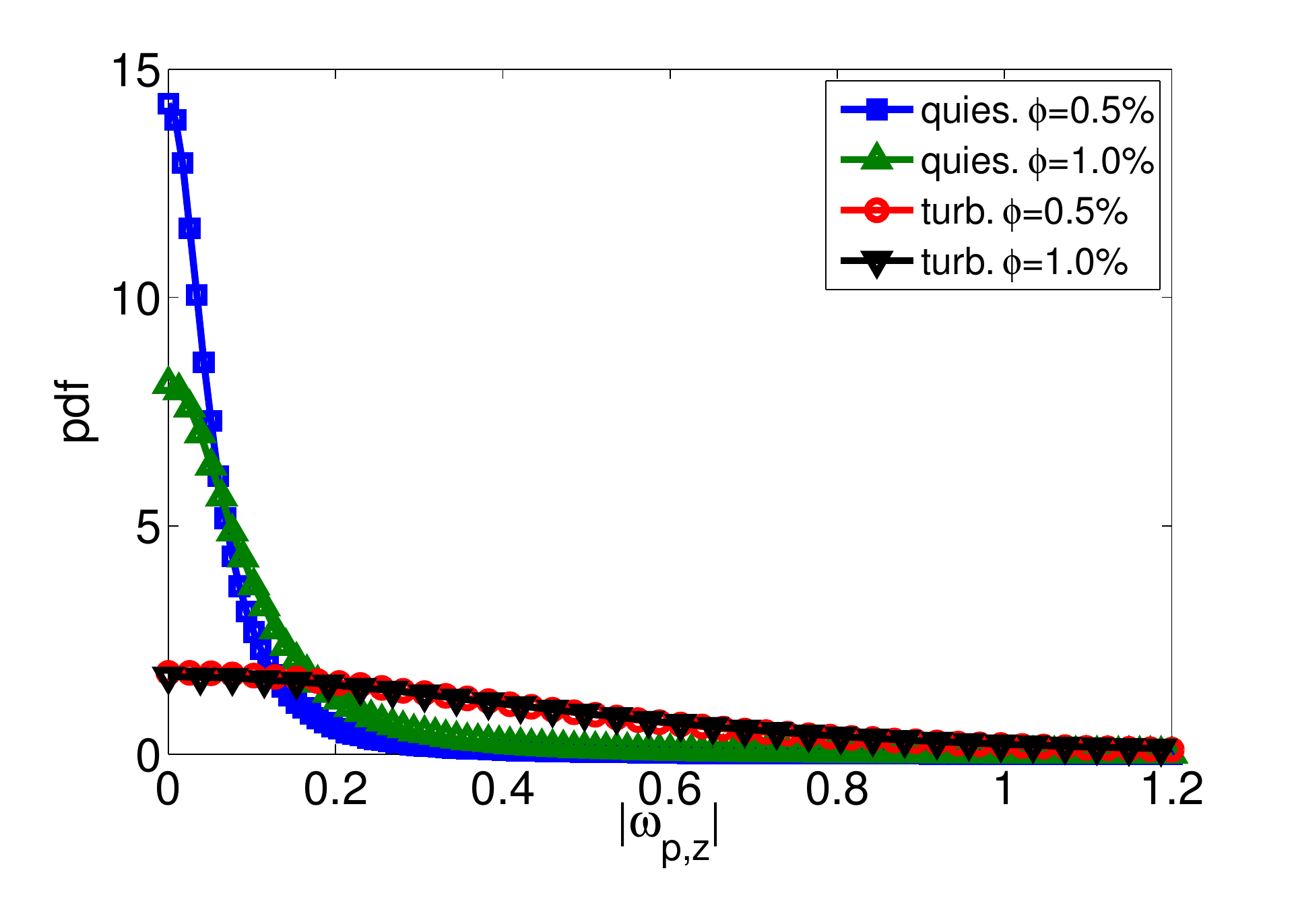}  \put(-190,110){{\large a)}}}%
  \subfigure{%
    \includegraphics[scale=0.35]{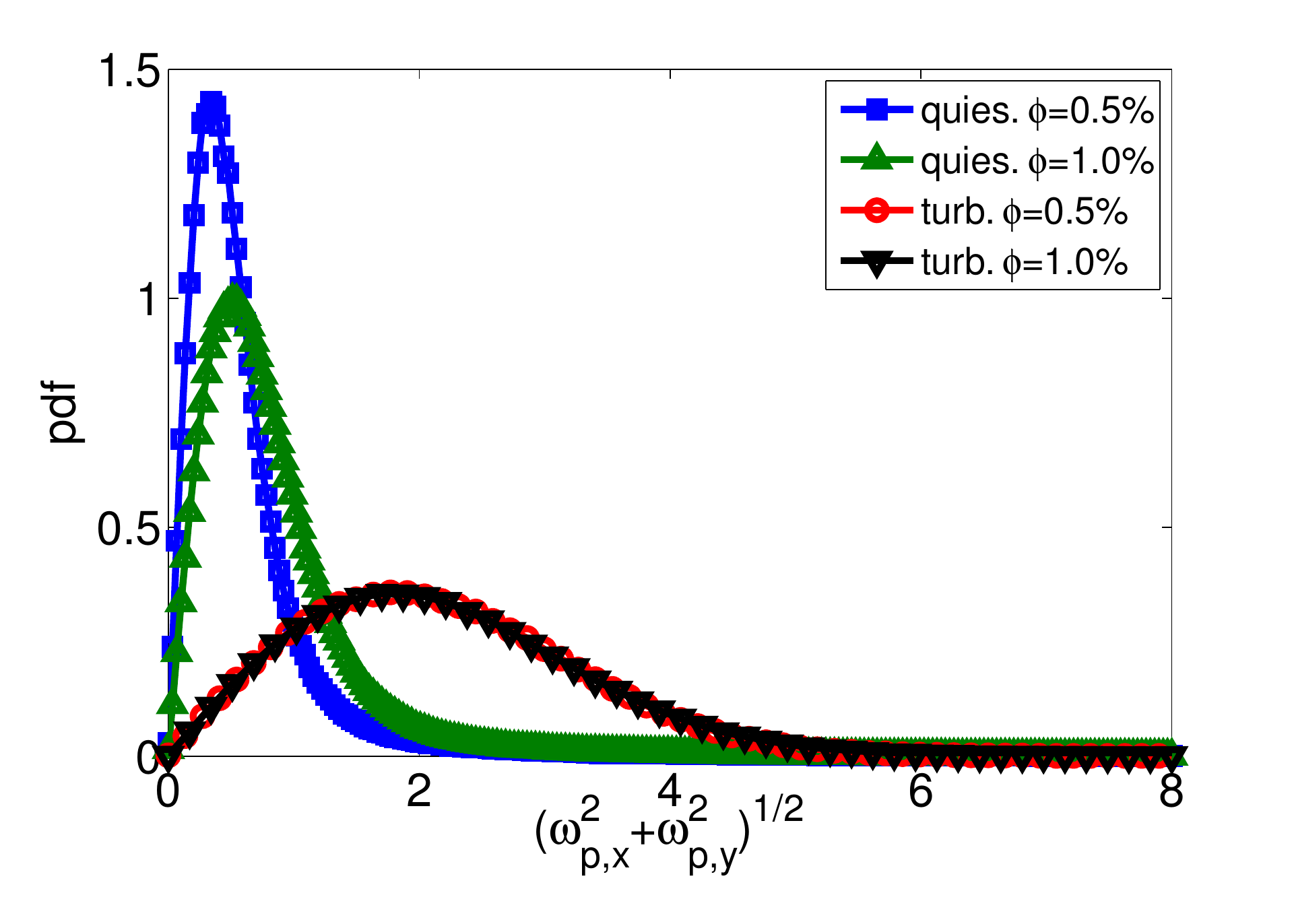} \put(-190,110){{\large b)}}}%
\caption{Probability density functions of a) $|\omega_{p,z}|$ and b) $\sqrt{\omega_{p,x}^2 + \omega_{p,y}^2}$ for $\phi=0.5\%$ and $1\%$. 
The angular velocities are normalized by $v_t/(2a)$, the settling velocity of a single particle in a quiescent environment and its diameter. 
Quiescent fluid cases denoted by blue curve with squares for $\phi=0.5\%$ and green curve with triangles for $\phi=1\%$, 
turbulent flow data by red curve with circles for $\phi=0.5\%$ and black curve with downward-pointing triangles for $\phi=1\%$.}
\label{fig:pdf_op}
\end{figure}

The probability density function of the particle angular velocities are also different in quiescent and turbulent flows. 
These are shown in
figures~\ref{fig:pdf_op}a) and b) 
for rotations about an axis parallel to gravity, $|\omega_{p,\tau}|=|\omega_{p,z}|$, and orthogonal to it, $\sqrt{\omega_{x}^2+\omega_{y}^2}$. 
In the settling direction the peak of the $pdf$ is always at $|\omega_{p,z}|=0$. As for the translational 
velocities, the $pdf$s are broader in the turbulent cases. Due to the interaction with turbulent eddies, particles tend also to spin faster around axis perpendicular to gravity. 
From figure~\ref{fig:pdf_op}b) we see that the modal value slightly increases in the quiescent cases when increasing the volume fraction. In the turbulent cases, 
the modal value is more than $3$ times the values of the quiescent cases and the variance is also increased. Unlike the quiescent cases, the curves 
almost perfectly overlap for the two different $\phi$, meaning that turbulent fluctuations dominate the particle dynamics. Turbulence hinders particle hydrodynamic interactions.

Figure~\ref{fig:corr_p} shows the temporal correlations of the particle velocity fluctuations,
\begin{align}
\label{corr_ut}
R_{v_{\tau}v_{\tau}}(\Delta t) = \frac{\langle V_{p,\tau}'(p,t) V_{p,\tau}'(p,t+\Delta t) \rangle}{\sigma_{V_{p,\tau}}^2}\\
\label{corr_un}
R_{v_nv_n}(\Delta t) = \frac{\langle V_{p,n}'(p,t) V_{p,n}'(p,t+\Delta t) \rangle}{\sigma_{V_{p,n}}^2}
\end{align}
for the turbulent and quiescent cases at $\phi=0.5\%$ and $\phi=1\%$. 
Focusing on the data at the lower volume fraction, we observe that the particle settling velocity decorrelates much faster in the turbulent
environment, within $\Delta t\sim 50$, while it takes around one order of magnitude longer in a quiescent fluid.
Falling particles may encounter intense vortical structures that change their settling velocity. 
The turbulence strongly alters the fluid velocity field seen by the particles, which in the
quiescent environment is only constituted by coherent long particle wakes. This results in a faster decorrelation of the velocity fluctuations along the settling direction in the turbulent environment. 
Moreover, $R_{v_nv_n}$ crosses the null value earlier than for the settling
velocity component. This result is not surprising since the particle wakes develop only in the settling direction. 

\begin{figure}
  \centering
    \includegraphics[scale=0.329]{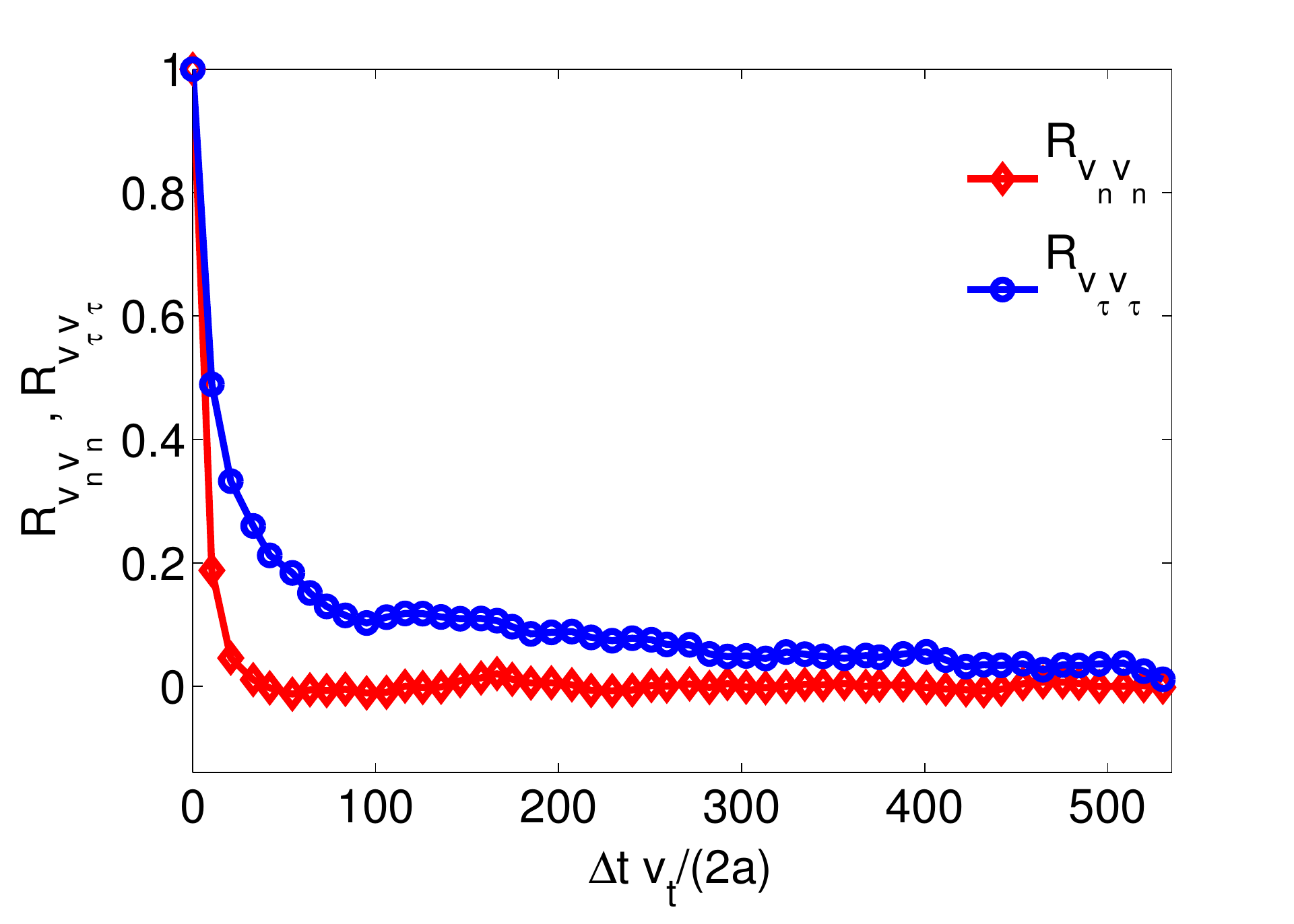}
    \put(-185,100){{\large a)}}
    \includegraphics[scale=0.329]{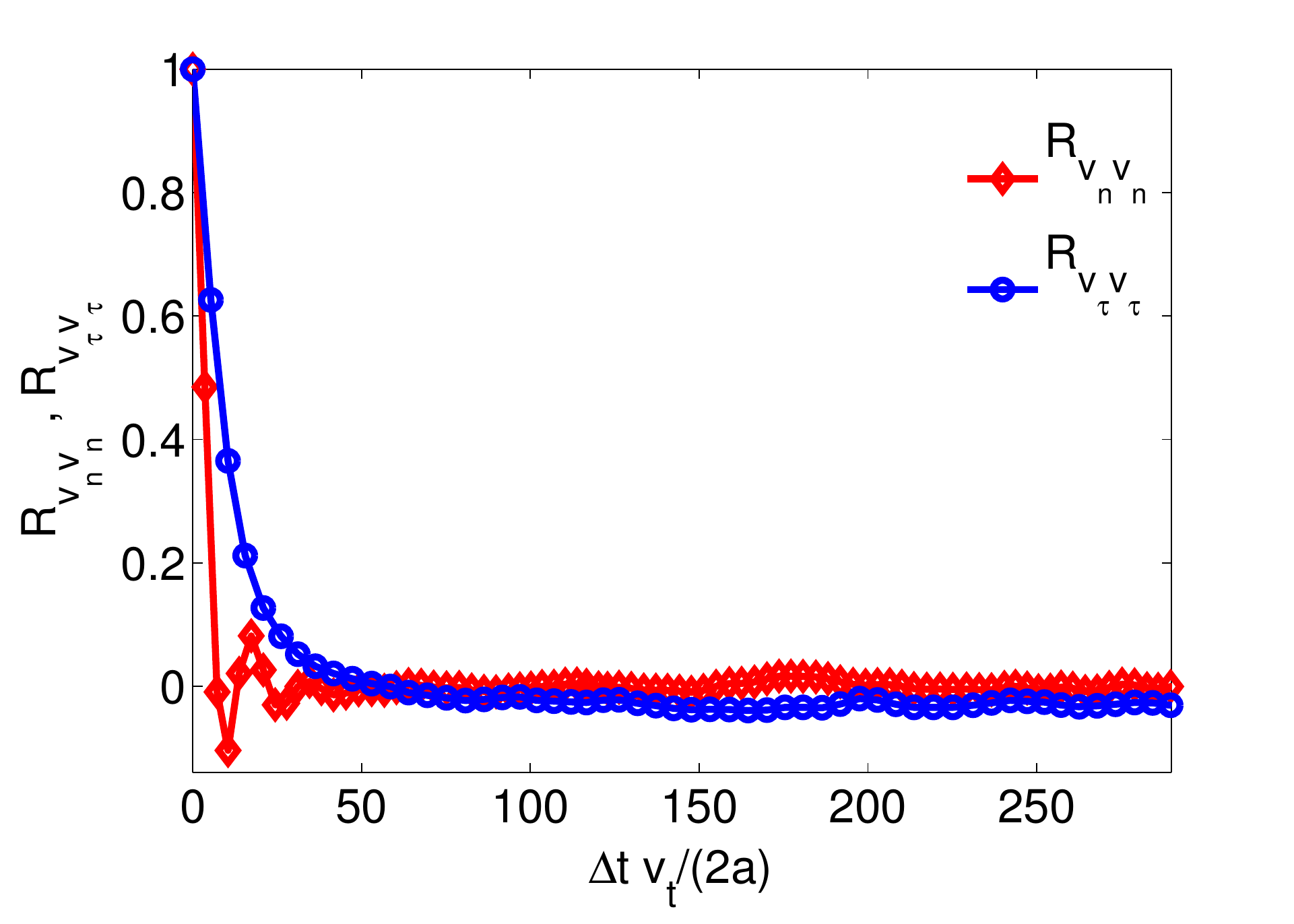}%
    \put(-185,100){{\large b)}}\\
    \includegraphics[scale=0.329]{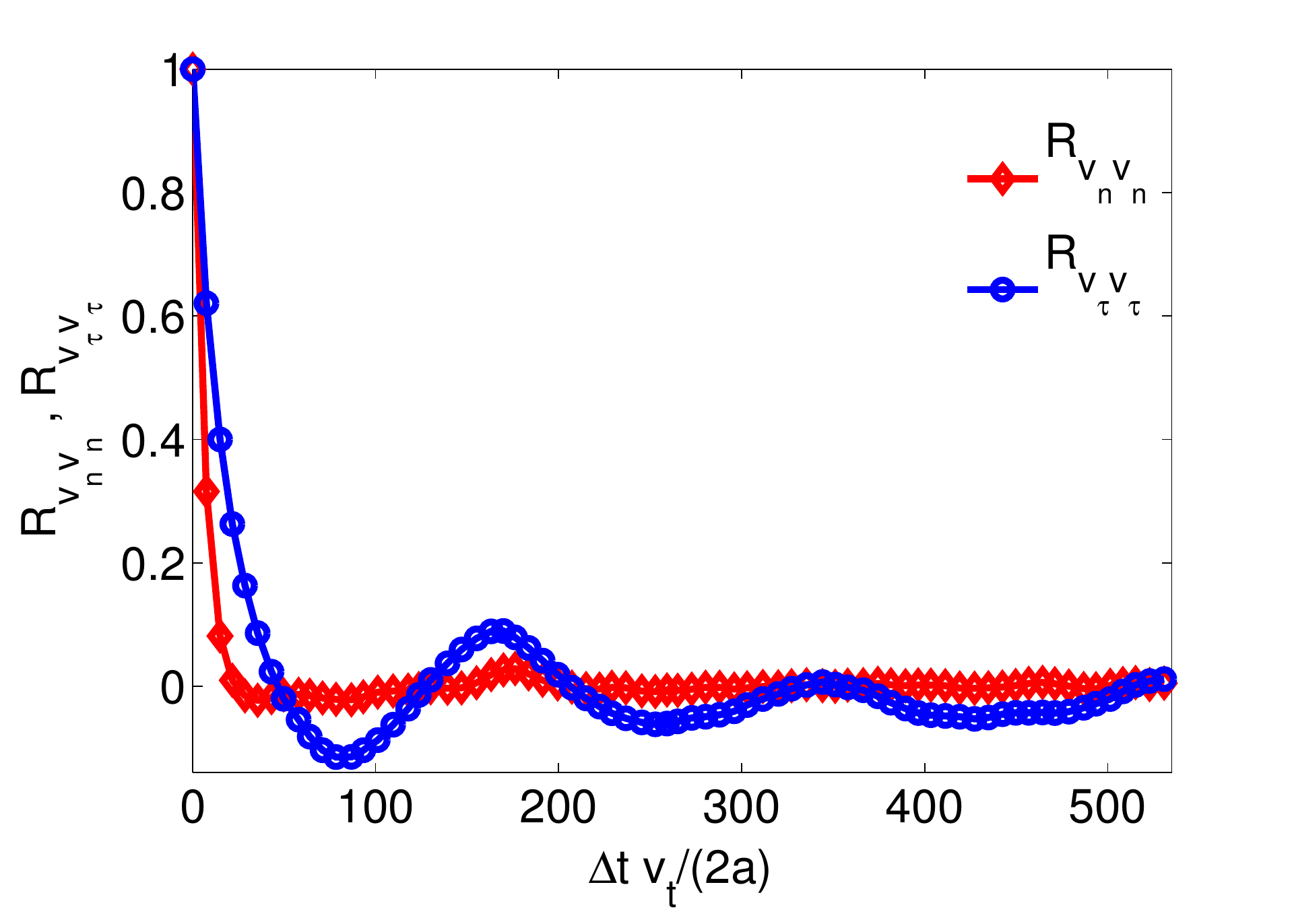}
    \put(-185,100){{\large c)}}
    \includegraphics[scale=0.329]{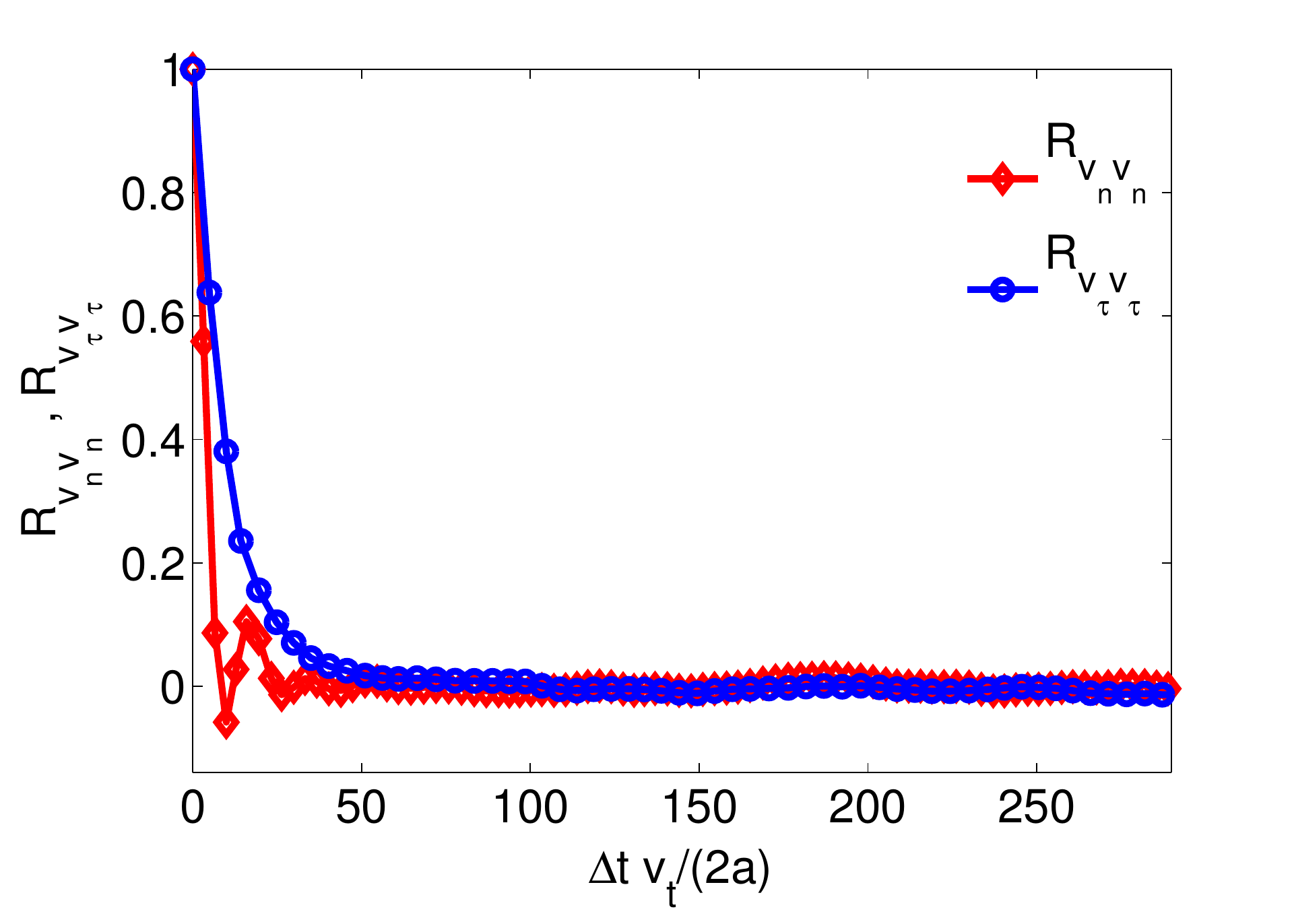}
    \put(-185,100){{\large d)}}
  \caption{Time correlations of the particle velocity fluctuations. 
  The blue curve with circles represents the correlation 
of $V_{p,\tau}'$ ($R_{v_{\tau}v_{\tau}}$), while the red curve with diamonds is used 
for the correlation of $V_{p,n}'$ ($R_{v_n v_n}$). a) Quiescent case with $\phi=0.5\%$, b) Turbulent case with $\phi=0.5\%$, c) Quiescent case with $\phi=1.0\%$ and d) 
Turbulent case with $\phi=1.0\%$.} 
\label{fig:corr_p}
\end{figure}

The normal 
velocity correlation $R_{v_nv_n}$ of the turbulent case oscillates around zero before vanishing at longer times.
We 
attribute this to the presence of the large-scale turbulent eddies. 
As a settling particle encounters sufficiently strong and large eddies, its trajectory is swept on planes normal to 
gravity in an oscillatory way. 
To provide an approximate estimate of this effect, we consider as a first approximation the turbulent flow seen by the particles 
as frozen since the particles fall at a higher velocity than the turbulent fluctuations (3.4 times).
Since the strongest eddies are of the order of the transversal 
integral scale we can presume that these structures are responsible for this behavior. 
In particular, the transversal integral 
scale $L_T=L_0/2\simeq8$ and $u'_{rms}=0.3$, so we expect a typical period of $t=L_T/u'_{rms}\simeq26$ 
which is of the order of the oscillations  found for both turbulent cases, i.e.\ $t \approx 20$. Note that a similar behavior has been observed by \citet{wang1993} for  sufficiently small and heavy particles, termed the preferential sweeping phenomenon.

The same process can be interpreted in terms of crossing trajectories and continuity effects as described by \citet{csanady1963}. An inertial particle 
falling in a turbulent environment changes continuously its fluid-particle neighborhood. It will fall out from the eddy where it was at an earlier instant and will therefore 
rapidly decorrelate from the flow. In order to accommodate the back-flow necessary to satisfy continuity, the normal correlations must then contain negative loops (as those seen in figures~\ref{fig:corr_p}b and d). Following \citet{csanady1963} we define the period of oscillation of the fluctuations as the ratio of the typical 
eddy diameter in the direction of gravity (i.e. the longitudinal integral scale $L_0$), and the particle terminal velocity $v_t$ obtaining 
$t=L_0/v_t \simeq 16$. This value is similar to the period of oscillations in the correlations in figure~\ref{fig:corr_p}.

\begin{figure}
  \centering
  \subfigure{%
    \includegraphics[scale=0.45]{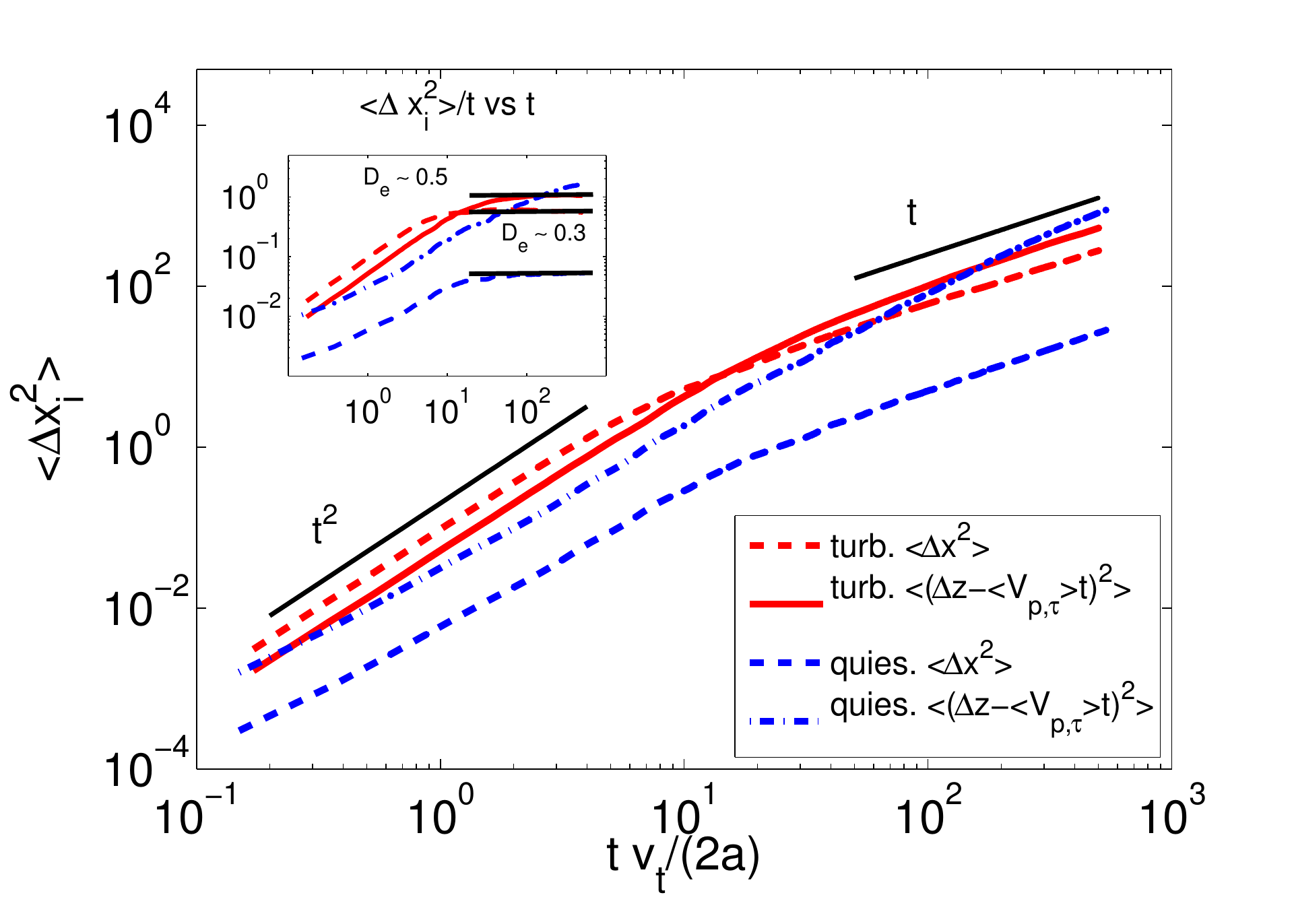} \put(-255,140){{\large a)}}}%
\\
  \subfigure{%
    \includegraphics[scale=0.45]{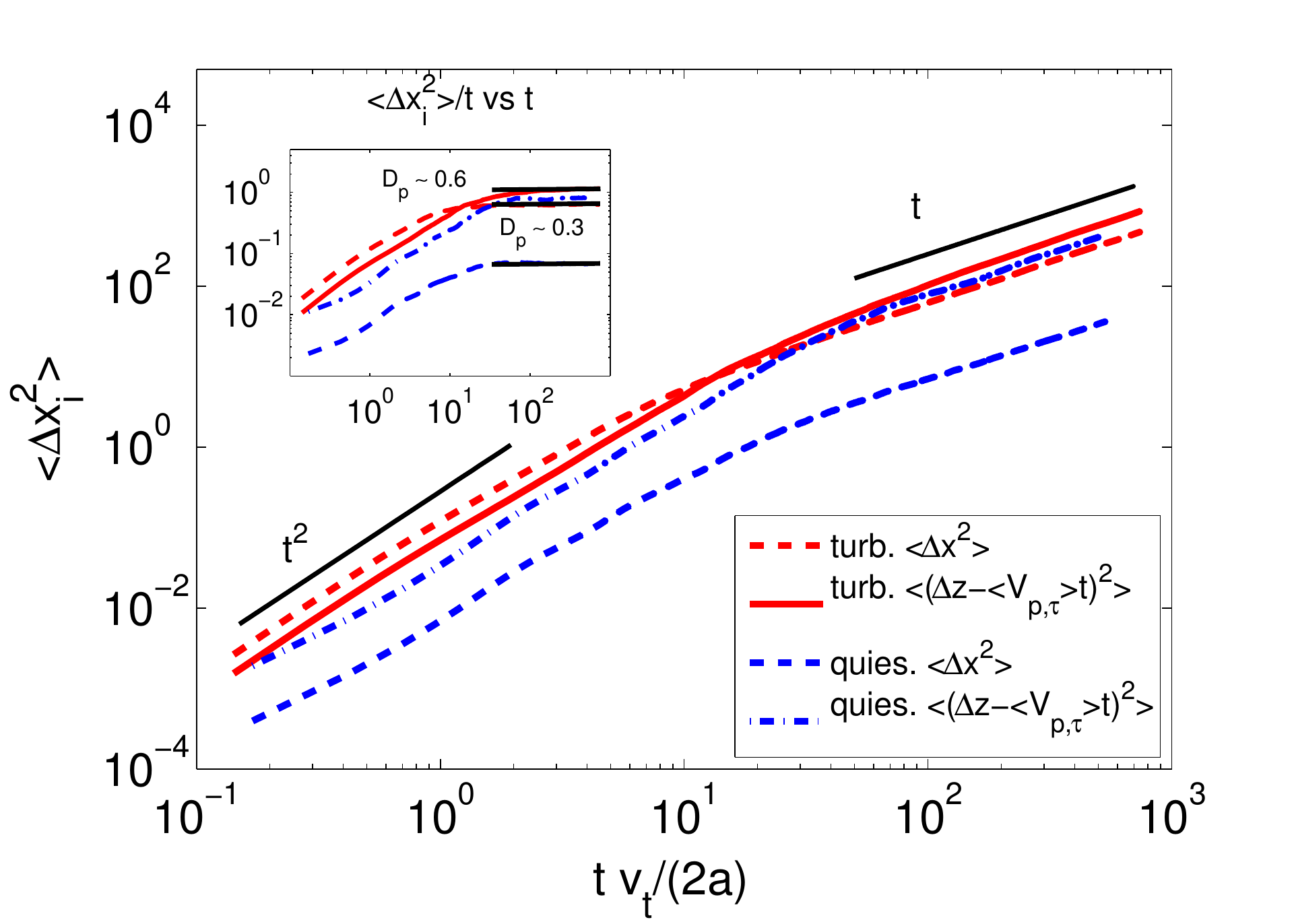} \put(-255,140){{\large b)}}}
  \caption{Mean square particle displacement in the directions parallel and perpendicular to gravity for both quiescent and turbulent cases for a) $\phi=0.5\%$ and b) $\phi=1\%$. 
The mean particle displacement $\langle V_{p,\tau}\rangle t$ in the settling direction
is subtracted from the instantaneous displacement when computing the statistics. In the inset, we show $\langle \Delta x_i^2 \rangle/t$ as a function of time as well as the 
diffusion coefficients $D_e$ for the turbulent cases.}
\label{fig:disper}
\end{figure}

As shown in figure~\ref{fig:corr_p}c),
the quiescent environment presents a peculiar behavior of the settling velocity correlation at $\phi=1\%$. 
In particular we observe
oscillations around zero of long period, $T=160-180$. 
From the analysis of particle snapshots at different times (not shown), we observe that
these seem correlated
to the formation of regions of different density of the particle concentration.
Hence 
a particle crossing regions with different local particle concentration may experience a varying settling velocity.

To further understand the particle dynamics, we display in figure~\ref{fig:disper}a) and b) the single particle dispersion, i.e.\ the mean square displacement, for both quiescent and turbulent cases at $\phi=0.5\%$ and $1.0\%$. 
The mean displacement, $\langle V_{p,\tau} \rangle t$, is 
subtracted from the instantaneous displacement in the settling direction $\Delta z(t)$ to highlight the fluctuations with respect to the mean motion. 
For all cases we found initially a quadratic scaling in time ($\langle \Delta x_i^2 \rangle \sim t^2$) typical of correlated motions while the linear diffusive behavior takes over at longer times ($\langle \Delta x_i^2 \rangle \sim 2D_e t$ with $D_e$ the diffusion coefficient).

The turbulent cases show a similar behavior for both volume fractions. The crossover time when the initial quadratic scaling is lost and the linear one takes place is about 
$t\simeq10 $ and  $t\simeq 50$ for the normal and tangential component, respectively. 
This difference is consistent with the correlation timescales previously discussed. The dispersion rates are similar in all directions in a turbulent environment. 
The quiescent cases present different features. First of all, dispersion is much more effective in the settling direction than
in the normal one. 
The dispersion rate is smaller than the turbulent cases in the horizontal directions, while,  surprisingly,
the mean square displacement in the settling direction is similar to that of the turbulent cases being even higher at $\phi=0.5\%$, something we relate to the drafting-kissing-tumbling behaviour discussed above.
The crossover time scale is similar to that of turbulent cases with the exception of the most dilute case which does not reach a
fully diffusive behavior at $t\simeq500$. This long correlation time  makes  
the mean square displacement of this case higher than the corresponding turbulent case at long times.\\
In the insets of figure~\ref{fig:disper}a) and b) we show $\langle \Delta x_i^2 \rangle/t$ as a function of time. In all cases except the quiescent one at $\phi=0.5\%$, the diffusive regime is reached 
and it is possible to calculate the diffusion coefficients $D_e = \langle \Delta x_i^2 \rangle/(2t)$. For the turbulent case with $\phi=0.5\%$ we obtain $D_e=0.52$ and $0.28$ in the directions parallel 
and perpendicular to gravity while for $\phi=1\%$ we obtain $D_e=0.57$ and $0.32$. In the quiescent cases the diffusion coefficients in the horizontal directions  
are approximately $0.03$, whereas the coefficient in the gravity direction at $\phi=1\%$ is about $0.40$. \citet{csanady1963} proposed a theoretical estimate of 
the diffusion coefficients for pointwise particles. Using these estimates, we obtain approximately $D_e=1.4$ and $0.7$ in the directions parallel and perpendicular to gravity. 
These are about 
$2.5$ times larger than those found here for finite-size particles.

\subsection{Fluid statistics}

Table~\ref{tab:stat_fl} reports the fluctuation intensities of the fluid velocities for all cases considered. 
These are calculated by 
excluding the volume occupied by the spheres at each time step and  averaging over the number of samples  
associated with the fluid phase volume. 
As expected, the fluid velocity 
fluctuations are smaller in the quiescent cases than in the turbulent regime. In the quiescent 
case
 the rms of the velocity fluctuations is about $50\%$ larger in the settling direction than in the normal direction 
 because of the long range disturbance induced by the particles wakes. The increase of the volume fraction enhances 
 the fluctuations in both directions. 
 Fluctuations are always larger in the turbulent case, with the most significant differences compared to the quiescent cases in  the
 normal direction, where 
 the presence of the buoyant solid phase brakes the isotropy of the turbulent velocity fluctuations.

\begin{table}
  \begin{center}
\def~{\hphantom{0}}
  \begin{tabular}{ccccc}
                   & Quiescent $\phi=0.5\%$ & Turbulent $\phi=0.5\%$ & Quiescent $\phi=1\%$ & Turbulent $\phi=1$ \\
  $\sigma_{V_{f,\tau}}$           & $+0.18$                & $+0.28$                & $+0.25$                & $+0.29$ \\
  $\sigma_{V_{f,n}}$              & $+0.04$                & $+0.27$                & $+0.06$                & $+0.27$ \\
  \end{tabular}
  \caption{Fluctuation rms of the fluid velocities parallel $\sigma_{V_{f,\tau}}$ and perpendicular $\sigma_{V_{f,n}}$ to gravity.
  The turbulent fluid velocity undisturbed rms is $\sim0.3$ }
  \label{tab:stat_fl}
  \end{center}
\end{table}

Hence, the solid phase clearly affects the turbulent flow field. Although the present study focuses on the
settling dynamics, it is interesting to briefly discuss how turbulence is modulated. 
Modulation of isotropic turbulence by neutrally buoyant particles is examined in \cite{lucci2010}; however the results change due to buoyancy as investigated here. 
Typical turbulent quantities are reported 
in table~\ref{tab:conf_turb} where they are compared with the unladen case at $\phi=0$. The energy dissipation $\epsilon$ increases 
with $\phi$ becoming almost double at $\phi=1\%$.  This behavior is expected since 
 the buoyant particles inject energy in the system that is transformed into  kinetic energy of the fluid phase that has to be dissipated. 
 The higher energy flux, i.e.\ dissipation, is reflected in a reduction of the 
Kolmogorov length $\eta$.
The particles  reduce the velocity fluctuations, decreasing the turbulent kinetic energy level. The combined effect on $k$ and $\epsilon$
result 
 in a decrease of the Taylor microscale $\lambda$ and of $Re_{\lambda}$;  
likewise the
integral length $L_0$ and $Re_{L_0}$ also decrease. The reduction of the large and small turbulence scales 
 is associated to the additional energy injection from the settling particles. Energy injection
 occurs at the size of the particles, which is below the unperturbed integral scale $L_0$ explaining the
lowering of the effective integral $L_0$ and of Taylor $\lambda$ length-scales. This additional 
energy is transferred to the bulk flow in the particle wake. Associated to this energy input there is a new mechanism for dissipation that is the interaction of the flow with the no-slip surface
of the particles. 
The mean energy dissipation field in the particle reference frame  for the turbulent case 
with $\phi=0.5\%$ is therefore shown in figure~\ref{fig:eps_f}. After a statistically steady state is reached, the norm of the symmetric part of the velocity gradient tensor 
$E_{ij}$ and the dissipation $\epsilon = 2 \nu E_{ij}E_{ij}$ are calculated at each time step on a cubic mesh centered around each particle; the dissipation is calculated on the grid points outside 
the particle volume. The data presented have been averaged over all particles and time to get the mean dissipation field displayed in the figure.
The maximum $\langle \varepsilon \rangle$ is found around the particle surface with maximum values in the front; the mean dissipation drops down to the values found in the rest of the domain on the particle rear.
The overall energy dissipation is therefore made up of two parts:  the first associated to the dissipative eddies far from the 
particle surfaces and the second associated to the mean and fluctuating flow field 
near the particle surface. 
To conclude, the settling strongly alters the typical turbulence features  via an anisotropic energy injection and dissipation, thus
breaking the isotropy of the unladen turbulent flow. The energy is injected by the fluctuations in the particle wake whereas stronger energy dissipation occurs in the front of each particle. As a consequence, the 
fluid velocity fluctuations change in the directions parallel and perpendicular to gravity as shown in table~\ref{tab:stat_fl}.

\begin{table}
  \begin{center}
\def~{\hphantom{0}}
  \begin{tabular}{cccccccc}
      $\phi$ & $\eta/(2a)$  &   $k$ & $\lambda/(2a)$ & $Re_{\lambda}$ & $\epsilon$ & $T_e$ & $Re_{L_0}$ \\[3pt]
      0.000  & 0.084   & 0.13 & 1.56 & 90 & 0.0026 & 47.86 & 1205 \\
      0.005  & 0.077   & 0.10 & 1.19 & 62 & 0.0037 & 27.97 & 570 \\
      0.010  & 0.069   & 0.11 & 1.01 & 54 & 0.0055 & 19.88 & 435 \\
  \end{tabular}
  \caption{Turbulent flow parameters in particle units for $\phi=0$, $\phi=0.5\%$ and $\phi=1\%$.}
  \label{tab:conf_turb}
  \end{center}
\end{table}
\begin{figure}
  \centerline{\includegraphics[scale=0.4]{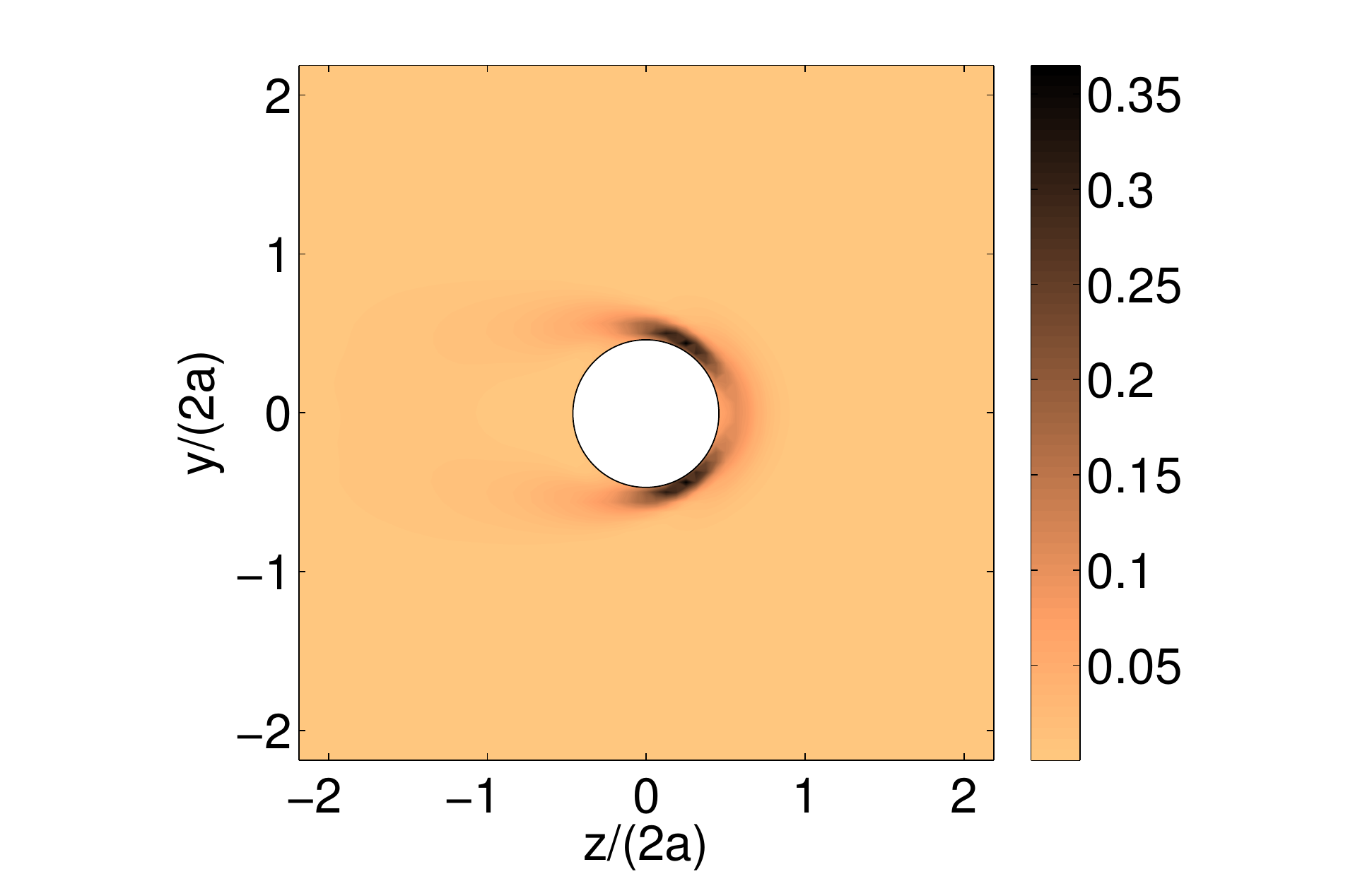}}% 
  \caption{Contours of the mean fluid  dissipation, $\langle \varepsilon \rangle$ ($v_t^3/(2a)$), averaged in the particle frame of reference.}
\label{fig:eps_f}
\end{figure}

\subsection{Relative velocity}

An important quantity to understand and model the settling dynamics is the particle to fluid relative motion.  
Although it is still unclear 
how to properly calculate the slip velocity between the two phases,
 we consider spherical shells around each particle, centered on the 
particles centroids, inspired by the works of 
\citet{bellani2012} and \citet{cisse2013}. 
We calculate the mean difference between the particle and fluid velocities in each shell as
$$\langle \vec U_{rel} \rangle_{\vec x,t,NP}=\left \langle \, \vec u_p- \frac{1}{\Omega(\Delta)} \int_{\Omega(\Delta)}
 \vec u_f d{\cal V} \,\right \rangle_{t,NP}$$ 
where $\Omega(\Delta)$ is the volume of a shell of inner radius $\Delta$. 
 A parametric study on the slip 
velocity is performed by changing the inner radii of these spherical shells from $\Delta=0.75$ particle diameters to $\Delta=5.0$ particle diameters, 
while keeping the shell thickness $\delta$ constant and equal to 0.063 in units of $2a$.
\begin{figure}
  \centerline{\includegraphics[scale=0.4]{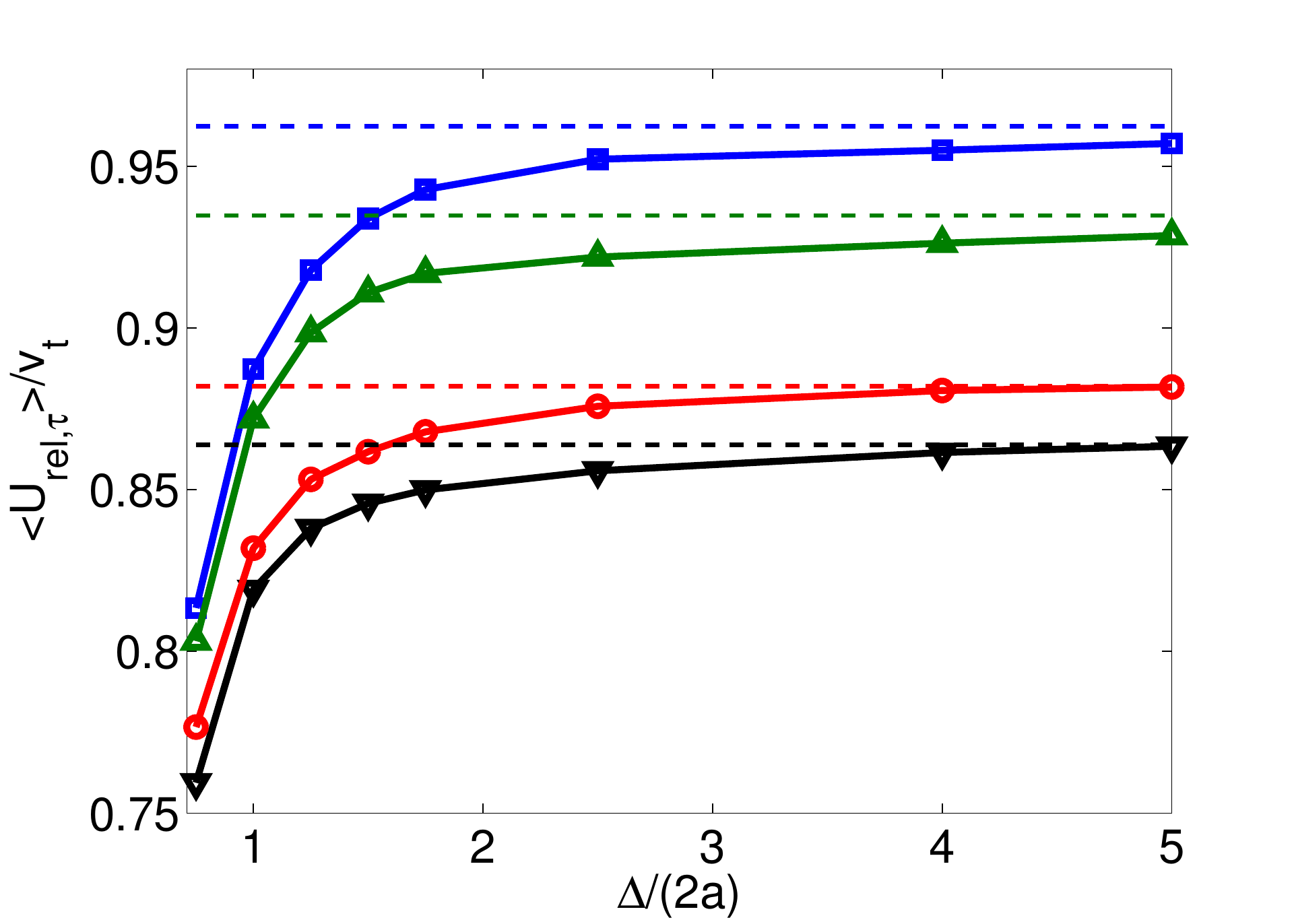}}% 
  \caption{The component of $\vec U_{rel}$ aligned with gravity as a function of the shell inner radii $\Delta/(2a)$ 
  for the four cases studied. The dashed lines represent instead the mean particles velocities in the direction of gravity, $\langle V_{p,\tau} \rangle$.}
\label{fig:conf_v_rel}
\end{figure}
In figure~\ref{fig:conf_v_rel} we 
report the component of $\vec U_{rel}$ parallel to gravity as a function of the shells inner radii $\Delta/(2a)$. As the shell inner radii 
increase, $|U_{rel,\tau}|$ tends exponentially to an asymptotic value which corresponds to the mean particle velocity in the same direction, $\langle V_{p,\tau} \rangle$. This is expected since the
correlation between the fluid and particle velocity goes to zero at large distances. 
The quiescent cases still show a $0.5\%$ difference between $|U_{rel,\tau}|$ and $\langle V_{p,\tau} \rangle$ at $\Delta/(2a)=5$. 
This difference  is again attributed to the long coherent wakes of the particles.

\begin{table}
  \begin{center}
\def~{\hphantom{0}}
  \begin{tabular}{ccccc}
      $\Delta/(2a)$           & $\langle U_{rel,\tau} \rangle$ & $\sigma_{U_{rel,\tau}}$ & $S_{U_{rel,\tau}}$ & $K_{U_{rel,\tau}}$ \\
    $0.75$ & $+0.81$              & $0.06$                & $+8.040$              & $123.29$ \\
    $1.00$ & $+0.88$              & $0.09$                & $+5.655$              & $68.26$ \\
    $1.25$ & $+0.92$              & $0.10$                & $+4.547$              & $ 47.66$ \\
    $1.50$ & $+0.93$              & $0.11$                & $+3.912$              & $ 37.69$ \\
    $1.75$ & $+0.94$              & $0.12$                & $+3.482$              & $ 31.83$ \\
    $2.50$ & $+0.95$              & $0.13$                & $+2.919$              & $ 24.34$ \\
    $4.00$ & $+0.95$              & $0.14$                & $+2.18$              & $ 16.25$ \\
    $5.00$ & $+0.96$              & $0.14$                & $+2.22$              & $ 17.09$ \\
  \end{tabular}
  \caption{Moments of the $pdf(U_{rel,\tau})$ for $\phi=0.5\%$ in the quiescent case. The thickness of the shell used to compute the slip velocity is $\delta/(2a)=0.063$.}
  \label{tab:vrel_dr_l}
  \end{center}
\end{table}
\begin{table}
  \begin{center}
\def~{\hphantom{0}}
  \begin{tabular}{ccccc}
    $\Delta/(2a)$            & $\langle U_{rel,\tau} \rangle$ & $\sigma_{U_{rel,\tau}}$ & $S_{U_{rel,\tau}}$ & $K_{U_{rel,\tau}}$ \\
    $0.75$ & $+0.76$              & $0.07$                & $+0.072$              & $ 3.90$ \\
    $1.00$ & $+0.83$              & $0.09$                & $+0.010$              & $ 4.28$ \\
    $1.25$ & $+0.85$              & $0.10$                & $+0.087$              & $ 4.56$ \\
    $1.50$ & $+0.86$              & $0.11$                & $+0.128$              & $ 4.54$ \\
    $1.75$ & $+0.87$              & $0.11$                & $+0.117$              & $ 4.32$ \\
    $2.50$ & $+0.87$              & $0.13$                & $+0.054$              & $ 3.91$ \\
    $4.00$ & $+0.88$              & $0.16$                & $+0.046$              & $ 3.32$ \\
    $5.00$ & $+0.88$              & $0.18$                & $+0.047$              & $ 3.10$ \\
  \end{tabular}
  \caption{Moments of the $pdf(U_{rel,\tau})$ for $\phi=0.5\%$ in a turbulent environment. The thickness of the shell used to compute the slip velocity is $\delta/(2a)=0.063$.}
  \label{tab:vrel_dr_t}
  \end{center}
\end{table}

\begin{figure}
  \centering
  \subfigure{%
    \includegraphics[scale=0.35]{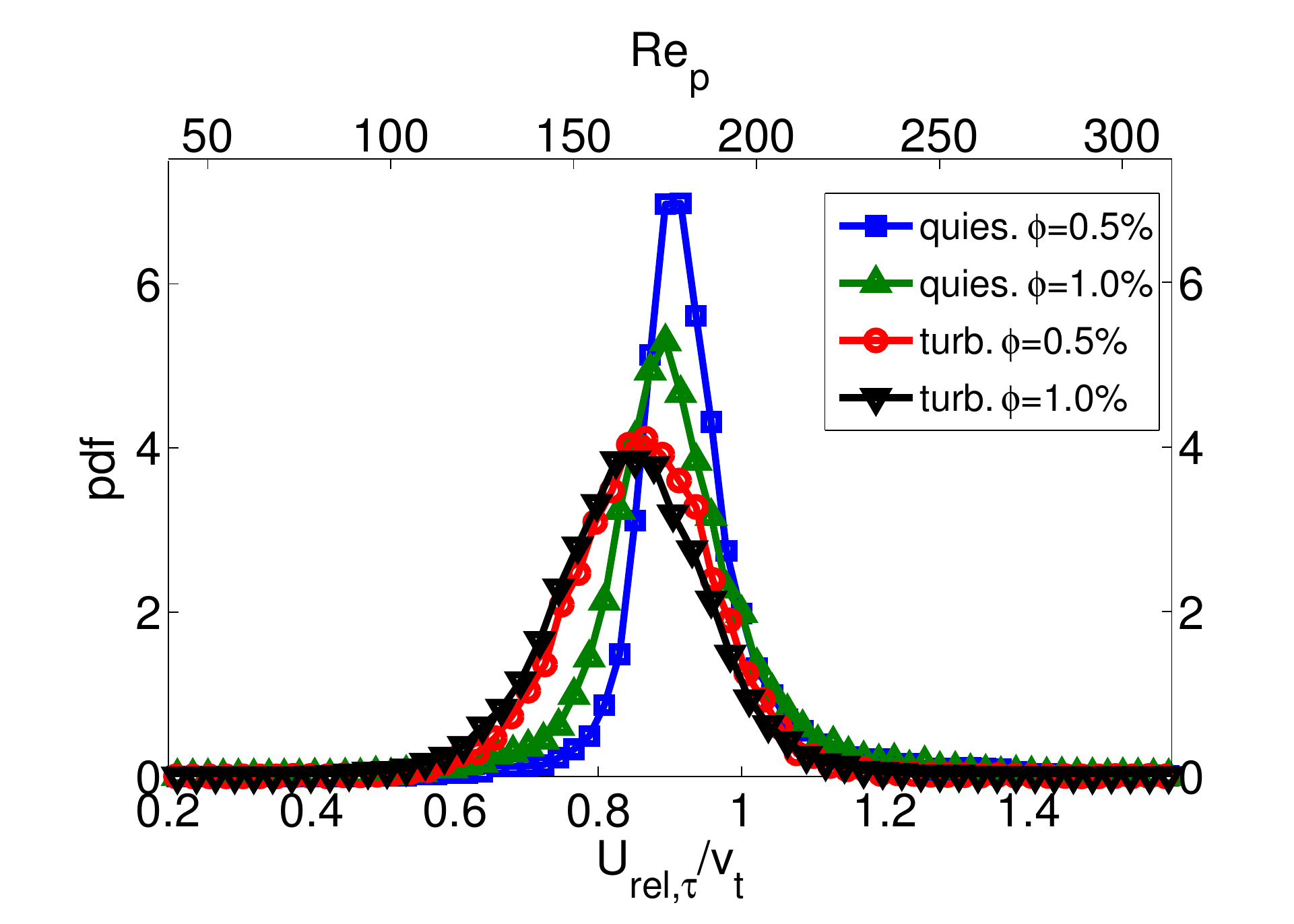}  \put(-193,100){{\large a)}}}% 
  \subfigure{%
    \includegraphics[scale=0.35]{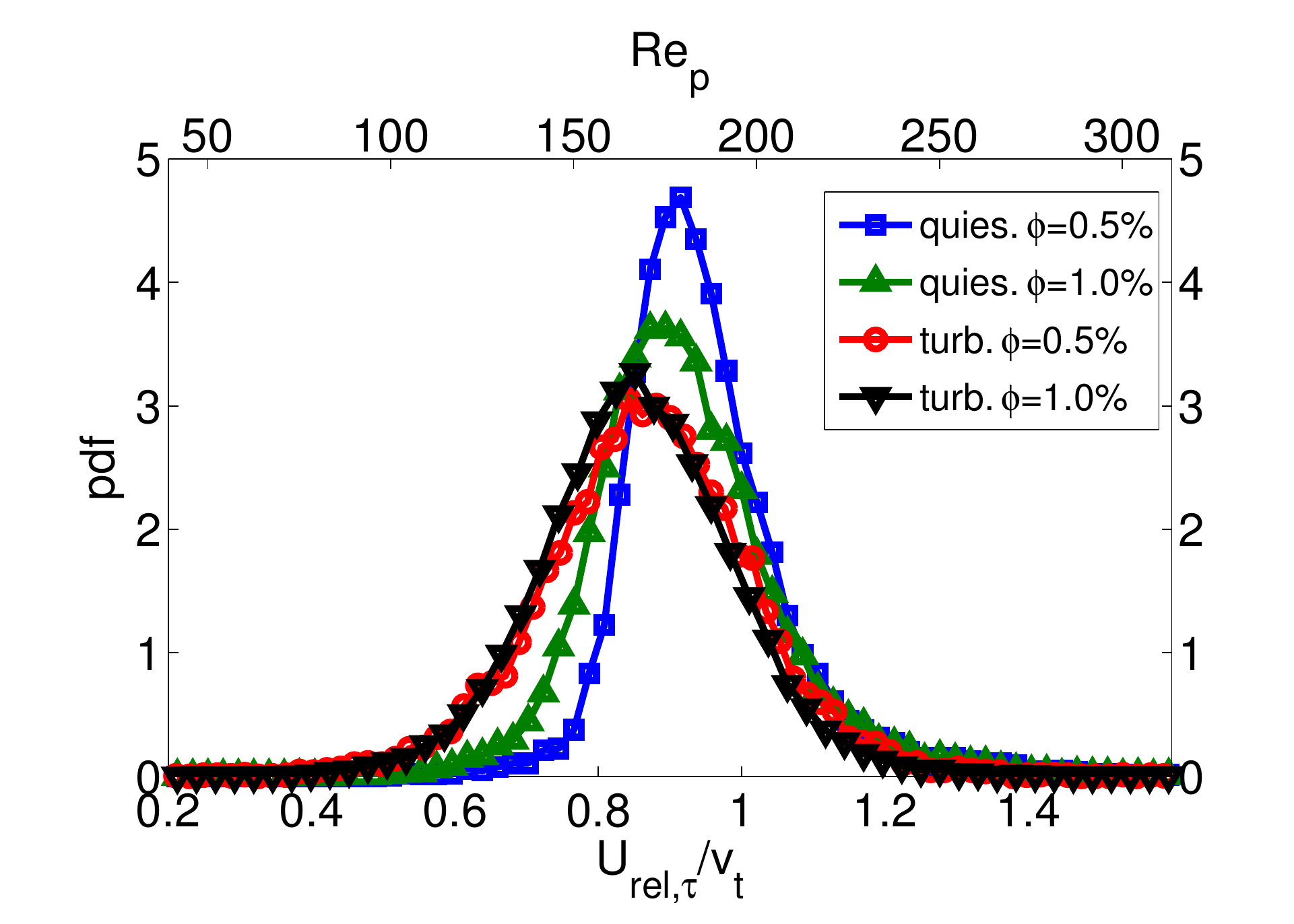} \put(-193,100){{\large b)}}}% 
\caption{Comparison of the probability density functions of tangential relative velocity and corresponding particle Reynolds number $Re_p$. Colors and symbols are the same 
as previous figures. In Panel a) the cases for $\Delta/(2a)=1.5$ are 
reported while in panel b) we show the results for $\Delta/(2a)=2.5$.}
\label{fig:pdf_urel}
\end{figure}

The probability density function of these relative velocities, $U_{rel,\tau}$, and their first four central moments are computed and reported in tables~\ref{tab:vrel_dr_l} and~\ref{tab:vrel_dr_t} 
as a function of $\Delta/(2a)$ for the quiescent fluid and turbulent cases at $\phi=0.5\%$. In the 
turbulent case, the moments approach those of a  Gaussian distribution with vanishing skewness and 
flatness close to $3$, especially at large $\Delta$. In the quiescent case, the third and fourth moments display higher values that decrease as $\Delta$ is increased, tending 
to the values of the particle velocities. 
The $pdf$s pertaining the four cases considered, calculated in spherical shells with an inner radius of $1.5$ and $2.5$ particle diameters, are compared in figure~\ref{fig:pdf_urel}. 
A second axis, reporting the the particle Reynolds number $Re_p$ based on $U_{rel,\tau}$, is also displayed in each figure.
In the former case, $\Delta/ (2a)=1.5$, the shell radius is of the order 
of the Taylor scale to highlight  the particle dynamics, while the relative velocity is approaching the asymptotic values for the larger shell.
The $pdf$s of the relative velocity appear narrower
than those of the particle absolute velocity, indicating that the particles tend to be transported by
the large-scale motions, filtering the smallest scales.

The distributions pertaining the simulations in a turbulent environment
 are nearly
Gaussian with modal values well below one. 
The quiescent cases show skewed distributions with long tails at high velocities, as observed for the particle velocities in figure~\ref{fig:pdf_vp}a). 
The particles settle on average with a velocity close to that of a single particle, with occasional events of higher velocity due
to the drafting-kissing-tumbling dynamics. The lower the volume fraction the more intermittent
is the dynamics. 

\begin{figure}
  \centering
  \subfigure{%
    \includegraphics[scale=0.34]{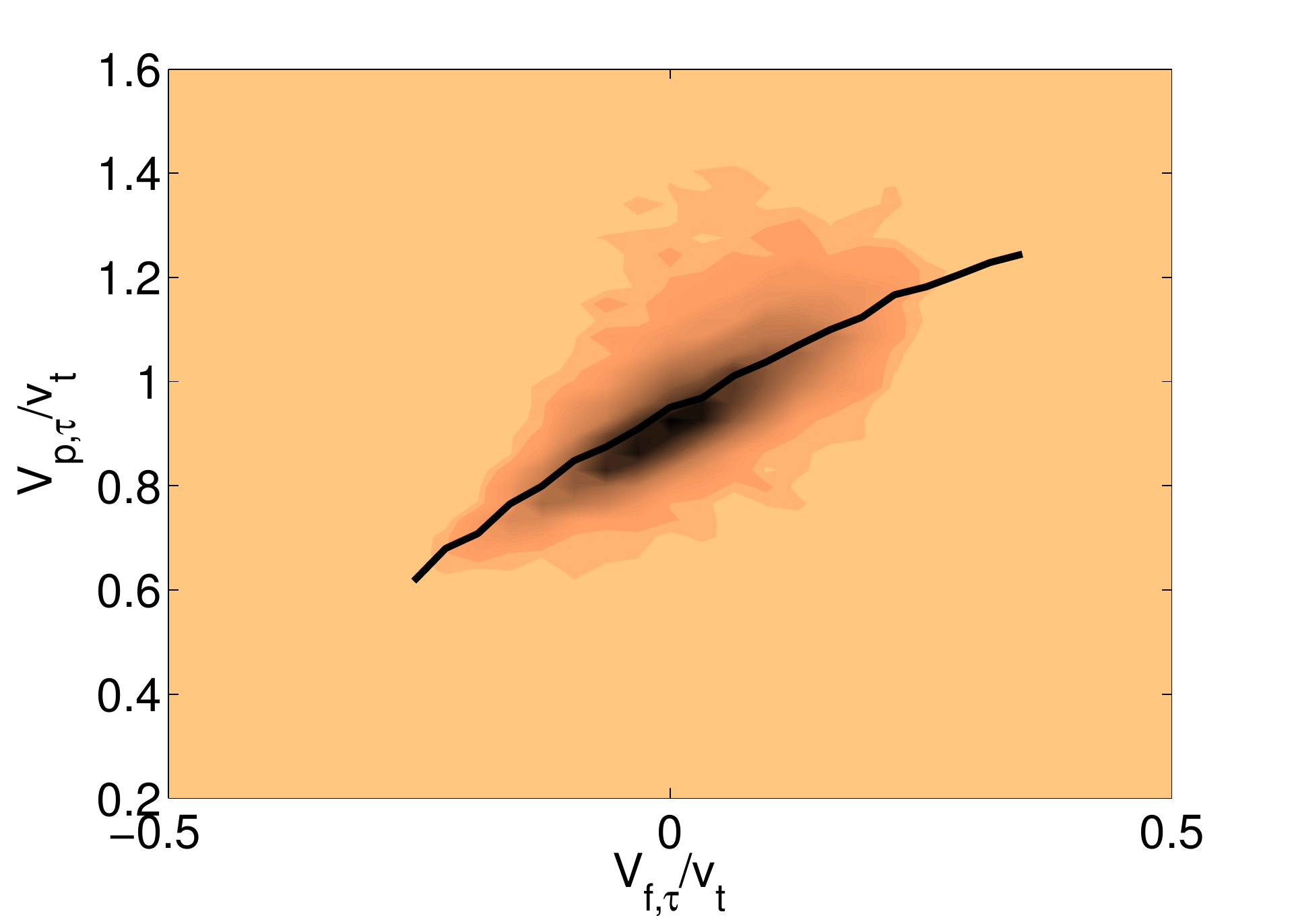} \put(-193,100){{\large a)}}}% 
%  \quad
  \subfigure{%
    \includegraphics[scale=0.34]{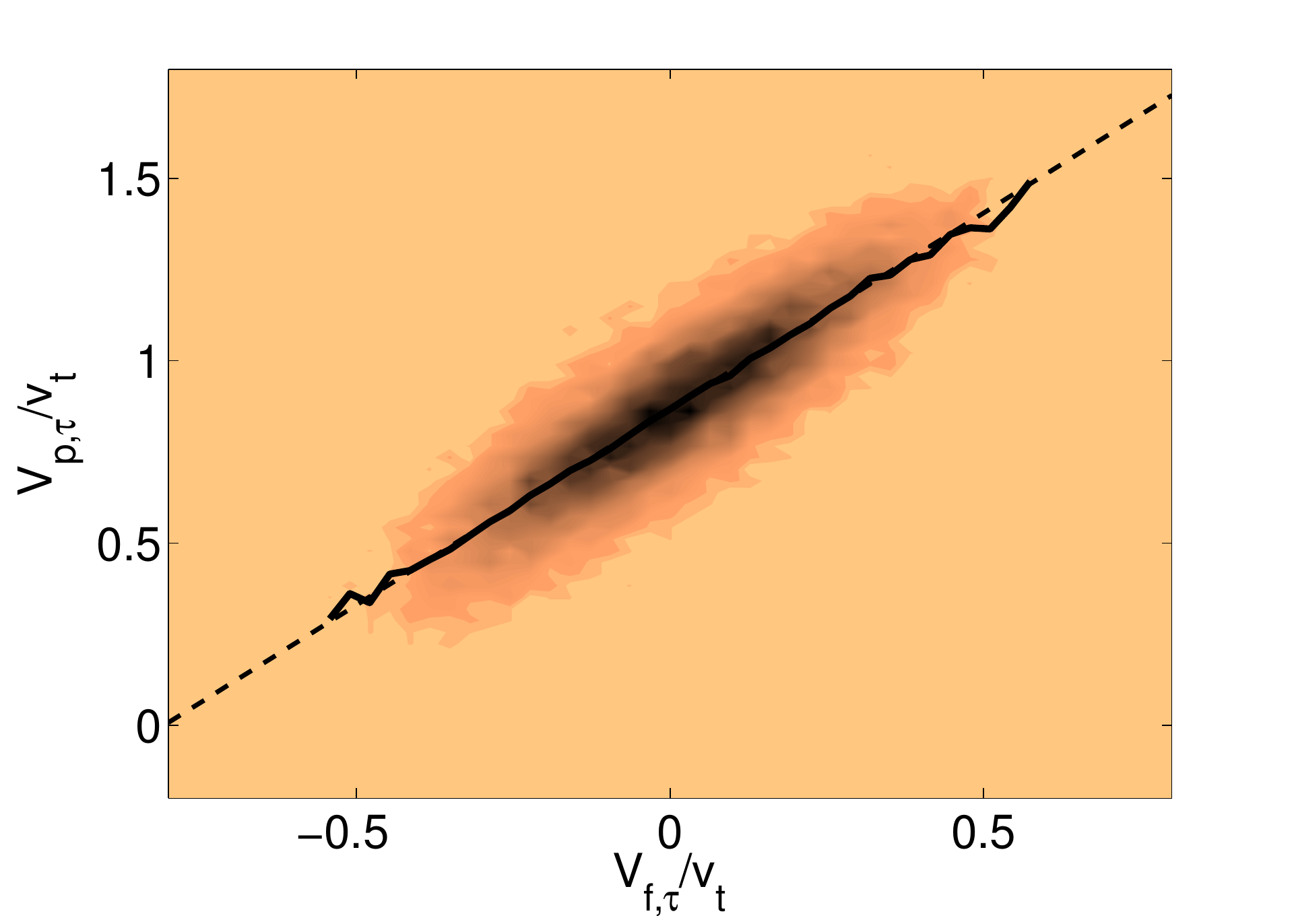} \put(-193,100){{\large b)}}} 
  \subfigure{%
    \includegraphics[scale=0.34]{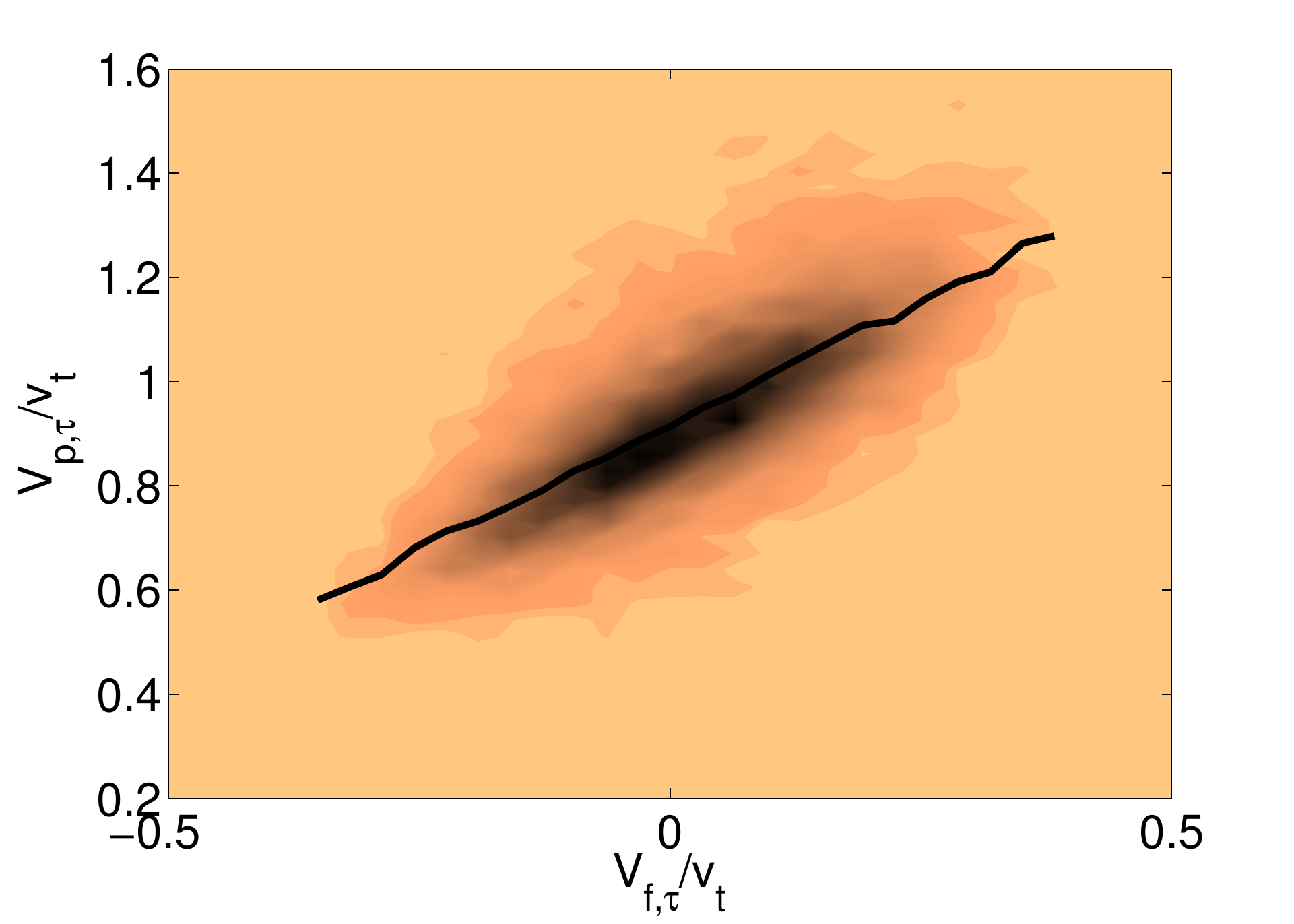} \put(-193,100){{\large c)}}} 
  \subfigure{%
    \includegraphics[scale=0.34]{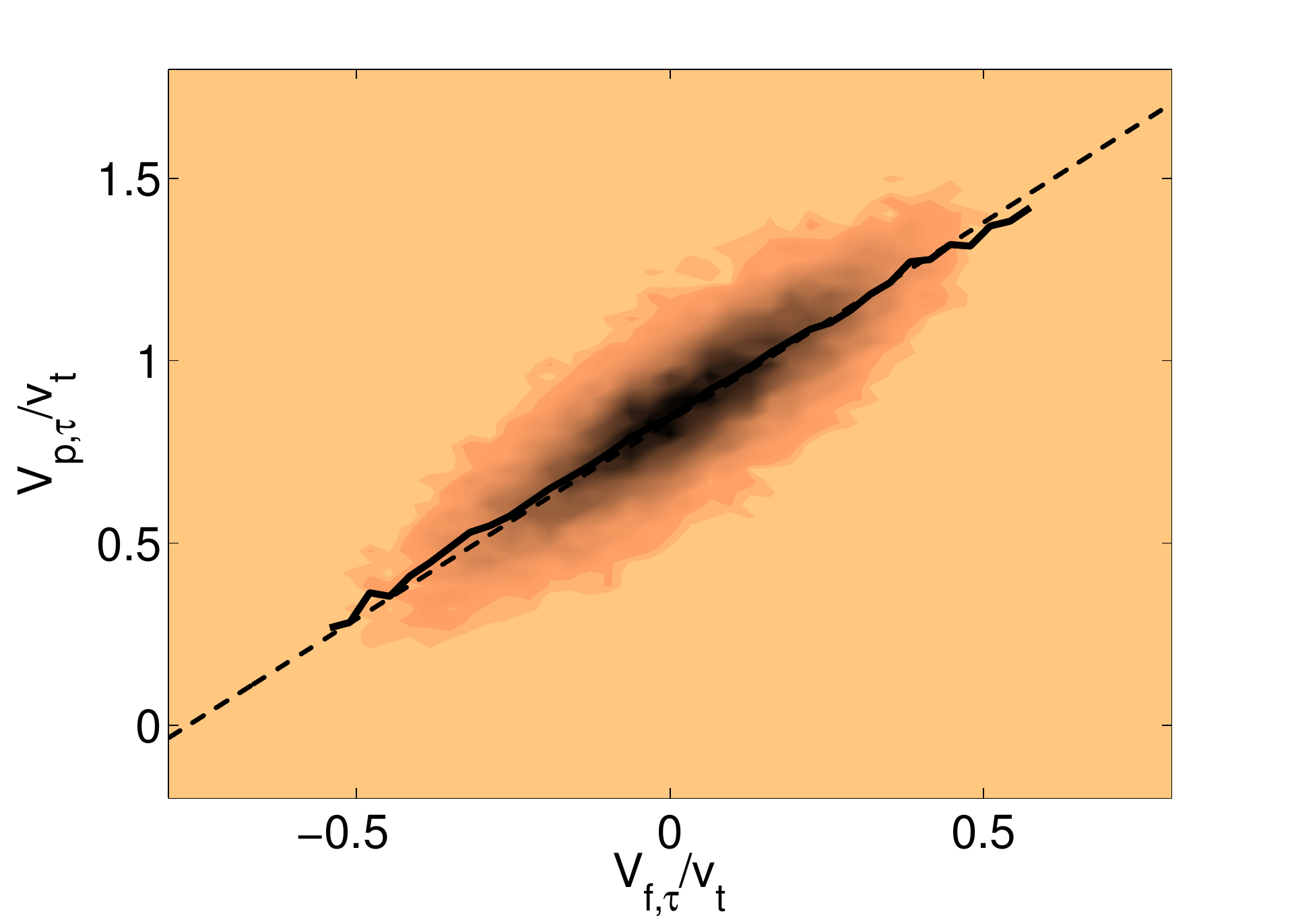}\put(-193,100){{\large d)}} } 
\caption{Joint probability distributions of $V_{f,\tau}$ and $V_{p,\tau}$ evaluated in spherical shells located at $1.75$ particle diameters from each particle. We report in panels a) and c) 
the quiescent cases for $\phi=0.5\%$ and $1\%$, while the respective turbulent cases are reported in panels b) and d). The continuous lines represent the integrals of the joint 
probability distributions. In the turbulent cases the dashed lines represent the best-fit of these integrals.}
\label{fig:jointp}
\end{figure}

Knowing the tangential fluid velocity $\langle V_{f,\tau} \rangle_{\vec r}(p,t)$, averaged in each shell and at each time step, and the corresponding tangential particle 
velocities $V_{p,\tau}(p,t)$, it is then possible to find their joint probability distribution $P(V_{f,\tau},V_{p,\tau})$ (for sake of simplicity we write 
$\langle V_{f,\tau} \rangle_{\vec r}$ as $V_{f,\tau}$). These are evaluated in shells at $\Delta/(2a)=1.75$ for each case studied and reported in figure~\ref{fig:jointp}. In each 
case, the integral of $P(V_{f,\tau},V_{p,\tau})$, 
\begin{equation}
\langle V_{p,\tau}|V_{f,\tau} \rangle = \int_{-\infty}^{\infty} P(V_{f,\tau},V_{p,\tau}) V_{f,\tau}\, dV_{f,\tau},
\label{int_joint}
\end{equation}
is also reported (continuous lines).
This represents the most probable particle velocity $V_{p,\tau}$ given a certain fluid velocity $V_{f,\tau}$, or, equivalently, the most probable fluid velocity surrounding a particle settling with velocity $V_{p,\tau}$. In the turbulent cases these integrals are well approximated by straight 
lines (displayed with dashed lines in the figure)
\begin{equation}
V_{p,\tau} = C_1 V_{f,\tau} + C_2.
\end{equation}
In both cases $C_1$ is approximately $1$ while $C_2$ is about $0.86$ for $\phi=0.5\%$ and $0.84$ for $\phi=1\%$. These values are in agreement with the values found for the 
average relative velocities of shells at $\Delta/(2a)=1.75$. In a quiescent flow, conversely, the integral in eq.~(\ref{int_joint}) gives a curved line and no best-fit is therefore reported. 
In these cases, the joint probability distribution is broader, particularly in the region of higher particle velocities, $V_{p,\tau}$. This is again due to the intense particle interactions and the drafting-kissing-tumbling behaviour described in figure~\ref{fig:kissing}, which  confirms the high flatness of the probability density functions of the relative particle velocities.

Further insight can be obtained by plotting isocontours of the average  particle relative velocities and their 
fluctuations, in both quiescent and turbulent flows. To this end, we follow the approach by \citet{garcia2012}. We place an uniform and 
structured rectangular mesh around each particle, with origin at the particle center. By means of trilinear interpolations we find 
the fluid and relative velocities on this local mesh and average 
over time and the number of particles to obtain the mean relative velocity field and its fluctuations.

\begin{figure}
  \centerline{\includegraphics[scale=0.4]{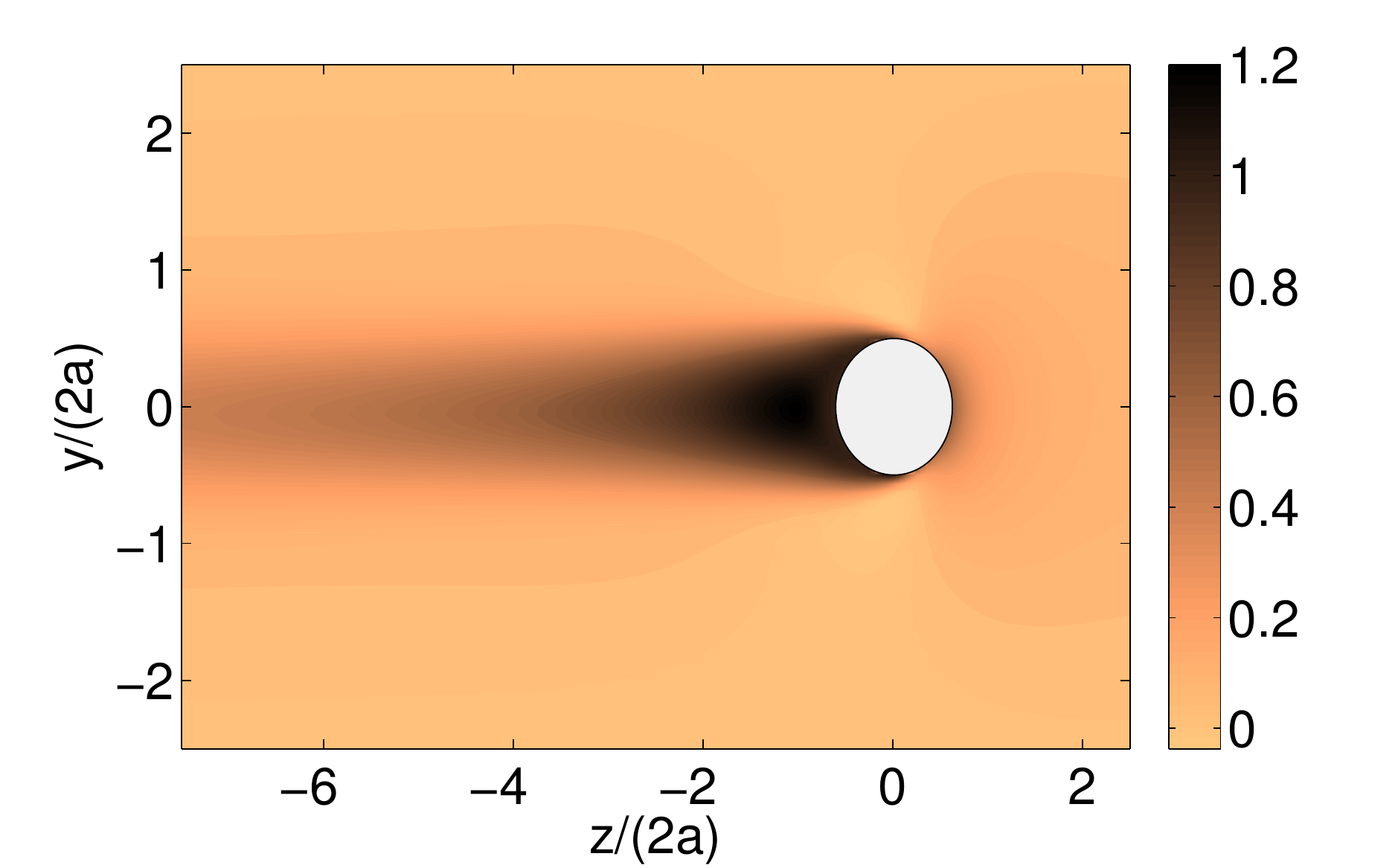}}% 
  \caption{Contour plot of the velocity component in the direction of gravity for the single sphere settling in a quiescent fluid.}
\label{fig:field_rel}
\end{figure}

The vertical velocity of the single sphere settling in a quiescent fluid is reported in figure~\ref{fig:field_rel}.
The relative normal and tangential 
velocities and their fluctuations are instead displayed in figure~\ref{fig:fields_rel} for suspensions with $\phi=0.5\%$ in both quiescent and turbulent environment. 
In the quiescent fluid simulations, a long wake forms behind the 
representative particle and, as seen from the single point particle velocity correlations, it takes a long time for this velocity fluctuations 
to decorrelate. 
In the turbulent case instead, the wakes are disrupted by the background fluctuations. 

Interesting observations can be drawn from the relative velocity fluctuation fields. 
Comparing figures~\ref{fig:fields_rel}c) 
and~\ref{fig:fields_rel}d) we note that intense vortex shedding occurs around the particles in the turbulent case, 
with important fluctuations of $U_{rel,\tau}$. From figures~\ref{fig:fields_rel}e) and~\ref{fig:fields_rel}f) we 
also see that the relative velocity fluctuations are drastically increased in the horizontal directions in a turbulent 
environment. Noteworthy, vortex shedding occurs at particle Reynolds numbers below the critical value above which this 
is usually observed \citep{bouchet2006}. Vortex shedding is unsteady in nature and unsteady effects may therefore play an 
important role in the increase of the overall drag, as further discussed below. Lower fluctuation intensities are found on the 
front part of the particles, where the energy dissipation is highest, and the immediate wake  in the 
recirculating region where the instability is found to develop.\\

\begin{figure}
  \centering
  \subfigure{%
    \includegraphics[scale=0.32]{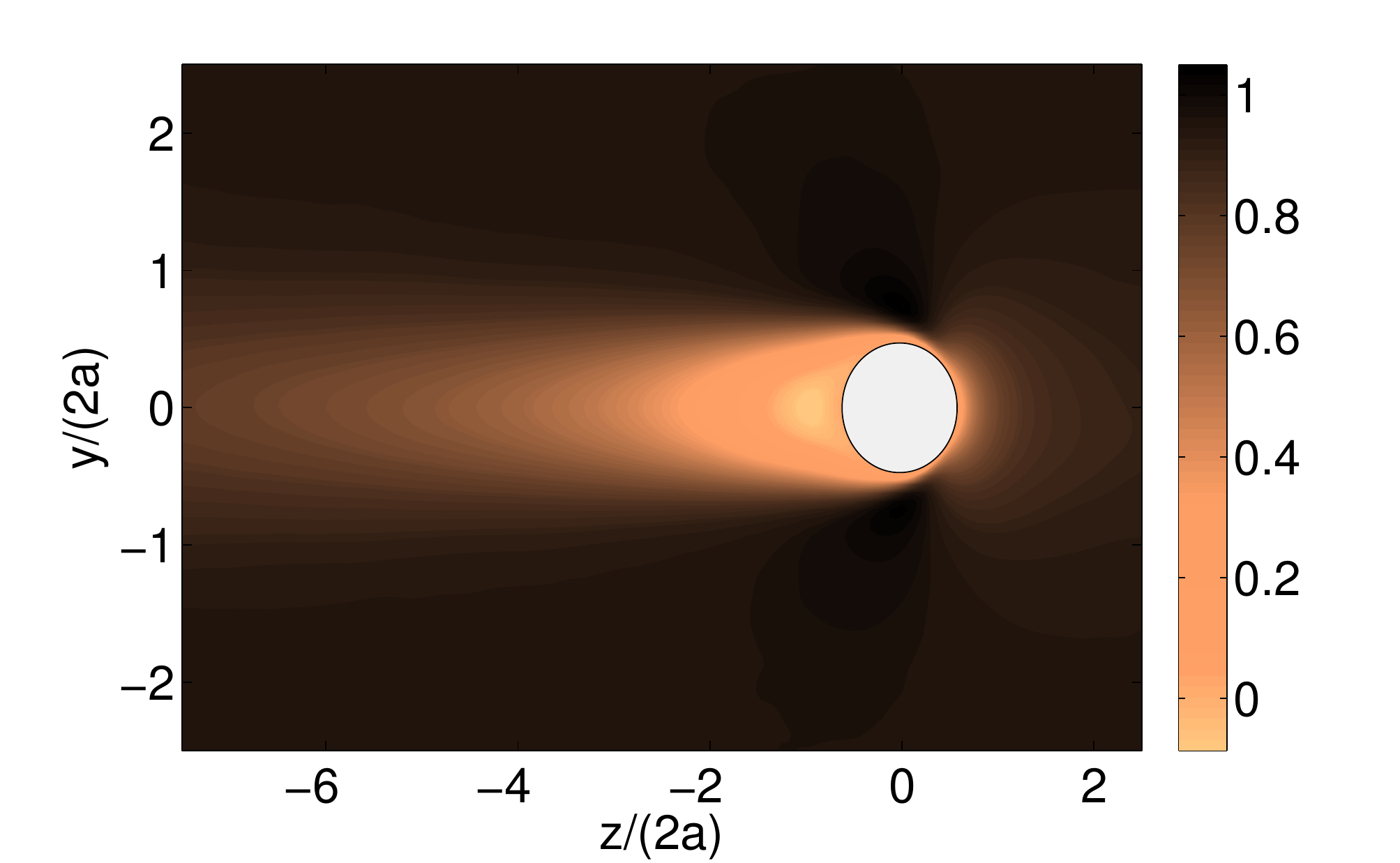} \put(-185,90){{\large a)}}}%
  \subfigure{%
    \includegraphics[scale=0.32]{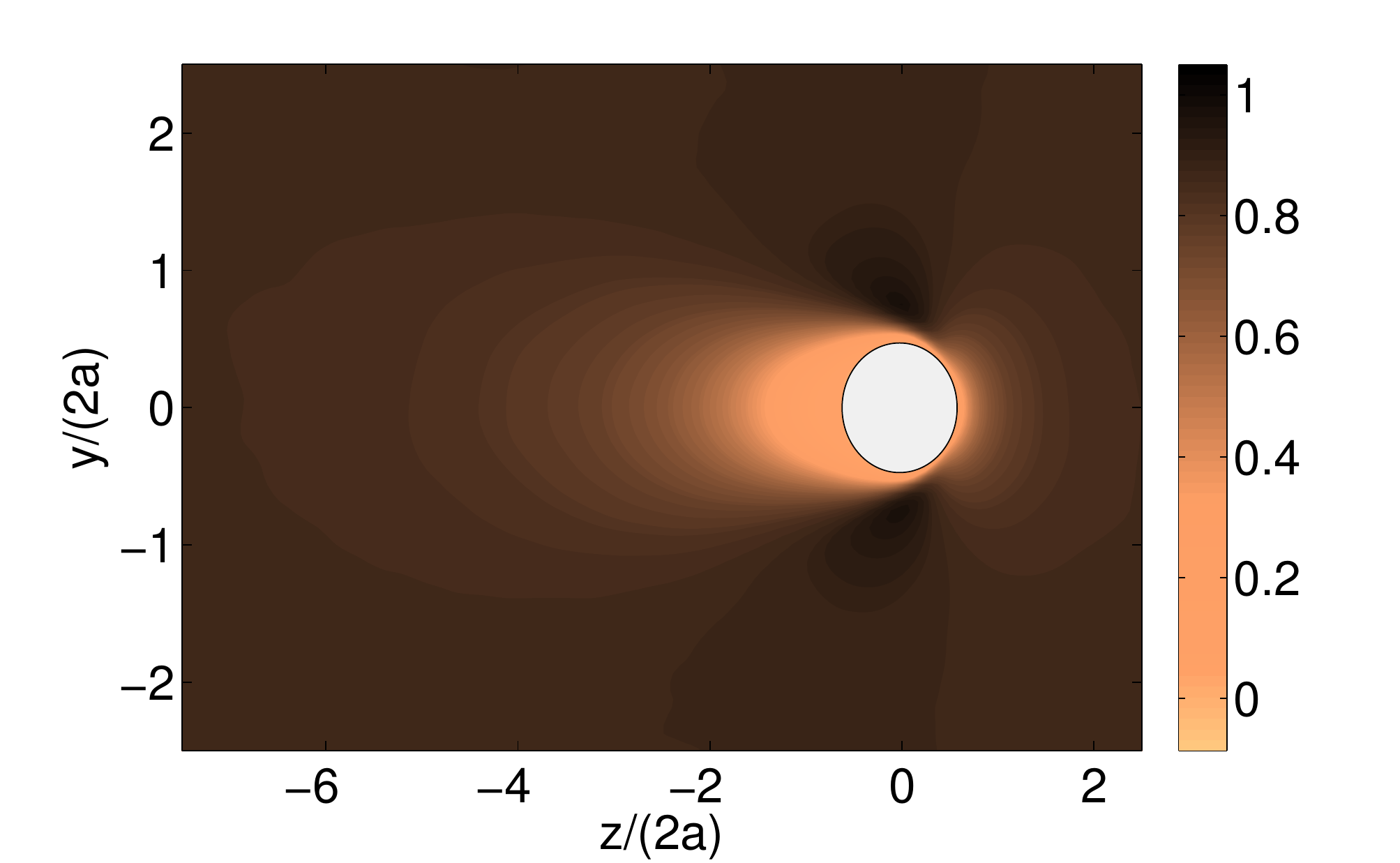} \put(-185,90){{\large b)}}}
  \subfigure{%
    \includegraphics[scale=0.32]{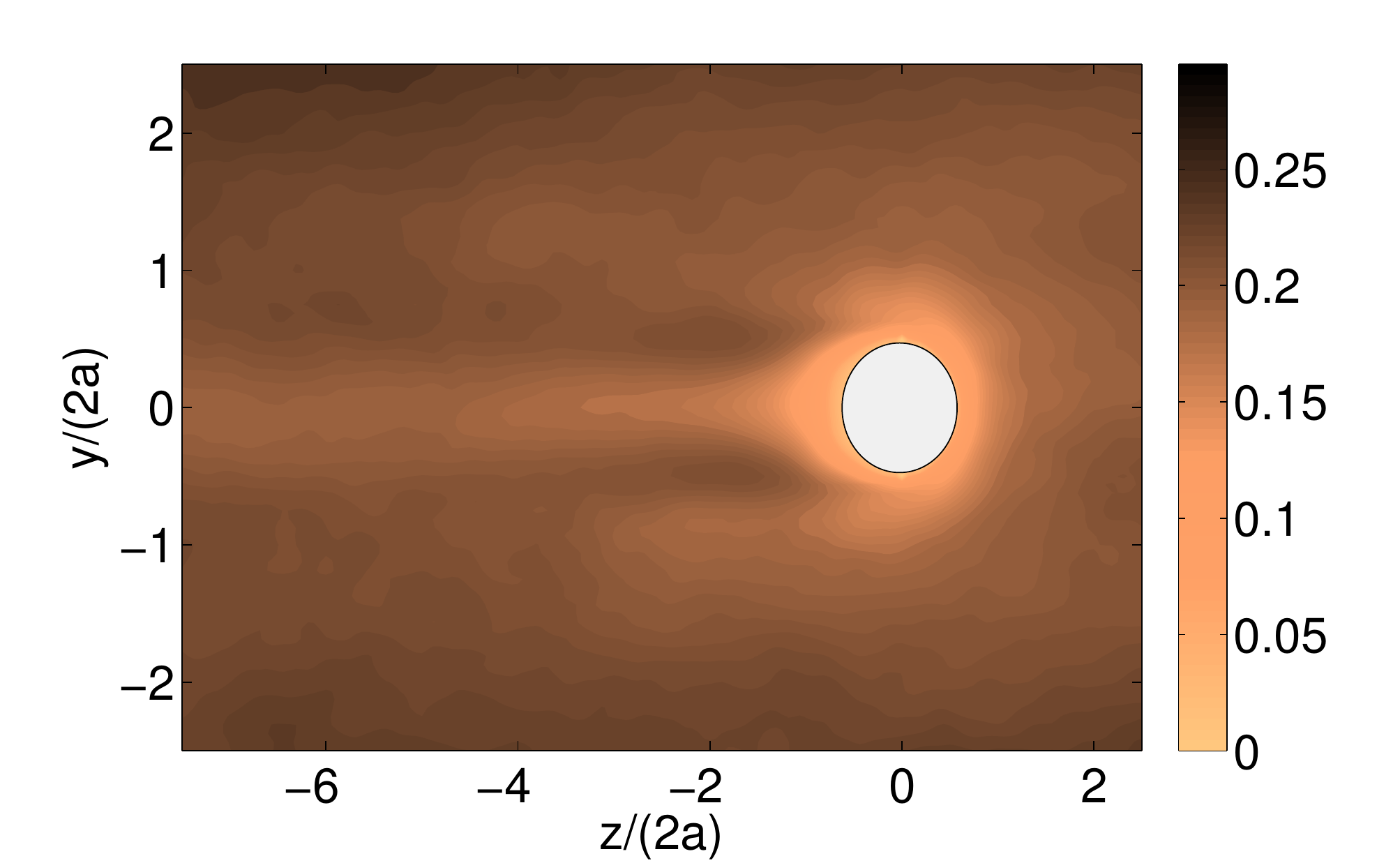} \put(-185,90){{\large c)}}}
  \subfigure{%
    \includegraphics[scale=0.32]{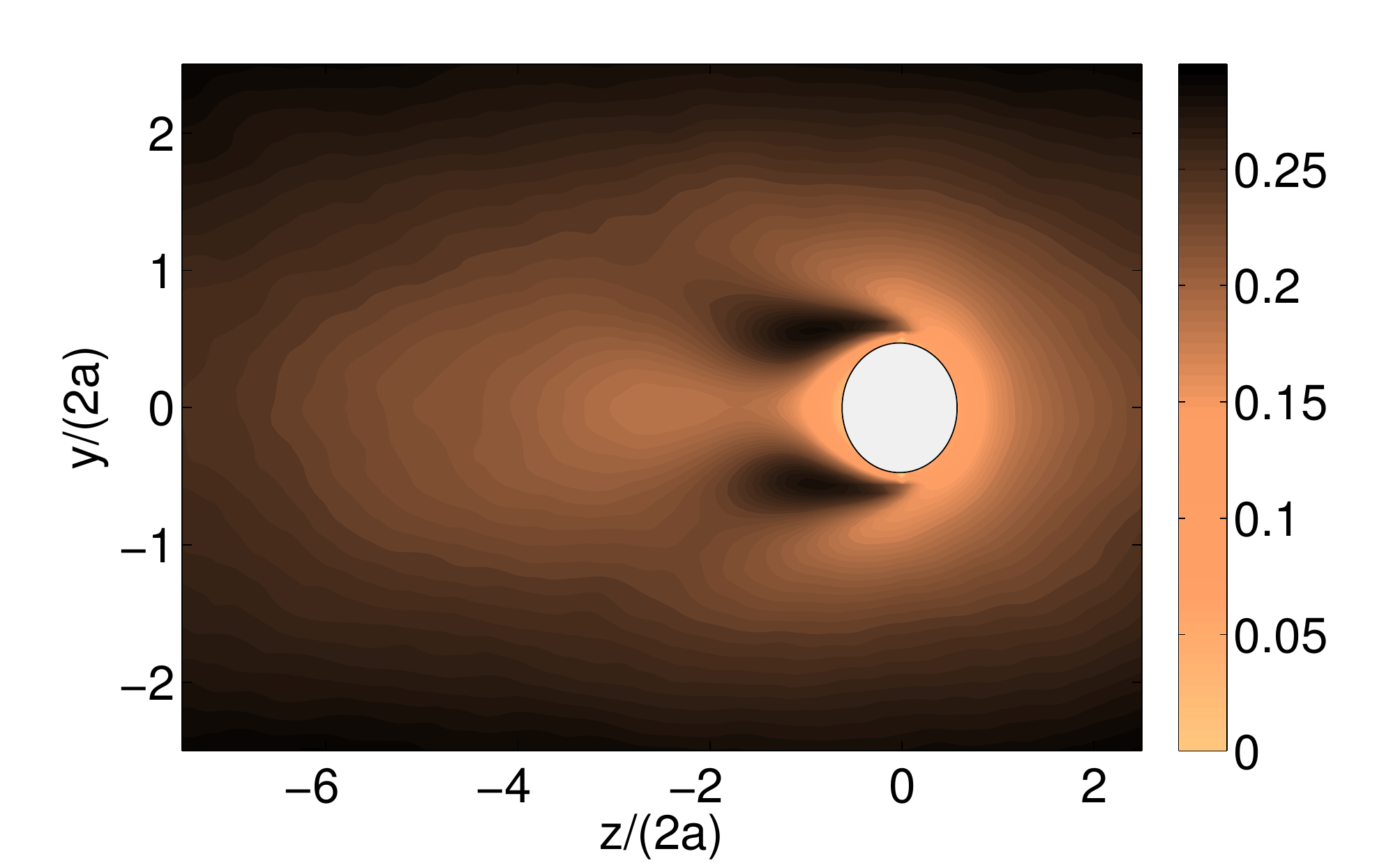}  \put(-185,90){{\large d)}}}
  \subfigure{%
    \includegraphics[scale=0.32]{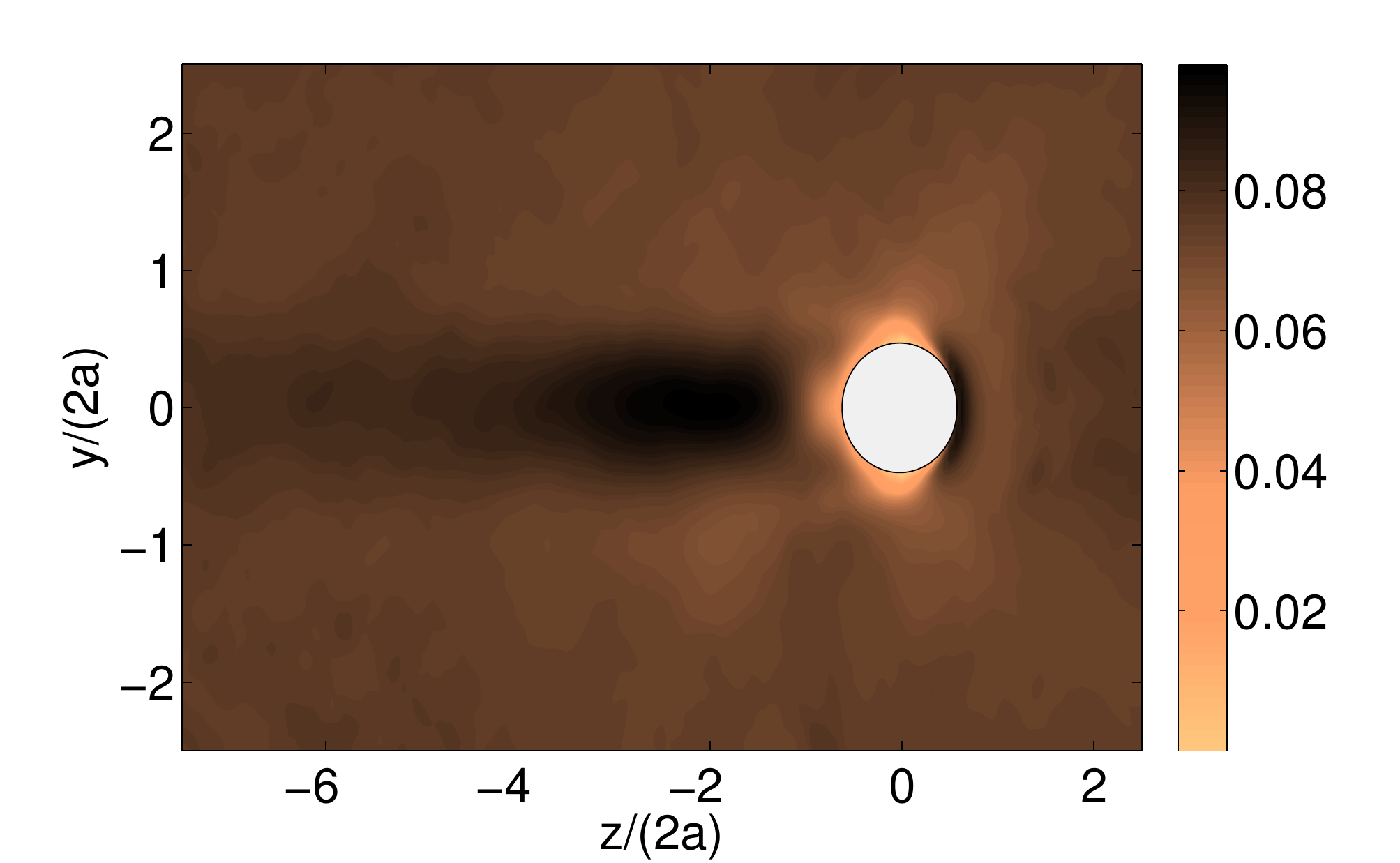} \put(-185,90){{\large e)}}}
  \subfigure{%
    \includegraphics[scale=0.32]{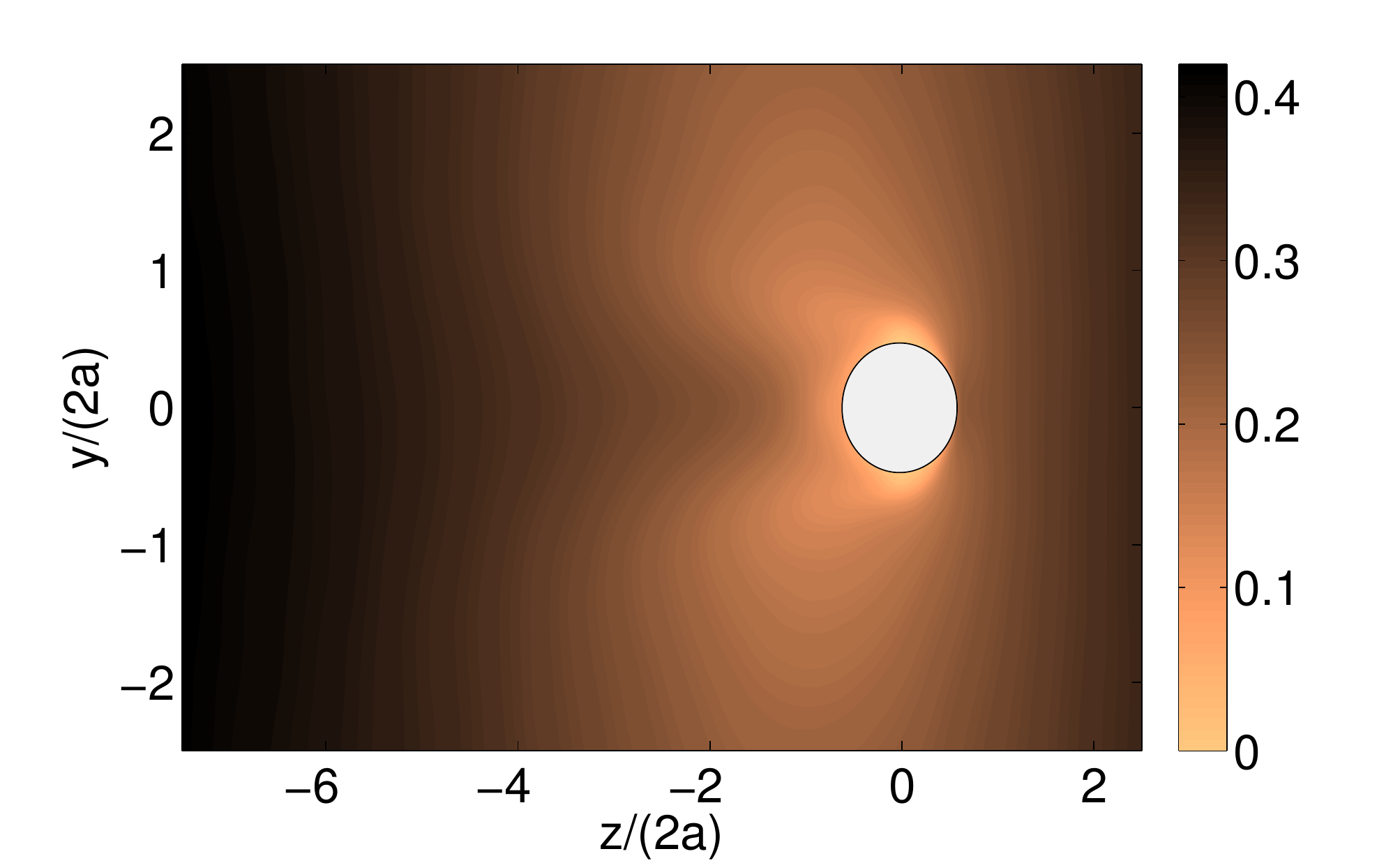} \put(-185,90){{\large f)}}}
\caption{Fields of $\langle U_{rel,\tau} \rangle$, $U_{rel,\tau}'$, $U_{rel,n}'$ for the quiescent (left column) and turbulent cases (right column).}
\label{fig:fields_rel}
\end{figure}

\subsection{Drag analysis}
As in \citet{maxey1983}, we write the balance of the forces acting on a single sphere settling through a turbulent flow. The equation of motion for a spherical particle reads
\begin{equation}
\label{eq_sph}
\frac{4}{3} \pi a^3 \rho_p \td{\vec V_p}{t} = \frac{4}{3} \pi a^3 (\rho_p - \rho_f) \vec g + \oint_{\partial \mathcal{V}_p}^{} \vec \taub \cdot \vec n\, dS
\end{equation}
where the integral is over the surface of the sphere $\partial \mathcal{V}_p$, $\vec n$ is the outward normal and $\vec \taub = -p \vec I + 2\mu \vec E$ is the fluid stress. As commonly done 
in aerodynamics, we replace the last term of equation~(\ref{eq_sph}) with a term depending on the relative velocity $\vec U_{rel}$ and a drag coefficient $C_D$
\begin{equation}
\label{eq_sph2}
\frac{4}{3} \pi a^3 \rho_p \td{\vec V_p}{t} = \frac{4}{3} \pi a^3 (\rho_p - \rho_f) \vec g - \frac{1}{2} \rho_f \pi a^2 |\vec U_{rel}| \vec U_{rel} C_D
\end{equation}
with $\pi a^2$ the reference area. Generally, the drag coefficient $C_D$ is a function of a Reynolds number and a Strouhal number which accounts for unsteady effects. In the present case, we consider 
it to be a function of the Reynolds number based on the relative velocity $Re_p = 2a|\vec U_{rel}|/\nu$ (in a turbulent field it is proper to define $Re_p$ in terms of the relative velocity between 
particle and fluid) and of the Strouhal number defined as 
\begin{equation}
St = \frac{\td{\vec V_p}{t} (2a)}{|\vec U_{rel}|^2}.
\end{equation}
The drag on the particle depends on both nonlinear and unsteady effects (such as the Basset history force and the added mass contribution) through these two non-dimensional numbers.

Commonly, the unsteady contribution is neglected and $C_D$ is assumed to depend only on the Reynolds number. Since we want to investigate both nonlinear and unsteady effects, we decide to express 
the drag coefficient as $C_D(Re_p,St) = C_{D_0}(Re_p) \left[1+\psi(Re_p,St)\right]$, yielding  
\begin{equation}
\label{eq_sph3}
\frac{4}{3} \pi a^3 \rho_p \td{\vec V_p}{t} = \frac{4}{3} \pi a^3 (\rho_p - \rho_f) \vec g - \frac{1}{2} \rho_f \pi a^2 |\vec U_{rel}| \vec U_{rel} C_{D_0}(Re_p) \left[1+\psi(Re_p,St)\right],
\end{equation}
where $\psi=0 \, \forall \, Re_p$ if $St=0$ (steady motion). 
We can therefore identify a quasi-steady term and the extra term which accounts for unsteady phenomena.

By ensemble averaging equation~(\ref{eq_sph3}) over time and the number of particles, and assuming the 
settling process to be at statistically steady state, we can find the most important contributions to the overall drag. The steady-state 
average equation reads
\begin{equation}
\label{eq_sph_av}
0 = \frac{4}{3} \pi a^3 (\rho_p - \rho_f) \vec g - \frac{1}{2} \rho_f \pi a^2 \left \langle |\vec U_{rel}| \vec U_{rel} C_{D_0}(Re_p) \left[1+\psi(Re_p,St)\right] \right \rangle.\\
\end{equation}
Denoting the entire time dependent term simply as $\Psi(t)$ and rearranging, we obtain the following balance
\begin{equation}
\label{eq_sph_av1}
\frac{4}{3} \pi a^3 \left(\frac{\rho_p}{\rho_f}-1\right) \vec g = \frac{1}{2} \pi a^2 \langle |\vec U_{rel}| \vec U_{rel} C_{D_0}(Re_p)\rangle + \Psi(t).
\end{equation}
The term on the left-hand side is known whereas the time-dependent term $\Psi(t)$ is of difficult evaluation. The nonlinear steady term can be 
 calculated using the approach described in section~\ref{sec:results}.2. At each time step we calculate the relative velocity in the spherical shells 
surrounding each particle. From these we compute the particle Reynolds number $Re_p=2a|U_{rel}|/\nu$ and using equation~(\ref{Cd_re}) (where we replace 
the terminal Reynolds number with the new particle Reynolds number) we obtain the drag coefficient. The first term on the right-hand side can therefore be evaluated by averaging over the number of particles and time steps considered. Finally we divide everything by the buoyancy acceleration to estimate the 
relative importance of each term
\begin{equation}
\label{eq_sph_av2}
100\% = \mathcal{S}(Re_p) + \Psi(t)
\end{equation}
where $\mathcal{S}(Re_p)$ represents the nonlinear steady term while the unsteady term has been denoted again as $\Psi(t)$ for simplicity.

This approach is applied to the results of the simulations of a 
single sphere and to the quiescent and turbulent cases with $\phi=0.5\%$ and 1\%. The  inner radius of the sampling shells is chosen to be $5$ particle radii and the 
results obtained are reported in table~\ref{tab:n_psi}. 
The single sphere simulation provides an estimate of the error of the method. Since our terminal 
Reynolds number is smaller than the critical Reynolds number above which unsteady effects become important and our velocity field is indeed steady, the  term $\Psi(t)$ 
should be negligible. The critical Reynolds number $Re_c$ has been found to be approximately $274$ by \citet{bouchet2006} while we recall that our terminal Reynolds number for 
the isolated particle is $200$.
The nonlinear  steady term provides in this case, however, an overestimated value of the drag, with a percentage error of about $+3\%$ with respect to the buoyancy term. 
The possible causes of this error are the long 
wake
and the fact that equation~(\ref{Cd_re}) is empirical. The data from the single-particle simulations are  used to correct the results from the other runs, i.e. the data are normalized with the total drag obtained in this case. 
In the quiescent case at $\phi=0.5\%$, 
unsteady effects are negligible (about 0.5\% of the total drag), while they increase to about 6\% when increasing the particle evolve fraction to 1\%.
 In a turbulent
flow, importantly, we notice that the contribution of $\Psi(t)$ adds up to approximately $10\%$ of the total at $\phi=0.5\%$ and to about $12\%$ of the total at $\phi=1\%$. 

Note that one can write the steady drag as mean and fluctuating component 
\[\mathcal{S}(Re_p) =  \langle |\vec U_{rel}| \vec U_{rel} C_{D_0}(Re_p) \rangle = \langle U_{rel} \rangle^2  C_{D_0}(\langle Re_p \rangle) + \mathcal{S}'(\langle Re_p\rangle). \]
The fluctuations  $\mathcal{S}'(\langle Re_p\rangle)$ would be responsible of the reduction of the settling velocity if this were to be attributed to nonlinear drag effects 
only \cite[see also][]{wang1993}.
We verified that for our results, the total and mean component differ by about  2\%, $\langle |\vec U_{rel}| \vec U_{rel} C_{D_0}(Re_p) \rangle \approx \langle U_{rel} \rangle^2  C_{D_0}(\langle Re_p \rangle)$.
This leads us to the conclusion that  
the main contribution to the overall drag is due to the steady term but the reduction of the mean settling velocity in a turbulent environment is almost entirely due to the various unsteady effects. These can be related to unsteady vortex shedding, see figure~\ref{fig:fields_rel}, as in the experiments of a single sphere by
\cite{mordant2000}
These observations are also in agreements with the results in \cite{homann2013}. These authors observe that the enhancement of the drag of a sphere towed in a turbulentt environment 
can be explained by the modification of the mean velocity and pressure profile, and thus of the boundary layer around the sphere, by the turbulent fluctuations.

\begin{table}
  \begin{center}
\def~{\hphantom{0}}
  \begin{tabular}{ccc}
      case  & $ \mathcal{S}(Re_p)$   &   $\Psi(t)$ \\[3pt]
      Quiescent $\phi=0.5\%$ & $99.5\%$  & $0.5\%$ \\
      Turbulent  $\phi=0.5\%$ & $90.4\%$  & $9.6\%$ \\
      Quiescent $\phi=1.0\%$ & $94.1\%$  & $5.9\%$ \\
      Turbulent  $\phi=1.0\%$ & $87.8\%$  & $12.2\%$ \\
  \end{tabular}
  \caption{Percentage contribution of the non-linear and unsteady terms for the quiescent and  
turbulent case with $\phi=0.5\%$. The data are normalized by the mean drag from the simulation of a single sphere in  quiescent fluid.}
  \label{tab:n_psi}
  \end{center}
\end{table}

%\appendix

\section{Final remarks}

We report numerical simulations of a suspension of rigid spherical slightly-heavy particles in a quiescent and turbulent environment using 
a direct-forcing immersed boundary method to capture the fluid-structure interactions.
Two dilute volume fractions, $\phi=0.5\%$ and 1\%, are investigated in quiescent fluid and homogeneous isotropic turbulence at $Re_\lambda=90$.
The particle diameter is of the order of the Taylor length scale and about 12 times the dissipative Kolmogorov scale. The ratio between the sedimentation velocity and the turbulent fluctuations is about 3.4, so that the strongest fluid-particle interactions occur at approximately  the Taylor  scale.

The choice of the parameters is inspired by the reduction in sedimentation velocity observed experimentally in a turbulent flow  by \cite{byron2015}  and in the group of Prof. Variano at UC Berkeley. In the experiment, the isotropic homogeneous turbulence is generated in a tank of  dimensions of several integral lengthscales by means of  two facing randomly-actuated synthetic jet arrays (driven stochastically). The Taylor microscale Reynolds number of the experiment is $Re_{\lambda}=260$.
Particle Image Velocimetry using Refractive-Index-Matched Hydrogel particles is used to measure the fluid velocity and the linear and angular velocities of  finite-size particles of diameter of about $1.4$ Taylor microscales and density ratios $\rho_p/\rho_f=1.02$, $1.006$ and 
$1.003$. 
The ratio between the terminal quiescent settling velocity $v_t$ and the turbulence fluctuating velocity $u'$ is about $1$,
higher than in our simulations where this ratio is $3.3$. \cite{byron2015} observes reductions of the slip velocity between  40\%  and 60 \% when varying the shape and density of the particles.
As suggested by \cite{byron2015}, the larger reduction in settling velocity observed in the experiments  is most likely explained by  the larger turbulence intensity. 

The new findings reported here can be summarized as follows: i) the reduction of settling velocity in a quiescent flow due to the hindering effect is reduced at finite inertia by pair-interactions, e.g. drafting-kissing-tumbling. ii) Owing to these particle-particle interactions, sedimentation in quiescent environment presents therefore significant intermittency. iii) The particle settling velocity is further reduced in a turbulent environment due to unsteady drag effects. iv) Vortex shedding and wake disruption is served also in subcritical conditions in an already turbulent flow.

In a quiescent environment, the mean settling velocity slightly decreases from the reference value pertaining few isolated particle
when the volume fraction  $\phi=0.5\%$ and  $\phi=1\%$. 
This limited reduction of the settling velocity with the volume fraction is in agreement with previous experimental findings in inertia-less and inertial flows.
The Archimedes number of our particle is 21000, in the steady vertical regime before  the occurrence of a first bifurcation to an asymmetric wake. In this regime, \cite{uhlmann2014} observe no significant particle clustering, which is is confirmed by the present data.

The skewness and flatness of the particle velocity reveal large positive values in a quiescent fluid,  and accordingly  the velocity probability distribution functions 
display evident positive tails. This indicates a highly intermittent behavior. In particular, it is most likely to see particles sedimenting at velocity significantly higher than the mean: this is 
caused by the close interactions between particle pairs (more seldom triplets). Particles approaching each other draft-kiss-tumble while falling 
faster than the average.

In a turbulent flow, the mean sedimentation velocity further reduces, to 0.88 and 0.86 at $\phi=0.5\%$ and $\phi=1\%$. 
The variance of both the linear and angular velocity increases in a turbulent environment and the single-particle time correlations decay faster due to the turbulence mixing. The velocity probability distribution function are almost symmetric and tend towards a Gaussian of corresponding variance. 
The particle lateral dispersion is, as expected, higher in a turbulent flow, whereas the vertical one is, surprisingly, of comparable magnitude in the two regimes; this can be explained  by the highly intermittent behavior observed in the quiescent fluid.

We compute the averaged relative velocity in the particle reference frame and  the fluctuations around the mean. We show that the wake behind each particle 
is on average significantly reduced in the turbulent flow and large-amplitude unsteady motions  appear on the side of the sphere in the regions of minimum 
pressure where vortex shedding is typically observed. The effect of a turbulent flow on the damping of the wake behind a rigid sphere has been 
discussed for example by \citet{bagchi2003}, while the case of a spherical bubble has been investigated by \citet{merle2005}.
Using the slip velocity between the particle and the fluid surrounding it, we estimate the nonlinear drag on each particle from empirical formulas  and quantify the relevance of non-stationary effects on the particle sedimentation.  We show that these become relevant  in the turbulent regime, amount to about 10-12\% of the total drag, and are responsible  for the reduction of settling velocity with the respect to the quiescent flow. This can be compared with the simulations  in \cite{good2014} who attribute the reduction of the sedimentation velocity of small ($2a < \eta$) heavy ($\rho_p/\rho_f \approx 900$) spherical particles in turbulence to the nonlinear drag. Here, we show that non-stationary effects become relevant for larger particles at lower density ratios.

The present investigation can be extended in a number of interesting directions. Preliminary simulations reveal that variations of the density ratio at constant Archimedes number do not significantly modify the results presented here. It would be therefore interesting to investigate the effect of the Galileo number on the particle dynamics and of the ratio between turbulent fluctuations and sedimentation velocity.

\begin{acknowledgments}
This work was supported by the European Research Council Grant No.\ ERC-2013-CoG-616186, TRITOS.
The authors acknowledge Prof. Variano for fruitful discussions and comments on the manuscript, computer time provided by SNIC (Swedish
National Infrastructure for Computing) and the support from the COST Action MP1305: \emph{Flowing matter}.
\end{acknowledgments}

%\bibliographystyle{jfm}

%\bibliography{jfm-instructions}

\begin{thebibliography}{61}
\expandafter\ifx\csname natexlab\endcsname\relax\def\natexlab#1{#1}\fi

\bibitem[Aliseda {\em et~al.\/}(2002)Aliseda, Cartellier, Hainaux \&
  Lasheras]{aliseda2002}
{\sc Aliseda, A, Cartellier, Alain, Hainaux, F \& Lasheras, Juan~C} 2002 Effect
  of preferential concentration on the settling velocity of heavy particles in
  homogeneous isotropic turbulence. {\em Journal of Fluid Mechanics\/} {\bf
  468}, 77--105.

\bibitem[Bagchi \& Balachandar(2003)]{bagchi2003}
{\sc Bagchi, Prosenjit \& Balachandar, S} 2003 Effect of turbulence on the drag
  and lift of a particle. {\em Physics of Fluids (1994-present)\/} {\bf
  15}~(11), 3496--3513.

\bibitem[Balachandar \& Eaton(2010)]{balach2010}
{\sc Balachandar, S \& Eaton, John~K} 2010 Turbulent dispersed multiphase flow.
  {\em Annual Review of Fluid Mechanics\/} {\bf 42}, 111--133.

\bibitem[Batchelor(1972)]{batchelor1972}
{\sc Batchelor, GK} 1972 Sedimentation in a dilute dispersion of spheres. {\em
  Journal of fluid mechanics\/} {\bf 52}~(02), 245--268.

\bibitem[Bec {\em et~al.\/}(2014)Bec, Homann \& Ray]{bec2014}
{\sc Bec, J{\'e}r{\'e}mie, Homann, Holger \& Ray, Samriddhi~Sankar} 2014
  Gravity-driven enhancement of heavy particle clustering in turbulent flow.
  {\em Physical review letters\/} {\bf 112}~(18), 184501.

\bibitem[Bellani \& Variano(2012)]{bellani2012}
{\sc Bellani, Gabriele \& Variano, Evan~A} 2012 Slip velocity of large
  neutrally buoyant particles in turbulent flows. {\em New Journal of
  Physics\/} {\bf 14}~(12), 125009.

\bibitem[Bergougnoux {\em et~al.\/}(2014)Bergougnoux, Bouchet, Lopez \&
  Guazzelli]{bergougnoux2014motion}
{\sc Bergougnoux, Laurence, Bouchet, Gilles, Lopez, Diego \& Guazzelli,
  Elisabeth} 2014 The motion of solid spherical particles falling in a cellular
  flow field at low stokes number. {\em Physics of Fluids (1994-present)\/}
  {\bf 26}~(9), 093302.

\bibitem[Bosse {\em et~al.\/}(2006)Bosse, Kleiser \& Meiburg]{bosse2006}
{\sc Bosse, Thorsten, Kleiser, Leonhard \& Meiburg, Eckart} 2006 Small
  particles in homogeneous turbulence: Settling velocity enhancement by two-way
  coupling. {\em Physics of Fluids (1994-present)\/} {\bf 18}~(2), 027102.

\bibitem[Bouchet {\em et~al.\/}(2006)Bouchet, Mebarek \&
  Du{\v{s}}ek]{bouchet2006}
{\sc Bouchet, G, Mebarek, M \& Du{\v{s}}ek, J} 2006 Hydrodynamic forces acting
  on a rigid fixed sphere in early transitional regimes. {\em European Journal
  of Mechanics-B/Fluids\/} {\bf 25}~(3), 321--336.

\bibitem[Brenner(1961)]{brenner1961}
{\sc Brenner, Howard} 1961 The slow motion of a sphere through a viscous fluid
  towards a plane surface. {\em Chemical Engineering Science\/} {\bf 16}~(3),
  242--251.

\bibitem[Breugem(2012)]{breugem2012}
{\sc Breugem, Wim-Paul} 2012 A second-order accurate immersed boundary method
  for fully resolved simulations of particle-laden flows. {\em Journal of
  Computational Physics\/} {\bf 231}~(13), 4469--4498.

\bibitem[Bush {\em et~al.\/}(2003)Bush, Thurber \& Blanchette]{bush2003}
{\sc Bush, John~WM, Thurber, BA \& Blanchette, F} 2003 Particle clouds in
  homogeneous and stratified environments. {\em Journal of Fluid Mechanics\/}
  {\bf 489}, 29--54.

\bibitem[Byron(2015)]{byron2015}
{\sc Byron, Margaret~L} 2015 The rotation and translation of non-spherical
  particles in homogeneous isotropic turbulence. {\em arXiv preprint
  arXiv:1506.00478\/} .

\bibitem[Ceccio(2010)]{ceccio2010}
{\sc Ceccio, Steven~L} 2010 Friction drag reduction of external flows with
  bubble and gas injection. {\em Annual Review of Fluid Mechanics\/} {\bf 42},
  183--203.

\bibitem[Cisse {\em et~al.\/}(2013)Cisse, Homann \& Bec]{cisse2013}
{\sc Cisse, Mamadou, Homann, Holger \& Bec, J{\'e}r{\'e}mie} 2013 Slipping
  motion of large neutrally buoyant particles in turbulence. {\em Journal of
  Fluid Mechanics\/} {\bf 735}, R1.

\bibitem[Climent \& Maxey(2003)]{climent2003}
{\sc Climent, E \& Maxey, MR} 2003 Numerical simulations of random suspensions
  at finite reynolds numbers. {\em International journal of multiphase flow\/}
  {\bf 29}~(4), 579--601.

\bibitem[Corrsin \& Lumley(1956)]{corrsin1956}
{\sc Corrsin, Se \& Lumley, J} 1956 On the equation of motion for a particle in
  turbulent fluid. {\em Applied Scientific Research\/} {\bf 6}~(2), 114--116.

\bibitem[Csanady(1963)]{csanady1963}
{\sc Csanady, GT} 1963 Turbulent diffusion of heavy particles in the
  atmosphere. {\em Journal of the Atmospheric Sciences\/} {\bf 20}~(3),
  201--208.

\bibitem[Di~Felice(1999)]{di1999}
{\sc Di~Felice, R} 1999 The sedimentation velocity of dilute suspensions of
  nearly monosized spheres. {\em International journal of multiphase flow\/}
  {\bf 25}~(4), 559--574.

\bibitem[Doostmohammadi \& Ardekani(2015)]{doost2015}
{\sc Doostmohammadi, A \& Ardekani, AM} 2015 Suspension of solid particles in a
  density stratified fluid. {\em Physics of Fluids (1994-present)\/} {\bf
  27}~(2), 023302.

\bibitem[Elgobashi(1991)]{elgo1991}
{\sc Elgobashi, S} 1991 Particle-laden turbulent flows: direct simulation and
  closure models. {\em Appl. Sci. Res\/} {\bf 48}~(3-4), 301--314.

\bibitem[Fortes {\em et~al.\/}(1987)Fortes, Joseph \& Lundgren]{fortes1987}
{\sc Fortes, Antonio~F, Joseph, Daniel~D \& Lundgren, Thomas~S} 1987 Nonlinear
  mechanics of fluidization of beds of spherical particles. {\em Journal of
  Fluid Mechanics\/} {\bf 177}, 467--483.

\bibitem[Garcia-Villalba {\em et~al.\/}(2012)Garcia-Villalba, Kidanemariam \&
  Uhlmann]{garcia2012}
{\sc Garcia-Villalba, Manuel, Kidanemariam, Aman~G \& Uhlmann, Markus} 2012 Dns
  of vertical plane channel flow with finite-size particles: Voronoi analysis,
  acceleration statistics and particle-conditioned averaging. {\em
  International Journal of Multiphase Flow\/} {\bf 46}, 54--74.

\bibitem[Garside \& Al-Dibouni(1977)]{garside1977}
{\sc Garside, John \& Al-Dibouni, Maan~R} 1977 Velocity-voidage relationships
  for fluidization and sedimentation in solid-liquid systems. {\em Industrial
  \& engineering chemistry process design and development\/} {\bf 16}~(2),
  206--214.

\bibitem[Good {\em et~al.\/}(2014)Good, Ireland, Bewley, Bodenschatz, Collins
  \& Warhaft]{good2014}
{\sc Good, GH, Ireland, PJ, Bewley, GP, Bodenschatz, E, Collins, LR \& Warhaft,
  Z} 2014 Settling regimes of inertial particles in isotropic turbulence. {\em
  Journal of Fluid Mechanics\/} {\bf 759}, R3.

\bibitem[Guazzelli \& Morris(2012)]{guazzelli2011}
{\sc Guazzelli, Elisabeth \& Morris, Jeffrey~F} 2012 {\em A physical
  introduction to suspension dynamics\/}, , vol.~45. Cambridge University
  Press.

\bibitem[Gustavsson {\em et~al.\/}(2014)Gustavsson, Vajedi \&
  Mehlig]{gustavsson2014}
{\sc Gustavsson, K, Vajedi, S \& Mehlig, B} 2014 Clustering of particles
  falling in a turbulent flow. {\em Physical Review Letters\/} {\bf 112}~(21),
  214501.

\bibitem[Hasimoto(1959)]{hasimoto1959}
{\sc Hasimoto, H} 1959 On the periodic fundamental solutions of the stokes
  equations and their application to viscous flow past a cubic array of
  spheres. {\em Journal of Fluid Mechanics\/} {\bf 5}~(02), 317--328.

\bibitem[Homann {\em et~al.\/}(2013)Homann, Bec \& Grauer]{homann2013}
{\sc Homann, Holger, Bec, J{\'e}r{\'e}mie \& Grauer, Rainer} 2013 Effect of
  turbulent fluctuations on the drag and lift forces on a towed sphere and its
  boundary layer. {\em Journal of Fluid Mechanics\/} {\bf 721}, 155--179.

\bibitem[Hwang \& Eaton(2006)]{hwang2006}
{\sc Hwang, Wontae \& Eaton, John~K} 2006 Homogeneous and isotropic turbulence
  modulation by small heavy ($ st \sim 50$) particles. {\em Journal of Fluid
  Mechanics\/} {\bf 564}, 361--393.

\bibitem[Johnson \& Tezduyar(1996)]{johnson1996}
{\sc Johnson, Andrew~A \& Tezduyar, Tayfun~E} 1996 Simulation of multiple
  spheres falling in a liquid-filled tube. {\em Computer Methods in Applied
  Mechanics and Engineering\/} {\bf 134}~(3), 351--373.

\bibitem[Kempe \& Fr{\"o}hlich(2012)]{kempe2012}
{\sc Kempe, Tobias \& Fr{\"o}hlich, Jochen} 2012 An improved immersed boundary
  method with direct forcing for the simulation of particle laden flows. {\em
  Journal of Computational Physics\/} {\bf 231}~(9), 3663--3684.

\bibitem[Ladd \& Verberg(2001)]{ladd2001}
{\sc Ladd, AJC \& Verberg, R} 2001 Lattice-boltzmann simulations of
  particle-fluid suspensions. {\em Journal of Statistical Physics\/} {\bf
  104}~(5-6), 1191--1251.

\bibitem[Ladd(1993)]{ladd1993}
{\sc Ladd, Anthony~JC} 1993 Dynamical simulations of sedimenting spheres. {\em
  Physics of Fluids A: Fluid Dynamics (1989-1993)\/} {\bf 5}~(2), 299--310.

\bibitem[Lambert {\em et~al.\/}(2013)Lambert, Picano, Breugem \&
  Brandt]{lambert2013}
{\sc Lambert, Ruth~A, Picano, Francesco, Breugem, Wim-Paul \& Brandt, Luca}
  2013 Active suspensions in thin films: nutrient uptake and swimmer motion.
  {\em Journal of Fluid Mechanics\/} {\bf 733}, 528--557.

\bibitem[Lashgari {\em et~al.\/}(2014)Lashgari, Picano, Breugem \&
  Brandt]{lashgari2014}
{\sc Lashgari, Iman, Picano, Francesco, Breugem, Wim-Paul \& Brandt, Luca} 2014
  Laminar, turbulent, and inertial shear-thickening regimes in channel flow of
  neutrally buoyant particle suspensions. {\em Physical review letters\/} {\bf
  113}~(25), 254502.

\bibitem[Lucci {\em et~al.\/}(2010)Lucci, Ferrante \& Elghobashi]{lucci2010}
{\sc Lucci, Francesco, Ferrante, Antonino \& Elghobashi, Said} 2010 Modulation
  of isotropic turbulence by particles of taylor length-scale size. {\em
  Journal of Fluid Mechanics\/} {\bf 650}, 5--55.

\bibitem[Maxey \& Riley(1983)]{maxey1983}
{\sc Maxey, Martin~R \& Riley, James~J} 1983 Equation of motion for a small
  rigid sphere in a nonuniform flow. {\em Physics of Fluids (1958-1988)\/} {\bf
  26}~(4), 883--889.

\bibitem[Merle {\em et~al.\/}(2005)Merle, Legendre \& Magnaudet]{merle2005}
{\sc Merle, Axel, Legendre, Dominique \& Magnaudet, Jacques} 2005 Forces on a
  high-reynolds-number spherical bubble in a turbulent flow. {\em Journal of
  Fluid Mechanics\/} {\bf 532}, 53--62.

\bibitem[Mordant \& Pinton(2000)]{mordant2000}
{\sc Mordant, N \& Pinton, J-F} 2000 Velocity measurement of a settling sphere.
  {\em The European Physical Journal B-Condensed Matter and Complex Systems\/}
  {\bf 18}~(2), 343--352.

\bibitem[Olivieri {\em et~al.\/}(2014)Olivieri, Picano, Sardina, Iudicone \&
  Brandt]{olivieri2014}
{\sc Olivieri, Stefano, Picano, Francesco, Sardina, Gaetano, Iudicone, Daniele
  \& Brandt, Luca} 2014 The effect of the basset history force on particle
  clustering in homogeneous and isotropic turbulence. {\em Physics of Fluids
  (1994-present)\/} {\bf 26}~(4), 041704.

\bibitem[Picano {\em et~al.\/}(2015)Picano, Breugem \& Brandt]{picano2015}
{\sc Picano, Francesco, Breugem, Wim-Paul \& Brandt, Luca} 2015 Turbulent
  channel flow of dense suspensions of neutrally buoyant spheres. {\em Journal
  of Fluid Mechanics\/} {\bf 764}, 463--487.

\bibitem[Pignatel {\em et~al.\/}(2011)Pignatel, Nicolas \&
  Guazzelli]{pignatel2011}
{\sc Pignatel, Florent, Nicolas, Maxime \& Guazzelli, Elisabeth} 2011 A falling
  cloud of particles at a small but finite reynolds number. {\em Journal of
  Fluid Mechanics\/} {\bf 671}, 34--51.

\bibitem[Pope(2000)]{pope2000}
{\sc Pope, Stephen~B} 2000 {\em Turbulent flows\/}. Cambridge university press.

\bibitem[Prosperetti(2015)]{AndreaFoF}
{\sc Prosperetti, Andrea} 2015 Life and death by boundary conditions. {\em
  Journal of Fluid Mechanics\/} {\bf 768}, 1--4.

\bibitem[Richardson \& Zaki(1954)]{richardson1954}
{\sc Richardson, JF \& Zaki, WN} 1954 The sedimentation of a suspension of
  uniform spheres under conditions of viscous flow. {\em Chemical Engineering
  Science\/} {\bf 3}~(2), 65--73.

\bibitem[Sangani \& Acrivos(1982)]{sangani1982}
{\sc Sangani, AS \& Acrivos, A} 1982 Slow flow past periodic arrays of
  cylinders with application to heat transfer. {\em International journal of
  Multiphase flow\/} {\bf 8}~(3), 193--206.

\bibitem[Siewert {\em et~al.\/}(2014)Siewert, Kunnen \&
  Schr{\"o}der]{siewert2014}
{\sc Siewert, C, Kunnen, RPJ \& Schr{\"o}der, W} 2014 Collision rates of small
  ellipsoids settling in turbulence. {\em Journal of Fluid Mechanics\/} {\bf
  758}, 686--701.

\bibitem[Sobral {\em et~al.\/}(2007)Sobral, Oliveira \& Cunha]{sobral2007}
{\sc Sobral, YD, Oliveira, TF \& Cunha, FR} 2007 On the unsteady forces during
  the motion of a sedimenting particle. {\em Powder Technology\/} {\bf
  178}~(2), 129--141.

\bibitem[Stout {\em et~al.\/}(1995)Stout, Arya \& Genikhovich]{stout1995}
{\sc Stout, JE, Arya, SP \& Genikhovich, EL} 1995 The effect of nonlinear drag
  on the motion and settling velocity of heavy particles. {\em Journal of the
  atmospheric sciences\/} {\bf 52}~(22), 3836--3848.

\bibitem[Sugiyama {\em et~al.\/}(2008)Sugiyama, Calzavarini \& Lohse]{sugi2008}
{\sc Sugiyama, Kazuyasu, Calzavarini, Enrico \& Lohse, Detlef} 2008 Microbubbly
  drag reduction in taylor--couette flow in the wavy vortex regime. {\em
  Journal of Fluid Mechanics\/} {\bf 608}, 21--41.

\bibitem[Tchen(1947)]{tchen1947}
{\sc Tchen, Chan-Mou} 1947 Mean value and correlation problems connected with
  the motion of small particles suspended in a turbulent fluid .

\bibitem[Toschi \& Bodenschatz(2009)]{toschi2009}
{\sc Toschi, Federico \& Bodenschatz, Eberhard} 2009 Lagrangian properties of
  particles in turbulence. {\em Annual Review of Fluid Mechanics\/} {\bf 41},
  375--404.

\bibitem[Tunstall \& Houghton(1968)]{tunstall1968}
{\sc Tunstall, EB \& Houghton, G} 1968 Retardation of falling spheres by
  hydrodynamic oscillations. {\em Chemical Engineering Science\/} {\bf 23}~(9),
  1067--1081.

\bibitem[Uhlmann(2005)]{uhlmann2005}
{\sc Uhlmann, Markus} 2005 An immersed boundary method with direct forcing for
  the simulation of particulate flows. {\em Journal of Computational Physics\/}
  {\bf 209}~(2), 448--476.

\bibitem[Uhlmann \& Doychev(2014)]{uhlmann2014}
{\sc Uhlmann, Markus \& Doychev, Todor} 2014 Sedimentation of a dilute
  suspension of rigid spheres at intermediate galileo numbers: the effect of
  clustering upon the particle motion. {\em Journal of Fluid Mechanics\/} {\bf
  752}, 310--348.

\bibitem[Vincent \& Meneguzzi(1991)]{vincent1991}
{\sc Vincent, A \& Meneguzzi, M} 1991 The satial structure and statistical
  properties of homogeneous turbulence. {\em Journal of Fluid Mechanics\/} {\bf
  225}, 1--20.

\bibitem[Wang \& Maxey(1993)]{wang1993}
{\sc Wang, Lian-Ping \& Maxey, Martin~R} 1993 Settling velocity and
  concentration distribution of heavy particles in homogeneous isotropic
  turbulence. {\em Journal of Fluid Mechanics\/} {\bf 256}, 27--68.

\bibitem[Yin \& Koch(2007)]{yin2007}
{\sc Yin, Xiaolong \& Koch, Donald~L} 2007 Hindered settling velocity and
  microstructure in suspensions of solid spheres with moderate reynolds
  numbers. {\em Physics of Fluids (1994-present)\/} {\bf 19}~(9), 093302.

\bibitem[Zhan {\em et~al.\/}(2014)Zhan, Sardina, Lushi \& Brandt]{zhan2014}
{\sc Zhan, Caijuan, Sardina, Gaetano, Lushi, Enkeleida \& Brandt, Luca} 2014
  Accumulation of motile elongated micro-organisms in turbulence. {\em Journal
  of Fluid Mechanics\/} {\bf 739}, 22--36.

\bibitem[Zhang \& Prosperetti(2010)]{zhang2010}
{\sc Zhang, Quan \& Prosperetti, Andrea} 2010 Physics-based analysis of the
  hydrodynamic stress in a fluid-particle system. {\em Physics of Fluids
  (1994-present)\/} {\bf 22}~(3), 033306.

\end{thebibliography}

\end{document}